\pdfoutput=1
\documentclass[runningheads]{llncs}
\usepackage[T1]{fontenc}

\usepackage[utf8]{inputenc}

\PassOptionsToPackage{hyphens}{url}%
\usepackage{hyperref}
\usepackage{graphicx}
\usepackage{xcolor}
\usepackage{makecell}
\usepackage{multirow,booktabs}

\usepackage{pifont}
\newcommand{\cmark}{\text{\ding{51}}}
\newcommand{\xmark}{\text{\ding{55}}}

\usepackage{algorithmicx}
\usepackage[noend]{algpseudocode}
\usepackage{algorithm}

\usepackage{IEEEtrantools}   %

\usepackage[shortlabels,inline]{enumitem}
\setlist[itemize]{leftmargin=1.25em}
\setlist[enumerate]{leftmargin=1.75em}

\usepackage{xspace}
\usepackage{xstring}
\usepackage{refstyle}

\usepackage{amsmath}

\usepackage{amsthm}
\usepackage{thm-restate}

\usepackage{amssymb}
\usepackage{amsfonts}
\allowdisplaybreaks

\usepackage{tikz}
\usetikzlibrary{calc}
\usetikzlibrary{arrows}
\usetikzlibrary{arrows.meta}
\usetikzlibrary{patterns}
\usetikzlibrary{positioning}
\usetikzlibrary{decorations.pathreplacing}
\usetikzlibrary{shapes.misc}
\usetikzlibrary{spy}

\usepackage{pgfplots}
\pgfplotsset{compat=1.17}

\usetikzlibrary{fillbetween}

\definecolor{myParula01Blue}{RGB}{0,114,189}
\definecolor{myParula02Orange}{RGB}{217,83,25}
\definecolor{myParula03Yellow}{RGB}{237,177,32}
\definecolor{myParula04Purple}{RGB}{126,47,142}
\definecolor{myParula05Green}{RGB}{119,172,48}
\definecolor{myParula06LightBlue}{RGB}{77,190,238}
\definecolor{myParula07Red}{RGB}{162,20,47}

\tikzset{myparula11/.style={color=myParula01Blue,solid,mark=+,mark options={solid}}}
\tikzset{myparula12/.style={color=myParula01Blue,densely dashed,mark=x,mark options={solid}}}
\tikzset{myparula13/.style={color=myParula01Blue,densely dotted,mark=o,mark options={solid}}}
\tikzset{myparula14/.style={color=myParula01Blue,dashdotted,mark=triangle,mark options={solid}}}
\tikzset{myparula15/.style={color=myParula01Blue,dashdotdotted,mark=square,mark options={solid}}}

\tikzset{myparula21/.style={color=myParula02Orange,solid,mark=+,mark options={solid}}}
\tikzset{myparula22/.style={color=myParula02Orange,densely dashed,mark=x,mark options={solid}}}
\tikzset{myparula23/.style={color=myParula02Orange,densely dotted,mark=o,mark options={solid}}}
\tikzset{myparula24/.style={color=myParula02Orange,dashdotted,mark=triangle,mark options={solid}}}
\tikzset{myparula25/.style={color=myParula02Orange,dashdotdotted,mark=square,mark options={solid}}}

\tikzset{myparula31/.style={color=myParula03Yellow,solid,mark=+,mark options={solid}}}
\tikzset{myparula32/.style={color=myParula03Yellow,densely dashed,mark=x,mark options={solid}}}
\tikzset{myparula33/.style={color=myParula03Yellow,densely dotted,mark=o,mark options={solid}}}
\tikzset{myparula34/.style={color=myParula03Yellow,dashdotted,mark=triangle,mark options={solid}}}
\tikzset{myparula35/.style={color=myParula03Yellow,dashdotdotted,mark=square,mark options={solid}}}

\tikzset{myparula41/.style={color=myParula04Purple,solid,mark=+,mark options={solid}}}
\tikzset{myparula42/.style={color=myParula04Purple,densely dashed,mark=x,mark options={solid}}}
\tikzset{myparula43/.style={color=myParula04Purple,densely dotted,mark=o,mark options={solid}}}
\tikzset{myparula44/.style={color=myParula04Purple,dashdotted,mark=triangle,mark options={solid}}}
\tikzset{myparula45/.style={color=myParula04Purple,dashdotdotted,mark=square,mark options={solid}}}

\tikzset{myparula51/.style={color=myParula05Green,solid,mark=+,mark options={solid}}}
\tikzset{myparula52/.style={color=myParula05Green,densely dashed,mark=x,mark options={solid}}}
\tikzset{myparula53/.style={color=myParula05Green,densely dotted,mark=o,mark options={solid}}}
\tikzset{myparula54/.style={color=myParula05Green,dashdotted,mark=triangle,mark options={solid}}}
\tikzset{myparula55/.style={color=myParula05Green,dashdotdotted,mark=square,mark options={solid}}}

\tikzset{myparula61/.style={color=myParula06LightBlue,solid,mark=+,mark options={solid}}}
\tikzset{myparula62/.style={color=myParula06LightBlue,densely dashed,mark=x,mark options={solid}}}
\tikzset{myparula63/.style={color=myParula06LightBlue,densely dotted,mark=o,mark options={solid}}}
\tikzset{myparula64/.style={color=myParula06LightBlue,dashdotted,mark=triangle,mark options={solid}}}
\tikzset{myparula65/.style={color=myParula06LightBlue,dashdotdotted,mark=square,mark options={solid}}}

\tikzset{myparula71/.style={color=myParula07Red,solid,mark=+,mark options={solid}}}
\tikzset{myparula72/.style={color=myParula07Red,densely dashed,mark=x,mark options={solid}}}
\tikzset{myparula73/.style={color=myParula07Red,densely dotted,mark=o,mark options={solid}}}
\tikzset{myparula74/.style={color=myParula07Red,dashdotted,mark=triangle,mark options={solid}}}
\tikzset{myparula75/.style={color=myParula07Red,dashdotdotted,mark=square,mark options={solid}}}

\definecolor{NiceRedColor}{HTML}{B1040E}
\definecolor{NiceBlueColor}{HTML}{006CB8}
\definecolor{NiceGreenColor}{HTML}{008566}

\definecolor{ColorMessageBuffering}{HTML}{620059}
\definecolor{ColorVoteExpiry}{HTML}{D1660F}

\newcommand{\cf}[0]{cf.\xspace}
\newcommand{\ie}[0]{\emph{i.e.}\xspace}
\newcommand{\eg}[0]{\emph{e.g.}\xspace}

\tikzset{blockchain/.style={
            x=1.25cm,
            y=1.25cm,
            node distance=0.5cm,
            block/.style = {
                    minimum width=0.75cm,
                    minimum height=0.75cm,
                    draw,
                    shade,
                    top color=white,
                    bottom color=black!10,
                },
            block-adv/.style = {
                    block,
                    bottom color=myParula07Red!50,
                    draw=myParula07Red!50!black,
                },
            block-hon/.style = {
                    block,
                    bottom color=myParula05Green!50,
                    draw=myParula05Green!50!black,
                },
            link/.style = {
                    -latex,
                },
            link-adv/.style = {
                    link,
                },
            link-hon/.style = {
                    link,
                },
        }
}

\newcommand{\GAT}[0]{\ensuremath{\mathsf{GAT}}}
\newcommand{\GST}[0]{\ensuremath{\mathsf{GST}}}

\newcommand{\thresholdVote}[0]{\ensuremath{\mathsf{thr}_{\mathrm{v}}}}
\newcommand{\thresholdBlock}[0]{\ensuremath{\mathsf{thr}_{\mathrm{b}}}}
\newcommand{\thresholdDUMMY}[0]{\ensuremath{\mathsf{thr}}}

\newcommand{\Thorizon}[0]{\ensuremath{T_{\mathrm{hor}}}}
\newcommand{\Tconf}[0]{\ensuremath{T_{\mathrm{conf}}}}
\newcommand{\Tafter}[0]{\ensuremath{T_\mathsf{sec}}}

\newcommand{\chain}[0]{\ensuremath{\mathsf{ch}}}
\newcommand{\chainava}[0]{\ensuremath{\chain_{\mathrm{ava}}}}
\newcommand{\chainacc}[0]{\ensuremath{\chain_{\mathrm{acc}}}}
\newcommand{\chainavaPARAM}[2]{\ensuremath{\chain^{#1}_{\mathrm{ava},#2}}}
\newcommand{\chainaccPARAM}[2]{\ensuremath{\chain^{#1}_{\mathrm{acc},#2}}}

\newcommand{\chainFast}[0]{\ensuremath{\chain_{\mathrm{fast}}}}

\newcommand{\textOurGhost}[0]{GHOST-Eph\xspace}
\newcommand{\funOurGhost}[0]{\ensuremath{\operatorname{\textsc{GHOST-Eph}}}}

\newcommand{\txs}[0]{\ensuremath{\mathsf{txs}}}
\newcommand{\id}[0]{\ensuremath{\mathsf{id}}}

\newcommand{\varBlock}[0]{\ensuremath{B}}
\newcommand{\varVote}[0]{\ensuremath{v}}
\newcommand{\varProposal}[0]{\ensuremath{P}}
\newcommand{\varPiece}[0]{\ensuremath{M}}
\newcommand{\varVoteBlock}[0]{\ensuremath{x}}

\newcommand{\varLotteryOpening}[0]{\ensuremath{\varrho}}

\newcommand{\V}[0]{\ensuremath{\mathcal{T}}}
\newcommand{\B}[0]{\ensuremath{\mathcal{B}}}

\newcommand{\ptrBlock}[1]{\ensuremath{{}^*[#1]}}

\newcommand{\Prob}[1]{\ensuremath{\operatorname{Pr}\left[#1\right]}}

\newcommand{\tx}[0]{\ensuremath{\mathsf{tx}}}

\newcommand{\Goldfish}[0]{\textsf{Goldfish}\xspace}

\newcommand{\piece}[0]{piece\xspace}
\newcommand{\pieces}[0]{pieces\xspace}

\newcommand{\Pieces}[0]{Pieces\xspace}

\newcommand{\proposal}[0]{proposal\xspace}
\newcommand{\proposals}[0]{proposals\xspace}

\newcommand{\bvset}[0]{bvset\xspace}

\newcommand{\bvtree}[0]{bvtree\xspace}
\newcommand{\Bvtree}[0]{Bvtree\xspace}
\newcommand{\bvtrees}[0]{bvtrees\xspace}

\newcommand{\phasePropose}[0]{\textsc{Propose}\xspace}
\newcommand{\phaseVote}[0]{\textsc{Vote}\xspace}
\newcommand{\phaseConfirm}[0]{\textsc{Confirm}\xspace}
\newcommand{\phaseFastConfirm}[0]{\textsc{Fast-Confirm}\xspace}

\newcommand{\tagGeneric}[0]{tag\xspace}
\newcommand{\tagBlock}[0]{\ensuremath{\mathtt{block}}}
\newcommand{\tagVote}[0]{\ensuremath{\mathtt{vote}}}
\newcommand{\tagProposal}[0]{\ensuremath{\mathtt{propose}}}
\newcommand{\tagPiece}[0]{\ensuremath{\mathtt{piece}}}
\newcommand{\tagLotteryBlock}[0]{\ensuremath{\mathtt{block}}}
\newcommand{\tagLotteryVote}[0]{\ensuremath{\mathtt{vote}}}
\newcommand{\tagLotteryDUMMY}[0]{\ensuremath{\mathtt{tag}}}

\newcommand{\concat}{\,\|\,}
\DeclareMathOperator*{\argmax}{arg\,max}
\DeclareMathOperator*{\argmin}{arg\,min}
\newcommand{\negl}[0]{\ensuremath{\operatorname{negl}}}
\newcommand{\poly}[0]{\ensuremath{\operatorname{poly}}}

\newcommand{\Merge}[0]{\ensuremath{\operatorname{\textsc{Merge}}}}
\newcommand{\isValidVote}[0]{\ensuremath{\operatorname{Valid}}}
\newcommand{\isValidBlock}[0]{\ensuremath{\operatorname{Valid}}}
\newcommand{\isValidPiece}[0]{\ensuremath{\operatorname{Valid}}}
\newcommand{\isValidProposal}[0]{\ensuremath{\operatorname{Valid}}}
\newcommand{\isConsistent}[0]{\ensuremath{\operatorname{Consistent}}}

\newcommand{\lotteryOpen}[0]{\ensuremath{\operatorname{Open}}}
\newcommand{\lotteryIsWinningTicket}[0]{\ensuremath{\operatorname{Wins}}}
\newcommand{\lotteryPriority}[0]{\ensuremath{\operatorname{Prio}}}

\newcommand{\TRUE}[0]{\ensuremath{\mathtt{1}}}
\newcommand{\FALSE}[0]{\ensuremath{\mathtt{0}}}

\newcommand{\sigSCHEME}[0]{\ensuremath{\mathsf{Sig}}}
\newcommand{\sigSk}[0]{\ensuremath{\mathsf{ssk}}}
\newcommand{\sigPk}[0]{\ensuremath{\mathsf{spk}}}
\newcommand{\sigGen}[0]{\ensuremath{\sigSCHEME.\mathsf{Gen}}}
\newcommand{\sigSign}[0]{\ensuremath{\sigSCHEME.\mathsf{Sign}}}
\newcommand{\sigVerify}[0]{\ensuremath{\sigSCHEME.\mathsf{Verify}}}
\newcommand{\sigGenBLANK}[0]{\ensuremath{\mathsf{Gen}}}
\newcommand{\sigSignBLANK}[0]{\ensuremath{\mathsf{Sign}}}
\newcommand{\sigVerifyBLANK}[0]{\ensuremath{\mathsf{Verify}}}

\newcommand{\vrfSCHEME}[0]{\ensuremath{\mathsf{Vrf}}}
\newcommand{\vrfSk}[0]{\ensuremath{\mathsf{vsk}}}
\newcommand{\vrfPk}[0]{\ensuremath{\mathsf{vpk}}}
\newcommand{\vrfGen}[0]{\ensuremath{\vrfSCHEME.\mathsf{Gen}}}
\newcommand{\vrfProve}[0]{\ensuremath{\vrfSCHEME.\mathsf{Eval}}}
\newcommand{\vrfVerify}[0]{\ensuremath{\vrfSCHEME.\mathsf{Verify}}}
\newcommand{\vrfGenBLANK}[0]{\ensuremath{\mathsf{Gen}}}
\newcommand{\vrfProveBLANK}[0]{\ensuremath{\mathsf{Eval}}}
\newcommand{\vrfVerifyBLANK}[0]{\ensuremath{\mathsf{Verify}}}

\newcommand{\hashBlock}[0]{\ensuremath{H}}

\algnewcommand{\CommentLine}[1]{\State $\triangleright$ #1}

\algnewcommand{\algorithmicat}{\textbf{at}}
\algdef{SE}[AT]{At}{EndAt}[1]{\algorithmicat\ $#1$\ \algorithmicdo}{\algorithmicend\ \algorithmicat}%
\algtext*{EndAt}%

\algnewcommand{\algorithmicforever}{\textbf{forever}}
\algdef{SE}[FOREVER]{Forever}{EndForever}[0]{\algorithmicforever\ \algorithmicdo}{\algorithmicend\ \algorithmicforever}%
\algtext*{EndForever}%

\algnewcommand{\Let}[2]{\State $#1 \gets #2$}
\algnewcommand{\Broadcast}[1]{\State Broadcast $#1$}
\algnewcommand{\Sleep}[1]{\State Sleep for $#1$ rounds}

\newcommand{\Theal}[0]{\ensuremath{T_\mathsf{heal}}}
\newcommand{\Tbft}[0]{\ensuremath{T_\mathsf{bft}}}
\newcommand{\Tcheckpoint}[0]{\ensuremath{T_{\mathrm{chkpt}}}}
\newcommand{\Tupper}[0]{\ensuremath{T_{\mathrm{upper}}}}
\newcommand{\Ttimeout}[0]{\ensuremath{T_{\mathrm{tmout}}}}

\newcommand{\LOGbft}[2]{%
\ifthenelse{\equal{#1}{}}{%
\ensuremath{\mathsf{LOG}_{\mathrm{bft}}^{#2}}%
}{%
\ensuremath{\mathsf{LOG}_{\mathrm{bft},#1}^{#2}}%
}%
}

\newcommand{\ld}[1]{%
    \ifthenelse{\equal{#1}{}}{%
        \ensuremath{\mathrm{L}^{(c)}}%
    }{%
        \ensuremath{\mathrm{L}^{(#1)}}%
    }%
}

\newcommand{\bprop}[1]{%
    \ifthenelse{\equal{#1}{}}{%
        \ensuremath{\Hat{b}}%
    }{%
        \ensuremath{\Hat{b}_{#1}}%
    }%
}
\newcommand{\Trecent}[0]{\ensuremath{T_{\mathrm{rcnt}}}}

\newcommand{\scriptspacing}[0]{\medmuskip=0mu\thickmuskip=0mu}

\pgfplotsset{
    mysimpleplot/.style = {
            every axis plot/.prefix style={thick},
            width=1.0\linewidth,
            height=0.75\linewidth,
            title style={font=\scriptsize,align=center},
            legend cell align=left,
            legend style={font=\scriptsize},
            legend columns=3,
            legend style={
                    at={(0.5,1)},
                    yshift=0.3em,
                    anchor=south,
                    draw=none,
                    /tikz/every even column/.append style={
                            column sep=0.3em
                        },
                    cells={
                            align=left
                        }
                },
            grid=both,
            minor tick num=4,
            major grid style={solid, thin, draw=black!15},
            minor grid style={solid, ultra thin,draw=black!15},
            label style={font=\scriptsize,align=center},
            tick label style={font=\scriptsize},
        },
}

\newcommand*\circledScriptBlackSlim[1]{\tikz[baseline=(char.base)]{
        \node[shape=circle,draw=none,fill=black,text=white,inner sep=1pt,minimum size=9pt] (char) {\scriptsize #1};}}

\newcommand{\Wp}[0]{\ensuremath{W_{\mathrm{p}}}}

\usepackage{color}

\title{\Goldfish: No More Attacks on Ethereum?!}

\newcommand{\gitSourceUrl}[0]{\url{https://github.com/tse-group/goldfish-experiments}}

\newcommand{\fullVersionRef}[1]{#1}

\author{%
Francesco D'Amato\inst{1}%
\and%
Joachim Neu\inst{2}%
\and%
Ertem Nusret Tas\inst{2}%
\and%
David Tse\inst{2}%
}%
\institute{%
Ethereum Foundation\\%
\email{francesco.damato@ethereum.org}%
\and%
Stanford University\\%
\email{\{jneu,nusret,dntse\}@stanford.edu}%
}%

\usepackage{cite}

\newcommand{\myparagraph}[1]{\smallskip\noindent\textbf{#1.}~}
\newcommand{\myparagraphlight}[1]{\smallskip\noindent\emph{#1.}~}

\setlist[itemize]{topsep=0.25em}
\setlist[enumerate]{topsep=0.25em}

\begin{document}
\maketitle
\begingroup%
\renewcommand\thefootnote{1,2}%
\footnotetext{The authors are listed alphabetically. Contact authors: FD, JN, ENT.}%
\endgroup%
\begin{abstract}
  The LMD GHOST consensus protocol is a critical component of proof-of-stake Ethereum.
  In its current form, this protocol is brittle, as evidenced by recent attacks and patching attempts.
  We propose \Goldfish, a new protocol that satisfies key properties required of a drop-in replacement for LMD GHOST:
  \Goldfish is \emph{secure} in the sleepy model, assuming a majority of the validators follows the protocol.
  \Goldfish is \emph{reorg resilient} so that honestly produced blocks are guaranteed inclusion in the ledger, and it supports \emph{fast confirmation} with expected confirmation latency independent of the desired security level.
  Subsampling validators can improve the \emph{communication efficiency} of \Goldfish, and \Goldfish is \emph{composable with finality/accountability gadgets}.
  Crucially, \Goldfish is structurally similar to LMD GHOST, providing a credible path to adoption in Ethereum.
  Attacks on LMD GHOST exploit lack of coordination among honest validators, typically provided by a locking mechanism in classical BFT protocols.
  However, locking requires votes from a quorum of all participants and is not compatible with fluctuating participation.
  \Goldfish is powered by a novel coordination mechanism to synchronize the honest validators' actions.
  Experiments with our prototype implementation of \Goldfish suggest practicality.
\end{abstract}

\section{Introduction}
\label{sec:introduction}

\vspace{-\smallskipamount}
\myparagraph{Ethereum's Consensus Protocol}
\begin{figure}[t]
    \centering%
    \begin{tikzpicture}[x=1.5cm,y=0.62cm]
    \scriptsize

    \node (txs) at (-1.75,0) {$\txs$};

    \node [draw,align=center,minimum width=2cm,minimum height=0.8cm] (lmdghost) at (-0.3,0) {\emph{Underlay:}\\LMD GHOST};

    \coordinate (logavailable) at (1,0) {};

    \node [draw,align=center,minimum width=2cm,minimum height=0.8cm] (casperffg) at (2.3,0) {\emph{Overlay:}\\Casper FFG};

    \coordinate (logaccountable) at (3.5,0) {};
    \coordinate (logaccountableOUT) at (5,0) {};

    \draw [-Latex] (txs) -- (lmdghost);
    \draw [-Latex] (lmdghost) -- (logavailable) node [above,pos=1,yshift=1pt] {Chain tip};
    \draw [-Latex] (logavailable) -- (casperffg);
    \draw [-Latex] (casperffg) -- (logaccountable);
    \draw [-Latex] (logaccountable) -- (logaccountableOUT) node [pos=1,above,align=right,anchor=south east,yshift=1pt] {Finalizations};
    \draw [-Latex] (logaccountable) -- ++(0,1.5) -- ++(-3.8,0) -- (lmdghost) node [pos=0.35,right] {Justifications};
    \draw [-Latex] (logavailable) -- ++(0,-0.9) -- ++(2,0) -- ++(2,0) node [pos=1,above,align=right,anchor=south east,yshift=1pt] {Chain tip};
\end{tikzpicture}%
%
    \vspace{-1em}%
    \caption{%
        Gasper \cite{gasper} consists of two sub-protocols:
        LMD GHOST (`fork choice rule') and Casper FFG \cite{casper} (`finality gadget').
        The desiderata for
        Gasper
        were formalized by
        \emph{ebb-and-flow}~\cite{ebbandflow,aadilemma,sankagiri_clc},
        which
        consists of
        security of the full chain
        under dynamic participation of validators,
        and accountable security
        of a finalized prefix.
    }%
    \label{fig:gasper-diagram}%
\end{figure}
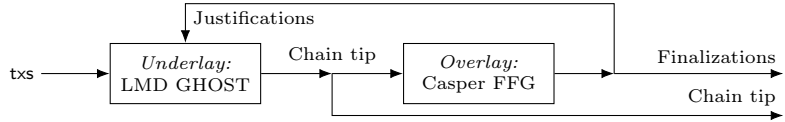
Ethereum's proof-of-stake (PoS) Byzantine fault tolerant (BFT) consensus protocol
(Gasper~\cite{gasper}, \figref{gasper-diagram})
consists of an \emph{overlay} finality gadget (Casper FFG~\cite{casper})
which provides safety under asynchrony, on top of an \emph{underlay}
chain (LMD GHOST, Latest Message Driven Greedy Heaviest Observed Sub-Tree~\cite{ghost,vb-blog-lmdghost})
which should be secure under synchrony and dynamic participation.
    Importantly,
    \emph{dynamic participation} here
    refers to the sleepy model~\cite{sleepy}
    for a large number of \emph{unexpected temporary crash faults},
    not to, for instance, stake shift.
This design works around the impossibility~\cite{cap,lewispye2020resource,ebbandflow,aadilemma,sankagiri_clc,accimplfin}
of having a single ledger
that is secure under both asynchrony and dynamic participation.
It is crucial that the underlay is both safe and live in the sleepy model,
because earlier works have shown~\cite{ethresearch-bouncing-attack,ethresearch-bouncing-attack-analysis,ebbandflow,aadilemma,sankagiri_clc} that otherwise
the whole protocol (\ie, underlay and overlay)
can stall indefinitely.
In particular, the overlay can only `checkpoint' transactions that are stable in the underlay.
If the underlay stalls (not live) or
keeps switching between different chains (not safe),
then the overlay
won't
make progress.
Thus, the underlay is not ``just some optional optimistic path'',
but on the critical path for any transaction to get confirmed.
Also, from a practical point of view,
overlay confirmation is typically slow
(\eg, min.\ $12\,\text{min}$ delay to finality in Ethereum),
while the underlay generates a block every few seconds.
As a result, in Ethereum today, most users don't wait for
Casper overlay confirmation, but de-facto already consider a transaction confirmed
when it enters the tip of the LMD GHOST underlay chain.
If the underlay does not provide at least some non-trivial safety guarantee,
transactions can be reverted
that have not yet been `checkpointed' by the overlay,
especially when overlay confirmation is delayed
due to many unexpected crash faults,
\eg, as happened on Ethereum mainnet in May 2023~\cite{finality_outage_coindesk,finality_outage_prysm}.

\myparagraph{Attacks and Patches for LMD GHOST}
But, LMD GHOST (\cf \figref{lmd-ghost})
is not secure in the sleepy model.
The initial version of LMD GHOST~\cite{gasper}
is
susceptible to the \emph{balancing attack}~\cite{ebbandflow,3attacks}.
In the attack, the adversary exploits the lack of a \emph{coordination mechanism} for synchronizing the views of honest validators; so that different validators vote for conflicting blocks at each slot, and the network fails to reach consensus indefinitely.
In response,
a patch called \emph{proposer boosting}
was added~\cite{mitigationlmdghostbalancingattacks}.
Proposer boosting gives a current proposal a temporary extra weight
in fork-choice.
This was supposed to coordinate voters towards honest proposals and break the balance.
However,
the LMD provision alone
can be exploited
to conduct a balancing-type attack
despite
boosting
\cite{2attacks},
and
LMD GHOST
without LMD would suffer
from a so called
\emph{avalanche attack}
\cite{2attacks}.
Again in response, a patch called \emph{equivocation discounting}
was added to the protocol.
Not least because of its complexity,
the protocol with these patches has so far defied security analysis---both in terms of giving a security proof
and further attacks.
Certainly, the extra weight from proposer boosting
gives an adversary much control over the chain,
especially when the number of votes is low.
Thus, proposer boosting renders LMD GHOST insecure
under dynamic participation.
\myparagraph{Quest for a Coordination Mechanism}
PBFT-style protocols~\cite{pbft,tendermint,streamlet,yin2018hotstuff}
coordinate validators using
locking and \emph{absolute quorums},
\ie, sets of votes from a large fraction of 
\emph{all}
validators.
However, absolute quorums cannot be reached
in the presence of many unexpected crash faults.
Thus, such protocols \emph{don't satisfy
    liveness under dynamic participation}
in the sleepy model,
which
is crucial
to withstand unforeseen
regulatory changes, or soft-/hardware failures or upgrades~\cite{ebbandflow}.
Tolerating dynamic participation is indeed one reason why LMD GHOST avoids absolute quorums and selects \emph{relatively} heavier blocks.

    Conversely,
PoS variants of Nakamoto's \emph{longest chain} (LC) protocol~\cite{kiayias2017ouroboros,david2018ouroboros,badertscher2018ouroboros,sleepy,snowwhite}
are secure in
the sleepy model~\cite{sleepy}.
In LC,
validators continuously build blocks extending the (relatively) longest chain.
Only after a while, honest validators reach coordination on a chain prefix.
Unfortunately, this entails \emph{slow confirmation}\footnote{Confirmation latency denotes the delay for a transaction to enter the ledgers of all validators.
    It is a random variable that depends on the sequence of block proposers.%
},
\ie,
expected confirmation latency linear in the
desired consensus failure probability.
Moreover,
in LC,
honestly produced blocks can be displaced (\emph{reorg'ed})
by adversaries~\cite{selfishmining},
and
validators have incentives to do so~\cite{mev-valuable,whaletx,bitcoininstability,flashboys}.

\myparagraph{Key Techniques of \Goldfish}
For \Goldfish's overall structure, failure
of LC protocols to satisfy reorg resilience and fast confirmation due to
``too few votes spread across too much time''
suggests
to employ a committee of voters that can create many votes supporting honest proposals soon after they are broadcast,
similar to committees in PBFT-style protocols.
However, as absolute quorums 
of PBFT-style protocols
are
incompatible with liveness under dynamic participation, rather than using the absolute number of votes,
a protocol for dynamic participation must use their relative weights to favor blocks with stronger support during fork choice,
similar to Nakamoto's longest-chain rule.
Together, these observations
vindicate
some structural elements of LMD GHOST
and suggest to retain them in
\Goldfish:
a succession of slots
with a proposer and a committee that votes,
all based on
blocks' relative vote weights
like in the GHOST rule.
    Key to
    \Goldfish is a novel coordination mechanism
    for honest voters to rally behind honest proposals.
    Unlike aforementioned mechanisms,
    this mechanism is secure in the sleepy model
    while also allowing for reorg resilience
    and fast confirmations.
    It is based on
    two techniques
    not commonly found in the literature:
\textbullet\hspace{0.3em}\textbf{Message buffering}\footnote{Message buffering was
        also called
        `view-merge' in a blog post~\cite{ethresearch-viewmerge} by one of the authors.
        We
        later noticed that a similar technique was used before
        in
        the unpublished Highway protocol~\cite{highway}.
        Message buffering (\cf \algref{protocol}) also bears some conceptual resemblance to
            the view-change sub-protocol of Sync~HotStuff~\cite[Fig.\ 2]{synchotstuff}.%
    } means each validator buffers votes received from the network and carefully times the inclusion of these votes into its local view, with priority given to votes relayed by the proposer.
Conceptually, a validator echoes received votes
    but processes them only after some time or as soon as the proposer relays them.
This ensures that in slots with an honest proposer, all honest validators adopt the view of the proposer and thus vote
for its proposed block.
\textbullet\hspace{0.3em}\textbf{Vote expiry}
means that during each slot, only votes from the immediately preceding slot influence honest validators' behavior.%
\footnote{Alleged forgetfulness of its animal namesake inspired \Goldfish's name.}
As a result, if in some slot all honest validators (which are assumed to outnumber adversary validators) vote to support a block (\ie, vote for the block or one of its descendants), then all honest validators will again vote to support that block in the next slot.

    Together, these two techniques allow for a simple inductive security argument:
    Because proposers are selected randomly among the majority-honest validators,
    slots with an honest proposer are frequent.
    By message buffering, all honest voters vote for the block proposed in such a slot (base case).
    By vote expiry, honest voters keep reaffirming this vote in perpetuity (induction step).
    Thus, honest proposals are guaranteed to remain in the canonical chain, implying reorg resilience.
    Since honest proposals contain fresh transactions and stabilize their prefix, and long streaks of adversary proposers are exponentially unlikely,
    liveness and safety
    follow readily.
    Details are provided
    in \secref{protocol-goldfish,analysis}.

A complementary perspective is that, conceptually,
    message buffering thwarts balancing-type attacks~\cite{ebbandflow,3attacks}
    (it ensures that honest voters rally behind honest proposals),
    while vote expiry thwarts avalanche-type attacks~\cite{2attacks}
    (it ensures that the adversary cannot reveal votes from long past slots---where the respective voter may still have been honest even).
    In this regard, LMD has a similar effect as vote expiry~\cite{2attacks},
    but LMD 
    does not recover reorg resilience under dynamic participation, and
    is rendered ineffective
    entirely
    if validators are subsampled
    to form small voter committees
    for
    reduced communication complexity.
    In contrast, vote expiry allows subsampling.
    With expiry and subsampling,
    vote expiry drastically reduces the number of votes
    validators need to buffer and consider at any point,
    greatly contributing to
    message buffering's practicality
    (see \secref{experiments}).
\myparagraph{\Goldfish's Contributions}
\Goldfish achieves the aforementioned desiderata:
\begin{enumerate*}[(a)]
    \item \Goldfish is \emph{provably secure}, \ie, safe and live, under \emph{dynamic participation}
        in the sleepy model
    assuming an honest majority of validators, and adversary network delay up to a known upper bound $\Delta$.

    \item \Goldfish is \emph{reorg resilient}, \ie, honest proposals
    eventually enter the ledger,
    with the proposal's prefix as determined at the time of block production 
    (thus, no selfish mining~\cite{selfishmining}).

    \item \Goldfish satisfies \emph{optimistic fast confirmation}: under optimistic conditions, \ie, when participation is high and $\frac{3}{4}$ fraction of validators are honest, it confirms transactions with constant expected latency independent of the consensus failure probability.
\end{enumerate*}

Additionally,
\Goldfish supports \emph{subsampling} of validators,
    which
    reduces communication and achieves resilience to adaptive corruption,
    since randomly selected validators send only a single protocol message
    (\cf player-replaceability~\cite{gilad2017algorand,algorand,DBLP:journals/iacr/ShengWNKV22}).
\Goldfish is also composable with \emph{finality and accountability gadgets} such as \cite{casper,ebbandflow,sankagiri_clc,aadilemma}.
This means it can indeed be used as a dynamically available underlay in conjunction with an overlay (\cf \figref{gasper-diagram}) that preserves safety
under asynchrony.
Since the construction and its security proof mostly reuses
techniques from~\cite{ebbandflow,sankagiri_clc,aadilemma},
we provide it in \secref{aa-analysis},
and focus in the following on \Goldfish's standalone behavior in the sleepy model assuming synchrony.

Crucially,
\Goldfish is intentionally simple, and similar to LMD GHOST as currently deployed in Ethereum,
to provide a credible path for adoption
(compare \figref{lmd-ghost,protocol-horizon,fast-confirmation-horizon}).
Message buffering and vote expiry can be realized with modest changes to the existing vote accounting logic of LMD GHOST.
\Goldfish is the first positive result (security proof) for a
variant of LMD GHOST,
strengthening confidence in this protocol family.
Simplicity of \Goldfish also makes it a good pedagogical example as a feature-rich consensus protocol for the sleepy model.

\myparagraph{Related Works}
\begin{table*}[t]
    \caption{%
        Comparison of \Goldfish with related works regarding key desiderata.
        Optimistic (`\textup{opt.}') fast confirmation requires high participation and less than $\frac{1}{4}$ adversary fraction.
        Dynamic participation
        `\textup{\textcolor{myParula05Green}{\cmark{}} (slow)}'
        indicates
        the protocol remains live only under slow fluctuations in
        participation.
        A number next to
        `\textcolor{myParula05Green}{\cmark{}}'
        for
        fast confirmation
        denotes the minimum confirmation latency.
        Responsive (`\textup{resp.}') confirmation
        means with delay of the actual network delay
        rather than delay bound $\Delta$.
    }%
    \vspace{-0.5em}%
    \centering
    \setlength{\tabcolsep}{2pt}
    \scriptsize
    \hspace*{-2.5cm}
    \begin{tabular}{l|ccccccccc|cc}
        \toprule
         & \multicolumn{9}{c|}{PoS/permissioned BFT consensus protocols ... for the sleepy model~\cite{sleepy}}
         & \multicolumn{2}{c}{... for other models}
        \\
        \midrule
         & \makecell{LC \cite{badertscher2018ouroboros}                                                         \\ \cite{snowwhite,prism,parallel}
        }
         & \makecell{Thunder-                                                                                   \\ella \cite{thunderella}}
         & \makecell{KW21                                                                                       \\\cite{wattenhoferDA}}
         & \makecell{GLR21                                                                                      \\\cite{goyalDA}}
         & \makecell{MR22                                                                                       \\\cite{lingDA}}
         & \makecell{MMR22                                                                                      \\\cite{malkhiDA_third_eprint}}
         & \makecell{GL23\textsuperscript{\textdagger}                                                          \\\cite{losaDA}}
         & \makecell{MMR23\textsuperscript{\textdagger}                                                         \\\cite{malkhiDA_half_eprint}}
         & \makecell{\Goldfish                                                                                  \\(this work)}
         & \makecell{PBFT-style
        \\\cite{pbft,tendermint,yin2018hotstuff,streamlet}}
         & \makecell{Highway                                                                                    \\\cite{highway}} \\
        \midrule
        Dynamic participation
         & \textcolor{myParula05Green}{\cmark{}}
         & \textcolor{myParula05Green}{\cmark{}} (slow)
         & \textcolor{myParula05Green}{\cmark{}} (slow)
         & \textcolor{myParula05Green}{\cmark{}}
         & \textcolor{myParula05Green}{\cmark{}}
         & \textcolor{myParula05Green}{\cmark{}}
         & \textcolor{myParula05Green}{\cmark{}}
         & \textcolor{myParula05Green}{\cmark{}}
         & \textcolor{myParula05Green}{\cmark{}}
         & \textcolor{myParula07Red}{\xmark{}}
         & \textcolor{myParula07Red}{\xmark{}}
        \\
        Reorg resilience
         & \textcolor{myParula07Red}{\xmark{}}
         & \textcolor{myParula07Red}{\xmark{}}
         & \textcolor{myParula05Green}{\cmark{}}
         & \textcolor{myParula05Green}{\cmark{}}
         & \textcolor{myParula05Green}{\cmark{}}
         & \textcolor{myParula05Green}{\cmark{}}
         & \textcolor{myParula05Green}{\cmark{}}
         & \textcolor{myParula05Green}{\cmark{}}
         & \textcolor{myParula05Green}{\cmark{}}
         & \textcolor{myParula05Green}{\cmark{}}
         & \textcolor{myParula05Green}{\cmark{}}
        \\
        Adversary resilience
         & $1/2$
         & $1/2$
         & $1/3$
         & $1/2$
         & $1/2$
         & $1/3$
         & $1/2$
         & $1/2$
         & $1/2$
         & $1/3$
         & flexible
        \\
        Fast confirmation
         & \textcolor{myParula07Red}{\xmark{}}
         & opt. (resp.)
         & \textcolor{myParula05Green}{\cmark{}}
         & \textcolor{myParula07Red}{\xmark{}}
         & \textcolor{myParula05Green}{\cmark{}} ($37\Delta$)
         & \textcolor{myParula05Green}{\cmark{}} ($3\Delta$)
         & \textcolor{myParula05Green}{\cmark{}} ($10\Delta$)
         & \textcolor{myParula05Green}{\cmark{}} ($4\Delta$)
         & opt. ($4\Delta$)
         & \textcolor{myParula05Green}{\cmark{}} (resp.)
         & \textcolor{myParula05Green}{\cmark{}}
        \\
        Similar to LMD GHOST
         & \textcolor{myParula07Red}{\xmark{}}
         & \textcolor{myParula07Red}{\xmark{}}
         & \textcolor{myParula07Red}{\xmark{}}
         & \textcolor{myParula07Red}{\xmark{}}
         & \textcolor{myParula07Red}{\xmark{}}
         & \textcolor{myParula07Red}{\xmark{}}
         & \textcolor{myParula07Red}{\xmark{}}
         & \textcolor{myParula07Red}{\xmark{}}
         & \textcolor{myParula05Green}{\cmark{}}
         & \textcolor{myParula07Red}{\xmark{}}
         & \textcolor{myParula05Green}{\cmark{}}
        \\
        \bottomrule
    \end{tabular}
    \textsuperscript{\textdagger}
    Appeared 
    in preprint
    after completion of \Goldfish~\cite{goldfishfull}.
    \label{tab:comp}
\end{table*}
The first secure consensus protocol for the sleepy model \cite{sleepy}
was Nakamoto's LC protocol,
first based on proof-of-work (PoW)~\cite{nakamoto_paper,backbone}, and subsequently on PoS~\cite{kiayias2017ouroboros,david2018ouroboros,badertscher2018ouroboros,sleepy,snowwhite} (see \tabref{comp} for a comparison of \Goldfish with related works).
Parallel composition of LC protocol instances
was suggested in \cite{prism,parallel} to overcome the scaling of LC protocols' confirmation latency with the security parameter $\kappa$.
For the same goal, Thunderella~\cite{thunderella} proposed combining a PBFT-style protocol achieving optimistic fast confirmation
with a slow LC protocol for when the adversary fraction is high.
(A similar idea was explored in Zyzzyva~\cite{zyzzyva,zyzzyva_attack}, where validators run a PBFT-style protocol, but optimistically confirm the primary's ordering.
Zyzzyva is not dynamically available.)
However, as observed above, LC protocols and Thunderella that builds on an LC protocol are not reorg resilient.
Moreover, under optimistic conditions, Thunderella recovers fast confirmation only after a period of LC confirmation delay, whereas \Goldfish can instantaneously resume fast-confirming.
Many classical PBFT-style consensus protocols \cite{pbft,yin2018hotstuff,streamlet} have constant (expected) confirmation latency and can be reorg resilient, but don't tolerate dynamic participation.
Highway \cite{highway}
enables confirming blocks using different absolute quorum sizes;
however it does not support dynamic participation.
An early `classical' BFT protocol for a model with unknown (but static) participation is
KW21 \cite{wattenhoferDA,wattenhoferDA2}.
A subsequent protocol
GLR21 \cite{goyalDA}
supports dynamic participation with
confirmation latency independent of the participation level, but still linear in the security parameter $\kappa$ \cite{lingDA}.
Confirmation latency independent of the security parameter is
achieved in the PoW setting
with omission and Byzantine faults by \cite{eyalDA} and \cite{eyalDA2}, respectively.

A recent work
MR22 \cite{lingDA} presents the first permissioned/PoS protocol that supports dynamic participation with confirmation latency independent of the security parameter
and participation level, with the caveats that temporary stability in the honest participation is necessary to ensure liveness, and a growing adversary cannot be tolerated.
Whereas \cite{lingDA} ensures fast confirmation
with adversary resilience $\frac{1}{2}$
without requiring high participation,
its confirmation latency is $37\Delta$,
considerably larger than the latency of \Goldfish ($4\Delta$) under optimistic conditions.
In the contemporary independent work
\cite{malkhiDA_third_blog,malkhiDA_third_eprint}, the prerequisites for liveness
were relaxed and latency was improved to $3\Delta$, at the expense of reduced adversary resilience (from $\frac{1}{2}$ down to $\frac{1}{3}$).
After completion of \Goldfish,
MMR23~\cite{malkhiDA_half_eprint} combines techniques from~\cite{lingDA} and~\cite{malkhiDA_third_eprint} to achieve resilience~$\frac{1}{2}$ and latency~$4\Delta$, while doing away with the stable participation requirement for liveness of~\cite{lingDA}.
GL23~\cite{losaDA} achieves a similar result independently and concurrently.
Both use expiring votes,
like \Goldfish, to tolerate increasing adversaries.

Increased communication complexity is a challenge with these later protocols.
In \Goldfish, each
voter only has to send a single message per slot, and messages can be relayed as-is to other validators, in practice efficiently implemented through gossip networks.
In MMR23~\cite{malkhiDA_half_eprint}, for instance, validators need to send tally messages for potentially many blocks, while
GL23~\cite{losaDA}
requires each validator to
attach its own signature on every received vote
before relaying.
Thus, \Goldfish always needs to gossip at most
linearly many
\emph{distinct} messages,
while GL23~\cite{losaDA} needs to at times gossip
quadratically many
distinct messages.
    The possibility of
many messages with different contents also makes these protocols less amenable to signature aggregation.
Together with their considerable deviations from LMD GHOST, these are concrete challenges faced by these protocols in their possible path to adoption
in Ethereum, as compared to \Goldfish.

\myparagraph{Follow-Ups \& Adoption}
A challenge
\Goldfish shares with
related works~\cite{lingDA,malkhiDA_half_eprint,losaDA}
is that a brief period of asynchrony
suffices to cause `deep' safety violations
(up to the last checkpoint if used with overlay,
\cf \figref{gasper-diagram}).
A \Goldfish variant in follow-up work~\cite{rlmdghost} addresses this issue,
by trading off a longer vote expiry period
for a less
dynamic participation model.
The current candidate protocol \cite{ssfproposal}
to provide
`single-slot finality' for Ethereum
is based on that \Goldfish variant.
\section{Preliminaries and Model}
\label{sec:boilerplate}

We recap the sleepy model~\cite{sleepy} for dynamic participation under synchrony.

\myparagraph{Preliminaries}
Let $\kappa, \lambda$ be the \emph{security parameters}
of \Goldfish itself
and of the cryptographic primitives it uses, respectively.
Specifically, $\kappa$ will be \Goldfish's (slow-path) confirmation latency, \cf~\myalgref{protocol}{confirmation}.
A function is \emph{negligible} in $\mu$,
denoted $\negl(\mu)$,
if it is $o(1/\mu^d)$ for all $d > 0$.
An event happens \emph{with overwhelming probability} (\emph{w.o.p.})
if it happens except with probability
(\emph{w.p.}) $\negl(\kappa)+\negl(\lambda)$.
\Goldfish uses a signature scheme $\sigSCHEME$
with key generation, 
sign, and verify
algorithms $\sigGen, \sigSign, \sigVerify$
(\cf \secref{crypto-details-boilerplate-sigs}).
A verifiable random function (VRF)~\cite{vrf} scheme $\vrfSCHEME$
with function generation, evaluation prove, and evaluation verify
algorithms $\vrfGen, \vrfProve, \vrfVerify$
(\cf \secref{crypto-details-boilerplate-vrfs})
is used for leader election and 
committee subsampling,
as in~\cite{gilad2017algorand,algorand}.
\label{sec:model}

\myparagraph{Validators}
\Goldfish is run among
$n$ validators,
with identities $\id \in [n] \triangleq \{1, ..., n\}$.
Each generates
a secret/public key pair
$(\sigSk_\id, \sigPk_\id)$ and $(\vrfSk_\id, \vrfPk_\id)$
for
$\sigSCHEME$ and $\vrfSCHEME$, respectively.
The public keys are commonly known (\ie, PKI).
As is customary to study new consensus protocols,
we
assume that every validator has one unit of stake throughout the execution (\ie, static homogeneous stake).
Gradual stake shift (\ie, dynamic stake)
can be supported 
using 
techniques 
that bootstrap
PoS protocols from static-stake protocols with PKI~\cite{snowwhite,david2018ouroboros,dynamicbft}.
\myparagraph{Environment and Adversary}
Time proceeds in discrete \emph{rounds} and the validators have synchronized clocks. (Bounded clock offsets can be lumped into the network delay upper bound $\Delta$ discussed below.)
Validators receive transactions ($\txs$) from the environment,
and can
\emph{broadcast} messages to each other.
The adversary
is a probabilistic poly-time (PPT) algorithm
that can control three aspects of the model
(corruption, sleepiness, network delay)
to attack consensus.
We first discuss these three aspects,
and then the adversary's powers and limits.

\myparagraphlight{Corruption}
The adversary
chooses
$f$ \emph{adversary}
validators (adaptively, subject to constraints
detailed below).
Non-adversary validators are \emph{honest}.
Naturally, the adversary learns the internal state
of its validators and
can make them deviate from the protocol
arbitrarily
(Byzantine faults)
for the rest of the execution
(permanent corruption).
We define the adversary fraction $\beta \triangleq f/n$.

\myparagraphlight{Sleepiness}
The adversary decides for each round and honest validator whether it is \emph{asleep} or not.
Asleep validators do not execute the protocol (temporary crash faults).
Messages delivered to an asleep validator are
picked up by it only once it is no longer asleep.
When a validator stops being asleep, it becomes \emph{dreamy}.
It then \emph{joins} the protocol,
possibly over multiple rounds, using a
\emph{joining procedure}
specified by the protocol.
Upon completion,
the
validator becomes \emph{awake} and follows the
protocol normally.
Adversary validators are always awake.
The number of awake validators is
bounded below by a constant $n_0$ across rounds.

\myparagraphlight{Network Delay}
Messages sent between validators are delivered
with an adversarially determined delay that can differ for each recipient.
Upon picking up messages
(\ie, once not asleep),
an honest validator re-broadcasts them.

\myparagraphlight{Adversary Powers and Limits}
For message delivery, the adversary has to obey a delay upper-bound of $\Delta$ rounds, which is
known to the validators (\emph{synchrony}).
Message delays and sleep schedule
are chosen by the adversary adaptively.
For sleepiness and corruption, \Goldfish supports two assumptions:
Either, we require \emph{mildly} adaptive corruption,
where it takes $3\Delta$ rounds for corruption to take effect,
together with the constraint that
for every round $r$,
the number of adversary validators at round $r$
must be less than
the number of honest awake validators at round $r-3\Delta$.
Or,
analogously to earlier works~\cite{algorand,kiayias2017ouroboros,badertscher2018ouroboros},
through the use of key evolving~\cite{keyevolving,keyevolving2} signature
and VRF schemes,
we allow for fully adaptive corruption,
together with the constraint that
for every round $r$,
the number of adversary validators at round $r$
must be less than
the number of honest awake validators at round $r$.
The precise technical assumptions are stated by \defref{compliant-execution}.

\myparagraph{Security}
Security is parameterized by $\kappa$, which for \Goldfish
affects the
confirmation latency.
We consider a finite execution horizon of $\Thorizon = \poly(\kappa)$ rounds.

\begin{definition}[Security]
    \label{def:security}
    Let $\chain \preceq \chain'$ express that ledger $\chain$ is
    a prefix of (or the same as) ledger $\chain'$.
    A consensus protocol,
    where
    at round $r$
    validator $\id$ outputs ledger $\chain^{\id}_{r}$,
    is \emph{secure with transaction confirmation time $\Tconf$}, iff w.o.p.:
    \begin{itemize}
        \item
              \textbf{Safety:} 
                $\forall r, r'\colon
                \forall \text{honest $\id, \id'$ awake at $r, r'$}\colon
                (\chain^{\id}_{r} \preceq \chain^{\id'}_{r'})
                \lor 
                (\chain^{\id'}_{r'} \preceq \chain^{\id}_{r})$.
        \item
              \textbf{Liveness:} If transaction $\tx$ was
              received by some awake honest validator by $r$, then 
              $\forall r' \geq r+\Tconf\colon \forall \text{honest $\id$ awake at $r'$}\colon \tx \in \chain^{\id}_{r'}$.
    \end{itemize}

\end{definition}
The protocol satisfies \emph{$\bar{f}$-safety} (\emph{$\bar{f}$-liveness}) if it is safe (live)
if $f < \bar{f}$.
It satisfies \emph{$1/2$-safety} (\emph{$1/2$-liveness}) if
it is safe (live) if $\beta < 1/2-\varepsilon$ for some $\varepsilon>0$.

\section{Protocol}
\label{sec:protocol}

We describe the basic \Goldfish protocol
in \secref{protocol-goldfish}
and its optimistic fast confirmation extension
in \secref{fastconfirmations}.
The composition of \Goldfish as underlay chain
with an overlay gadget (\cf \figref{gasper-diagram})
is described and analyzed in \secref{aa-analysis},
due to space constraints and since this mostly reuses orthogonal techniques from~\cite{sankagiri_clc,aadilemma,ebbandflow},

\subsection{The \Goldfish Protocol}
\label{sec:protocol-goldfish}

The basic \Goldfish protocol (\cf~\algref{helpers,protocol,ghost}) proceeds in \emph{slots} of $3\Delta$ rounds.

\myparagraph{VRF-based Lotteries}
\begin{algorithm}[tb]
    \caption{
        Interface of VRF-based lotteries and validity of data structures.
    }
    \label{alg:helpers}
    \begin{algorithmic}[1]
        \scriptsize
        \CommentLine{\textbf{VRF-based lotteries}}
        \Let
        {\varLotteryOpening \triangleq (y, \pi)}
        {\lotteryOpen_\id^{(\tagLotteryDUMMY, \thresholdDUMMY)}(t) \triangleq \vrfProve(\vrfSk_\id, \tagLotteryDUMMY \concat t)}
        \label{line:helpers-lottery-opening}
        \Let
        {\{\TRUE,\FALSE\}}{\lotteryIsWinningTicket^{(\tagLotteryDUMMY, \thresholdDUMMY)}((\id, t), \varLotteryOpening)
        \triangleq
        \begin{aligned}[t]
             & (\varLotteryOpening.y \leq \thresholdDUMMY\,2^\lambda)                                                   
             \land \vrfVerify(\vrfPk_\id, \tagLotteryDUMMY \concat t, (\varLotteryOpening.y, \varLotteryOpening.\pi))
        \end{aligned}}
        \label{line:helpers-lottery-iswinningticket}
        \Let
        {[0,1]}
        {\lotteryPriority(\varLotteryOpening) \triangleq \frac{\varLotteryOpening.y}{2^\lambda}}
        \label{line:helpers-lottery-priority}
        \CommentLine{\textbf{Data structures}}
        \Let
        {\{\TRUE,\FALSE\}}
        {\isValidBlock(\varBlock)
            \triangleq
            \begin{aligned}[t]
                 & (\varBlock = \varBlock_0)                                                                                                          \\[-2pt]
                 & \lor \begin{aligned}[t]
                             & (\lotteryIsWinningTicket^{(\tagLotteryBlock, \thresholdBlock)}((\varBlock.\id, \varBlock.t), \varBlock.\varLotteryOpening) \\[-2pt]
                             & \land{} \sigVerify(\sigPk_{\varBlock.\id}, \tagBlock \concat \varBlock.h \concat \varBlock.\txs, \varBlock.\sigma)         \\[-2pt]
                             & \land{} \isValidBlock(\ptrBlock{\varBlock.h}) \land (\varBlock.t > \ptrBlock{\varBlock.h}.t))
                        \end{aligned}
            \end{aligned}}
        \Comment{Block $\varBlock$}
        \label{line:helpers-block-isvalid}
        \Let
        {\{\TRUE,\FALSE\}}
        {\isValidVote(\varVote)
            \triangleq
            \begin{aligned}[t]
                 & \lotteryIsWinningTicket^{(\tagLotteryVote, \thresholdVote)}((\varVote.\id, \varVote.t), \varVote.\varLotteryOpening) 
                 \land{} \sigVerify(\sigPk_{\varVote.\id}, \tagVote \concat \varVote.h, \varVote.\sigma)                             
                 \\[-2pt]
                 &
                 \land{} \isValidBlock(\ptrBlock{\varVote.h})                                                                        
                 \land (\varVote.t \geq \ptrBlock{\varVote.h}.t)
            \end{aligned}}
        \Comment{Vote $\varVote$}
        \label{line:helpers-vote-isvalid}
        \Let
        {\{\TRUE,\FALSE\}}
        {\isValidPiece(\varPiece)
            \triangleq
            \isValidVote(\varPiece.\varVoteBlock)}
        \Comment{Piece $\varPiece$}
        \label{line:helpers-piece-isvalid}
        \Let
        {\{\TRUE,\FALSE\}}
        {\isValidProposal(\varProposal)
            \triangleq
            \begin{aligned}[t]
                 & \isValidBlock(\varProposal.\varBlock)                                                                                                  \\[-2pt]
                 & \land
                \isConsistent(\varProposal.\V \cup \{ \varProposal.\varBlock \})                                                                          \\[-2pt]
                 & \land{}
                \sigVerify(\sigPk_{\varProposal.\varBlock.\id}, \tagProposal \concat \varProposal.\V \concat \varProposal.\varBlock, \varProposal.\sigma) \\[-2pt]
                 & \land{}
                (\forall \varVoteBlock\in\varProposal.\V\colon \isValidVote(\varVoteBlock) \land (\varVoteBlock.t < \varProposal.\varBlock.t))
            \end{aligned}}
        \Comment{Proposal $\varProposal$}
        \label{line:helpers-proposal-isvalid}
    \end{algorithmic}
\end{algorithm}

The VRF PKI enables
cryptographic lotteries.
A \emph{lottery} $(\tagLotteryDUMMY, \thresholdDUMMY)$
is defined by a
fixed
$\tagLotteryDUMMY$
and threshold $\thresholdDUMMY \in [0,1]$.
Each validator $\id$ receives for each slot $t$
a lottery \emph{ticket} $(\id, t)$.
A ticket can be \emph{opened},
\myalgref{helpers}{helpers-lottery-opening}.
An \emph{opened ticket}
with \emph{opening} $\varLotteryOpening$ can be \emph{winning} for $(\tagLotteryDUMMY, \thresholdDUMMY)$,
\myalgref{helpers}{helpers-lottery-iswinningticket},
and winning opened tickets are totally ordered
by increasing \emph{precedence},
\myalgref{helpers}{helpers-lottery-priority}.

\myparagraph{Data Structures}
\emph{Blocks} and \emph{votes} are central to \Goldfish.
A block
$\varBlock \triangleq (\tagBlock, (\id, t), \varLotteryOpening, h, \txs, \sigma)$
consists of
\tagGeneric `$\tagBlock$',
ticket $(\id, t)$ and opening $\varLotteryOpening$
to the $(\tagLotteryBlock, \thresholdBlock)$ block production lottery,
hash $h$ committing to the new block's parent block
and transactions $\txs$
(as block `content'),
and
signature $\sigma$
binding together block production opportunity
and the block's content.
A special \emph{genesis block} $\varBlock_0 \triangleq (\tagBlock, (\bot, 0), \bot, \bot, \emptyset, \bot)$ is known to all validators.
A block $\varBlock$ is \emph{valid}
following \myalgref{helpers}{helpers-block-isvalid},
where $\ptrBlock{\varBlock.h}$ means the parent block that $\varBlock.h$ commits to (namely, ${}^*[x]$ represents the block committed by hash $x$).
The context within which these references get resolved
is detailed with the different network message types below.
A vote
$\varVote \triangleq (\tagVote, (\id, t), \varLotteryOpening, h, \sigma)$
consists of
\tagGeneric `$\tagVote$',
ticket $(\id, t)$ and opening $\varLotteryOpening$
to the $(\tagLotteryVote, \thresholdVote)$ voting lottery,
hash $h$ committing to the block voted for (as vote `content'),
and
signature $\sigma$
binding together
voting opportunity and the vote's content.
Every vote $\varVote$ is tied to its slot $\varVote.t$ via the lottery ticket $(\id, t)$.
A vote $\varVote$ is \emph{valid}
following \myalgref{helpers}{helpers-vote-isvalid}.

We call \emph{block-vote-set} (short \emph{\bvset})
a set of blocks and votes.
Commitments to blocks for the purpose of
the references
$\varVote.h$ or $\varBlock.h$ are
computed using $\hashBlock(.)$.
For a \bvset $\V$ we denote by $\V[h]$ the block
$\varBlock\in\V$ with $\hashBlock(\varBlock) = h$,
and $\bot$ if non-existent.
In \Goldfish, votes and blocks
are encapsulated and exchanged
in two network message types,
\emph{\pieces} and \emph{\proposals}.
A \piece
$\varPiece \triangleq (\tagPiece, \varVoteBlock)$
consists of
\tagGeneric `$\tagPiece$'
and for payload $x$ either a vote or a block,
and is \emph{valid}
following \myalgref{helpers}{helpers-piece-isvalid}.
\Pieces are used to propagate blocks and votes
and abstract
peer-to-peer broadcast object propagation.
In determining a \piece's validity,
block references $\ptrBlock{.}$
are resolved with respect to the \bvset $\V$
each validator maintains as part of its state,
see below.
If a validator does not have any matching block in $\V$,
it cannot currently determine the \piece's validity.
It then queues the \piece `in limbo' for re-examination
until its (in-)validity is established.
A \proposal
$\varProposal \triangleq (\tagProposal, \V, \varBlock, \sigma)$
consists of
\tagGeneric `$\tagProposal$',
\bvset $\V$ and block $\varBlock$ (as \proposal content),
and
signature $\sigma$
tying the \proposal to the
block production opportunity
of $\varBlock$.
Thus, a \proposal $\varProposal$ is \emph{valid}
following \myalgref{helpers}{helpers-proposal-isvalid},
where $\isConsistent(\V)$ is
satisfied on a \bvset $\V$
iff
$\varBlock_0\in\V$ and for every vote and block in $\V$ the referenced target/parent block is also in $\V$.
In determining the validity of \proposal $\varProposal$,
block references $\ptrBlock{.}$
are resolved with respect to $\varProposal.\V$.
We call a \bvset $\V$ with $\isConsistent(\V)$
a \emph{block-vote-tree} (short \emph{\bvtree}).
$\Merge(\V, \B)$
returns the largest \bvtree $\V'$
that is a subset of the union of $\V$ and the \pieces in $\B$.

\begin{algorithm}[tb]
    \caption{
        \Goldfish executed by
        validator $\id$
        with
        signature keys $(\sigSk_\id, \sigPk_\id)$,
        VRF keys $(\vrfSk_\id, \vrfPk_\id)$,
        \bvtree $\V$ and buffer $\B$.
        Here,
        notation `$\mathbf{at}$' means executing the code block at the specified round,
        $\chain^{\id}$ denotes the \Goldfish chain
        momentarily confirmed
        at $\id$.
        For $\funOurGhost(\V,t)$, see \algref{ghost}.%
    }
    \label{alg:protocol}
    \begin{algorithmic}[1]
        \scriptsize
        \Let{(\B, \V, t)}{(\emptyset, \{ \varBlock_0 \}, 0)}
        \Comment{Initialize buffer $\B$ and \bvtree $\V$}
        \CommentLine{As is customary, only \emph{valid} messages \emph{with time slot numbers at most $t$} are re-broadcast and put into $\B$. Invalid messages are discarded. Messages of unknown validity are queued. Blocks and votes contained in \proposals are also re-broadcast and added to $\B$ as individual \pieces.}
        \For{$t=1,2,\ldots$}
            \Comment{Slots}
            \At{3\Delta t}
                \Comment{\textbf{\phasePropose phase}}
                \Let{\varLotteryOpening}{\lotteryOpen_\id^{(\tagLotteryBlock, \thresholdBlock)}(t)}
                \label{line:prop-obtain-vrf}
                \Comment{Check if eligible to propose}
                \If{$\lotteryIsWinningTicket^{(\tagLotteryBlock, \thresholdBlock)}((\id, t), \varLotteryOpening)$}
                    \label{line:vrf-check-proposer}
                    \Let{\V'}{\Merge(\V, \B)}   \label{line:goldfish-simple-merge1} \Comment{\Bvtree to propose}
                    \Let{\varBlock}{\funOurGhost(\V',t-1)}   \label{line:fcr-1}   \Comment{Parent block}
                    \Let{\sigma}{\sigSign(\sigSk_\id, \tagBlock \concat \hashBlock(\varBlock) \concat \txs)}
                    \Let{\varBlock}{(\tagBlock, (\id, t), \varLotteryOpening, \hashBlock(\varBlock), \txs, \sigma)}   \Comment{New block}
                    \Let{\sigma}{\sigSign(\sigSk_\id, \tagProposal \concat \V' \concat \varBlock)}
                    \Broadcast{(\tagProposal, \V', \varBlock, \sigma)}
                    \label{line:broadcast-prop} \Comment{Propose}
                    \EndIf
            \EndAt
            \At{3\Delta t+\Delta}
                \Comment{\textbf{\phaseVote phase}}
                \CommentLine{Filter for proposals from slot $t$}
                \Let{\B'}{\{(\V',\varBlock) \mid (\tagProposal, \V', \varBlock, .) \in \B \land \varBlock.t = t\}}
                \CommentLine{Identify the leader of slot $t$ and its proposal}\label{line:vote-slot-leader}
                \Let{(\V'^*,\varBlock^*)}{ \argmin_{(\V',\varBlock) \in \B'} \lotteryPriority(\varBlock.\varLotteryOpening)}
                \label{line:new-proposal-received}
                \CommentLine{Merge own buffer and that of the leader into own \bvtree}
                \Let{\V}{\Merge(\V, \V'^* \cup \{\varBlock^*\})}
                \label{line:view-merge-leader}
                \label{line:goldfish-simple-merge2}
                \Let{\varLotteryOpening}{\lotteryOpen_\id^{(\tagLotteryVote, \thresholdVote)}(t)}
                \label{line:voter-obtain-vrf}
                \Comment{Check if eligible to vote}
                \If{$\lotteryIsWinningTicket^{(\tagLotteryVote, \thresholdVote)}((\id, t), \varLotteryOpening)$}   \label{line:vrf-check-voter}
                    \Let{\varBlock}{\funOurGhost(\V,t-1)}   \label{line:fcr-2}   \Comment{Target block}
                    \Let{\sigma}{\sigSign(\sigSk_\id, \tagVote \concat \hashBlock(\varBlock))}
                    \Let{\varVote}{(\tagVote, (\id, t), \varLotteryOpening, \hashBlock(\varBlock), \sigma)}
                    \Comment{New vote}
                    \Broadcast{(\tagPiece, \varVote)}   \label{line:broadcast-vote}   \Comment{Vote}
                \EndIf
            \EndAt
            \At{3\Delta t+2\Delta}
                \Comment{\textbf{\phaseConfirm phase}}
                \Let{\V}{\Merge(\V,\B)}   \label{line:view-merge-validator}
                \label{line:goldfish-simple-merge3}   \Comment{Merge buffer and \bvtree}
                \Let{\varBlock}{\funOurGhost(\V,t)}   \label{line:fcr-3}
                \Comment{Canonical \textOurGhost chain}%
                \Let{\chain^\id}{\varBlock^{\lceil \kappa}}   \label{line:confirmation}
                \Comment{Output ledger: $\varBlock$'s $\kappa$-deep prefix in terms of slots}
            \EndAt
        \EndFor
    \end{algorithmic}
\end{algorithm}

\begin{algorithm}[tb]
    \caption{\textOurGhost fork-choice rule.}
    \label{alg:ghost}
    \begin{algorithmic}[1]
        \scriptsize
        \State $\operatorname{\textsc{Children}}(\V,\varBlock) \triangleq \{ \varBlock' \in \V \mid \varBlock'.h = \hashBlock(\varBlock) \}$
        \State $\operatorname{\textsc{Votes}}(\V,\varBlock,t) \triangleq |\{ \id' \mid (\tagVote, (\id', t), ., h, .) \in \V \land \varBlock \preceq \V[h] \}|$
        \Function{\sc GHOST-Eph}{$\V, t$}
            \Let{\varBlock}{\varBlock_0}
            \Comment{Start fork-choice at genesis block}
            \Forever
                \CommentLine{Choose the \emph{heaviest} subtree
                (breaking ties deterministically) rooted at one of the children blocks $\varBlock'$ of $\varBlock$, by number of validators that have cast a vote in slot $t$ for $B'$
                or one of its descendants;
                $\varBlock' = \bot$ if $\operatorname{\textsc{Children}}(\V,\varBlock) = \emptyset$}
                \Let{\varBlock'}{\argmax_{\varBlock'\in\operatorname{\textsc{Children}}(\V,\varBlock)} \textsc{Votes}(\V, \varBlock',t)}   \label{line:vote-count}
                \If{$\varBlock' = \bot$}
                    \Return $\varBlock$
                \EndIf
                \Let{\varBlock}{\varBlock'}
            \EndForever
        \EndFunction
    \end{algorithmic}
\end{algorithm}

\myparagraph{Protocol}
Each validator knows the current slot $t$,
and maintains a \emph{buffer} $\B$
and a bvtree $\V$.
On a high level, messages enter from the network into $\B$,
while votes are tallied on $\V$.
The `magic' of \Goldfish is in how blocks and votes
enter from $\B$ to $\V$ (\textcolor{ColorMessageBuffering}{\emph{message buffering}, purple})
and leave $\V$ (\textcolor{ColorVoteExpiry}{\emph{vote expiry}, orange}).

Valid messages received from the network
are re-broadcast and added to $\B$.
(As is customary, messages whose validity is unknown
due to missing references,
and messages with future slot numbers,
are queued `in limbo' outside the protocol.)
For \proposals, the blocks and votes contained therein
are additionally re-broadcast and added to $\B$ as individual \pieces.

We describe
the three phases (\phasePropose, \phaseVote, \phaseConfirm)
of each slot $t$ from the perspective of an awake honest validator $\id$ (\algref{protocol}, \figref{protocol-horizon}):

\textbullet\hspace{0.3em}\phasePropose: At round $3\Delta t$,
$\id$ checks if its
lottery ticket $(\id,t)$ is winning for $(\tagLotteryBlock, \thresholdBlock)$
(\myalgref{protocol}{vrf-check-proposer}).
If so, $\id$ \textcolor{ColorMessageBuffering}{temporarily merges its \bvtree with its buffer}
(\myalgref{protocol}{goldfish-simple-merge1}),
\textcolor{ColorVoteExpiry}{identifies the \textOurGhost chain tip using only slot $t-1$ votes}
(\myalgref{protocol}{fcr-1}, \algref{ghost}),
and proposes its temporary \bvtree and a new block based on it
(\myalgref{protocol}{broadcast-prop}).
Note that in a practical implementation, the \proposals need not contain the whole \bvtree, but merely the votes therein (see \secref{experiments}).

For \textOurGhost fork-choice (\algref{ghost}),
a validator walks its \bvtree, starting at the genesis block, and at each block $\varBlock$, the validator proceeds to the child of $\varBlock$ whose subtree is heaviest, \ie, received the plurality of \textcolor{ColorVoteExpiry}{non-expired} votes.

\textbullet\hspace{0.3em}\phaseVote: At
$3\Delta t+\Delta$,
$\id$ identifies as \emph{leader} for slot $t$
the \proposal
with smallest
precedence (\myalgref{protocol}{new-proposal-received}).
It \textcolor{ColorMessageBuffering}{merges the leading \proposal's \bvtree into its
    \bvtree
    $\V$}
(\myalgref{protocol}{view-merge-leader}).
Then it checks if its
lottery ticket $(\id,t)$ is winning for $(\tagLotteryVote, \thresholdVote)$
(\myalgref{protocol}{vrf-check-voter}).
If so, $\id$
\textcolor{ColorVoteExpiry}{identifies the \textOurGhost chain tip using only slot $t-1$ votes}
(\myalgref{protocol}{fcr-2}),
and votes for it (\myalgref{protocol}{broadcast-vote}).

\textbullet\hspace{0.3em}\phaseConfirm: At round $3\Delta t+2\Delta$,
$\id$ \textcolor{ColorMessageBuffering}{merges its buffer
    into its \bvtree}
(\myalgref{protocol}{view-merge-validator}).
It then
\textcolor{ColorVoteExpiry}{identifies the \textOurGhost chain tip using only slot $t$ votes}
(\myalgref{protocol}{fcr-3}),
and outputs as confirmed ledger $\chain^\id$
the transactions of those blocks in the \textOurGhost chain
that are from slots $\leq t-\kappa$ (`$\kappa$-deep in time',
\myalgref{protocol}{confirmation}).
Since the
ledger in view of an awake honest validator $\id$
is only updated at
this point,
we may view the ledger as indexed by slot $t$: $\chain^\id_t$.

\myparagraph{Key Mechanism}
\textcolor{ColorMessageBuffering}{\emph{Message buffering}} ensures that if in slot $t$
the leading proposal is honest,
then all honest voters in $t$ will vote for it
(\lemref{view-merge-property}),
because in \phasePropose, the leader's temporary \bvtree $\V'$ is a superset of
all honest validators' \bvtrees,
and thus in \phaseVote all honest voters adopt that leader's \bvtree.
\textcolor{ColorVoteExpiry}{\emph{Vote expiry}}
(and honest majority)
ensures that if in slot $t$
all honest voters vote into the subtree rooted at some block $B$,
then all honest voters in slot $t+1$ will also vote into the subtree rooted at $B$
(\lemref{all-honest-voting-together}).
Induction on $t$
readily
yields reorg resilience.
Furthermore, w.o.p., every interval of $\kappa$ slots has at least one honest
leading proposer
(\lemref{honest-leader-almost-everywhere}).
The prefix of that proposal stabilizes (by reorg resilience),
and the proposal includes unconfirmed transactions,
leading to safety and liveness (with $\Tconf = 2\kappa+2$) of the $\kappa$-deep confirmation rule.
    Without \textcolor{ColorMessageBuffering}{\emph{message buffering}},
    honest voters would no longer be guaranteed to rally behind honest \proposals.
    Instead, the adversary could
    induce inconsistent views among honest voters,
    leading to them no longer voting \emph{en bloc},
    which restores the balancing attack~\cite{ebbandflow,3attacks}.
    Without \textcolor{ColorVoteExpiry}{\emph{vote expiry}},
    honest voters would not be guaranteed to vote into the subtree of $B$
    in
    $t+1$ just because $B$ gathered
    votes from all honest validators in
    $t$.
    In fact, the adversary could
    use votes from longer ago to break the protocol.
    We give two examples.
    First, adversary validators could strategically release
    votes for long past slots,
    like in the
    avalanche attack~\cite{2attacks}.
    Second, in periods where participation is increasing,
    the adversary could, for validators that were asleep in the past but are now adversary, forge votes
    for these past slots
    for
    a block conflicting with $B$.
    Conceptually, thereby, the adversary gains control of a dishonest majority for past slots, which can break security.
    Vote expiry
    preserves security
    when the number of adversary validators
    increases together with the number of honest validators
    over time.

\myparagraph{Joining Procedure}
\label{sec:protocol-goldfish-joining-procedure}
At each round, a validator is either asleep, dreamy or awake
(\secref{model}).
Whenever a validator stops being asleep, it is \emph{dreamy}.
Dreamy validators don't follow \algref{protocol},
except for relaying messages.
With the next \phaseConfirm phase, the validator returns to being
\emph{awake} and fully resumes \algref{protocol}.
To allow for more time
to download messages missed during sleep, dreaminess can be extended accordingly, but should always end at a \phaseConfirm phase.

\subsection{Optimistic Fast Confirmations}
\label{sec:fastconfirmations}

Basic \Goldfish
(\secref{protocol-goldfish})
provides reorg resilience,
but its $\kappa$-deep confirmation rule
leads to $\Theta(\kappa)$ latency in worst and expected case.
We add a \phaseFastConfirm phase
and introduce a fast confirmation rule,
to
achieve constant expected confirmation latency \emph{under optimistic conditions}, \ie, under high participation and honest $3/4$-supermajority (\figref{fast-confirmation-horizon},
\algref{fast-protocol}).
In particular, validators can now confirm honest proposals
within the same slot,
under optimistic conditions.
The $\kappa$-deep confirmation rule (\myalgref{full-protocol}{confirmation}) (now called \emph{standard} confirmation rule),
still applies and guarantees security when optimistic conditions don't hold.

\begin{algorithm}[tb]
    \caption{
        \Goldfish executed by
        validator $\id$,
        \textcolor{NiceBlueColor}{modified (blue) to use (optimistic) fast and standard confirmation (\cf~\algref{protocol})}.
        See \algref{ghost} for
        $\operatorname{\textsc{Votes}}$.%
    }
    \label{alg:fast-protocol}
    \begin{algorithmic}[1]
        \scriptsize
        \State \emph{Same initialization and housekeeping as  \algref{protocol}}
        \For{$t=1,2,\ldots$}
            \Comment{Slots}
            \At{4\Delta t}
                \Comment{\textbf{\phasePropose phase}}
                \State \emph{Same as \phasePropose phase in \algref{protocol}}
            \EndAt
            \At{4\Delta t+\Delta}
                \Comment{\textbf{\phaseVote phase}}
                \State \emph{Same as \phaseVote phase in \algref{protocol}}
            \EndAt
            \At{4\Delta t+2\Delta}
                \Comment{\textcolor{NiceBlueColor}{\textbf{\phaseFastConfirm phase}}}
                \Let{\V}{\Merge(\V,\B)}   \label{line:fast-view-merge1}
                \Comment{Merge buffer and \bvtree}
                \Let{\chainFast^\id}{\argmax_{\varBlock\in\V\colon |\textsc{Votes}(\V, \varBlock,t)| \geq n (\frac{3}{4}+\frac{\epsilon}{2}) \thresholdVote} |\varBlock|}   \label{line:fast-confirmation}
            \EndAt
            \At{4\Delta t+3\Delta}
                \Comment{\textbf{\phaseConfirm phase}}
                \Let{\V}{\Merge(\V,\B)}   \label{line:fast-view-merge2}
                \Comment{Merge buffer and \bvtree}
                \Let{\varBlock}{\funOurGhost(\V,t)}   \label{line:fast-fcr-3}
                \Comment{Canonical \textOurGhost chain}%
                \Let{\chain^\id}{\argmax_{\chain \in \{\chainFast^\id, \varBlock^{\lceil \kappa}\}} |\chain|}
                \label{line:output-highest}\Comment{\textcolor{NiceBlueColor}{Output \Goldfish ledger}}
            \EndAt
        \EndFor
    \end{algorithmic}
\end{algorithm}

\myparagraph{Fast Confirmation Phase}
Slots are now $4\Delta$ rounds, with the insertion of phase \phaseFastConfirm at round $4\Delta t + 2\Delta$ (\figref{fast-confirmation-horizon},
\algref{fast-protocol}).
In \phaseFastConfirm, a validator $\id$ first merges its buffer into its \bvtree $\V$
(\myalgref{fast-protocol}{fast-view-merge1}).
It then marks a block $\varBlock$ as fast confirmed if $|\textsc{Votes}(\V,\varBlock,t)| \ge n(\frac{3}{4}+\frac{\epsilon}{2}) \thresholdVote$ for some $\epsilon>0$ that can be made arbitrarily small as $n \to \infty$, and updates $\chainFast^\id$ to the highest fast confirmed block (\myalgref{fast-protocol}{fast-confirmation}).
In \phaseConfirm (\myalgref{fast-protocol}{output-highest}),
validator $\id$ outputs the higher of $\chainFast^\id$ and the
standard-confirmed
$\kappa$-deep prefix $\varBlock^{\lceil \kappa}$.
For simplicity, we omit
a
mechanism to avoid ledger `roll back'
(to ensure $\forall\id, t'\geq t\colon \chain^\id_t \preceq \chain^\id_{t'}$).

Intuitively, the extra \phaseFastConfirm phase
guarantees that
when an honest validator fast confirms a block $\varBlock$ in slot $t$, all honest awake validators see the causative votes
by the time their \bvtrees are
last updated
in $t$.
The subtree rooted at $\varBlock$
will be heaviest in all steps of
\textOurGhost fork-choice
for all honest voters and forever
(\thmref{fast-strong-persistence}),
which implies that fast confirmations are safe (\thmref{fast-confirmation-safety}).
Security of \Goldfish with fast confirmations is proven in \secref{appendix-fast-confirmation-analysis}.

\myparagraph{Joining Procedure}
Once a validator stops being asleep, it is
dreamy until the next \phaseConfirm phase (\myalgref{fast-protocol}{fast-view-merge2}), when it turns awake and resumes execution.

\section{Security Proof}
\label{sec:analysis}

Due to space constraints,
we show the security proof for
`basic' \Goldfish (\secref{protocol-goldfish})
here.
The proofs for \Goldfish with fast confirmations
(\secref{fastconfirmations}),
and for \Goldfish when used as an underlay chain
composed
with an overlay finality/accountability gadget (as in \figref{gasper-diagram}),
are provided in \secref{appendix-fast-confirmation-analysis,aa-analysis}, respectively.

In the subsequent analysis,
a validator $\id$ is
\emph{eligible to vote at slot $t$}
if its ticket $(\id, t)$ is winning
for the lottery
$(\tagLotteryVote,\thresholdVote)$.
Recall that
awake honest validators
consider the \proposal with lowest precendence
received by $3\Delta t + \Delta$ from the leader
of slot $t$ (\myalgref{protocol}{vote-slot-leader}).
We
use blocks
and the chains
they induce via the parent relation
interchangeably.
A block $\varBlock_1$ is a \emph{descendant} (resp., \emph{ancestor}) of block $\varBlock_2$ iff the underlying chains satisfy $\varBlock_2 \preceq \varBlock_1$ (resp., $\varBlock_1 \preceq \varBlock_2$).

Let $A_r$ and $H_r$ denote the number of adversary and honest validators awake at round $r$, respectively.
Our theorems hold
for \emph{$(\frac{1}{2}, 3\Delta)$-compliant executions}
(\defref{compliant-execution}) that satisfy
the following relations on $A_r$ and $H_r$:
(i) in the absence of key-evolving cryptographic primitives, the adversary is mildly adaptive and $\forall r\colon \frac{A_r}{A_r+H_{r-3\Delta}} < \frac{1}{2} - \epsilon$, and (ii) with key-evolving primitives, $\forall r\colon \frac{A_r}{A_r+H_{r}} < \frac{1}{2} - \epsilon$.
\vspace{-1.9ex}%
\begin{restatable}{theorem}{restateStrongPersistenceOne}
    \label{thm:strong-persistence-1}
    Suppose a $(\frac{1}{2}, 3\Delta)$-compliant execution
    of \Goldfish
    in the synchronous sleepy network model
    of \secref{model},
    and
    validator $\id$ with \proposal $\varProposal^*$ is recognized as the leader of a slot $t$ by all awake honest validators at round $3\Delta t + \Delta$ (\myalgref{protocol}{vote-slot-leader}).
    Then, w.o.p., $\varProposal^*.\varBlock \preceq \varBlock$
    for any $\varBlock$ identified
    in \myalgref{protocol}{fcr-1,fcr-2,fcr-3}
    by any awake honest validator
    in any round
    $r \geq 3\Delta t + 2\Delta$.
\end{restatable}
\vspace{-1.9ex}%
\begin{restatable}[Security]{theorem}{restateSecurity}
    \label{thm:security}
    Suppose a $(\frac{1}{2}, 3\Delta)$-compliant execution
    of \Goldfish
    in the synchronous sleepy network model.
    Then, w.o.p.,
    \Goldfish is secure
    with transaction confirmation time
    $\Tconf = 2\kappa+2$ slots.
\end{restatable}
\vspace{-1.9ex}%
\begin{restatable}[Reorg resilience]{theorem}{restateReorgResilience}
    \label{thm:reorg-resilience}
    Suppose a $(\frac{1}{2}, 3\Delta)$-compliant execution
    of \Goldfish
    in the synchronous sleepy network model,
    and
    validator $\id$ with \proposal $\varProposal^*$ is recognized as the leader of a slot $t$ by all awake honest validators at round $3\Delta t + \Delta$ (\myalgref{protocol}{vote-slot-leader}).
    Then, w.o.p.,
    $\exists r'\colon
        \forall r \geq r'\colon
        \forall \id\colon
        \
        \varProposal^*.\varBlock \preceq \chain_r^\id,$
    where $\chain_r^\id$ denotes \Goldfish's
    ledger at validator $\id$ and round $r$.
    In particular, $r' = 3\Delta(t+\kappa) + 2\Delta$
    satisfies the above.
\end{restatable}
Due to space constraints,
a formal definition of $(\gamma, \tau)$-compliant executions
and formal proofs
of \thmref{strong-persistence-1,security,reorg-resilience} and the subsequent lemmas are given in \secref{appendix-analysis-of-goldfish}.
In the rest of this section, we focus on the intuition.
Proof of \thmref{strong-persistence-1} in follows from
\lemref{honest-leader-almost-everywhere,view-merge-property,all-honest-voting-together} below.
The structure of this argument is inductive:
\lemref{honest-leader-almost-everywhere} shows that in compliant executions, honest voters outnumber adversary voters;
and every long interval of slots contains at least one slot in which all honest validators recognize the same honest validator as the slot leader.
\lemref{view-merge-property} shows that in a slot $t$ with such recognized honest leader, all honest voters vote for the leader's proposal.
Finally, \lemref{all-honest-voting-together} shows that if in slot $t$, all honest voters have voted for a descendant of a certain block, then in slot $t+1$ all honest voters will vote for a descendant of that block.
This concludes the induction and \thmref{strong-persistence-1} follows.
Proofs of \thmref{security,reorg-resilience} follow readily
from \thmref{strong-persistence-1}
and \lemref{honest-leader-almost-everywhere} below.
\begin{restatable}{lemma}{restateHonestLeaderAlmostEverywhere}
    \label{lem:honest-leader-almost-everywhere}
    Suppose the \Goldfish execution is $(\frac{1}{2}, 3\Delta)$-compliant.
    Then, w.o.p., for every slot $t$,
    adversary validators
    at round $3\Delta (t+1)+\Delta$
    eligible to vote at slot $t$
    are less than
    honest validators
    awake at round $3\Delta t + \Delta$
    and eligible to vote at slot $t$.
    Also w.o.p., all slot intervals of length $\kappa$
    have at least one slot $t$ where an honest validator is recognized as the slot $t$ leader by all awake honest validators at round $3\Delta t + \Delta$.%
    \footnote{The proposer-lottery threshold $\thresholdBlock$ can be tuned following Algorand \cite[Appendix-B.1]{gilad2017algorand} so that each slot has at least one eligible proposer.}%
\end{restatable}
\lemref{honest-leader-almost-everywhere}'s proof uses
correctness, uniqueness and pseudorandomness
of VRF-based lotteries along with Chernoff bounds.
\begin{restatable}{lemma}{restateViewMergeProperty}
    \label{lem:view-merge-property}
    Suppose an
    execution
    of \Goldfish
    in the synchronous sleepy network model,
    and
    validator $\id^*$ with \proposal $\varProposal^*$ is recognized as 
    leader of a slot $t$ by all awake honest validators at round $3\Delta t + \Delta$ (\myalgref{protocol}{vote-slot-leader}).
    Then, all honest validators
    awake at round $3\Delta t + \Delta$
    and eligible to vote at 
    $t$
    vote for $\varProposal^*.\varBlock$ at 
    $t$.
\end{restatable}
By message buffering
and honest $\id^*$,
$\varProposal^*.\V$ is a superset of the \bvtrees of all honest validators awake at round $3\Delta t + \Delta$ and eligible to vote at slot $t$.
Hence, upon merging $\varProposal^*.\V$ into their \bvtrees (\myalgref{protocol}{view-merge-leader}) at round $3\Delta t + \Delta$, all of these validators vote
for the \textOurGhost tip in
$\varProposal^*.\V$, \ie, for $\varProposal^*.\varBlock$.
\begin{restatable}{lemma}{restateAllHonestVotingTogether}
    \label{lem:all-honest-voting-together}
    Suppose a $(\frac{1}{2}, 3\Delta)$-compliant execution
    of \Goldfish
    in the synchronous sleepy network model.
    Consider a slot $t$ where
    all honest validators
    awake at round $3\Delta t + \Delta$
    and eligible to vote at slot $t$,
    vote for a descendant of $\varBlock$.
    Then, w.o.p.,
    all honest validators
    awake at round $3\Delta (t+1) + \Delta$
    and eligible to vote at slot $t+1$,
    vote for a descendant of $\varBlock$.
\end{restatable}
By vote expiry, the eligible honest validators awake at round $3\Delta (t+1) + \Delta$ consider only the slot $t$ votes in
\textOurGhost fork-choice (\myalgref{protocol}{fcr-2}).
Due to honest majority, the subtree rooted at $\varBlock$
is heaviest in all steps of
\textOurGhost fork-choice.
Thus, all honest voters vote for a descendant of $\varBlock$, if all eligible honest validators awake at round $3\Delta t + \Delta$ voted for a descendant of $\varBlock$.

\section{Implementation and Experiments}
\label{sec:experiments}

We
discuss implementation aspects of \Goldfish
and
study its behavior under
dynamic participation.
We focus on
communication-efficient implementation of proposals
and 
message buffering,
and on
the interplay between the
block production lottery threshold $\thresholdBlock$,
communication load,
and 
behavior under low participation.
We have implemented
a prototype of 
\Goldfish
in Rust\footnote{Source code: \gitSourceUrl},
with BLAKE3 hashes \cite{blake3} and BLS signatures \cite{blssigs} over the BLS12-381 curve \cite{bls12381} for signatures and VRFs.
The network was simulated with delay $\Delta=4\,\text{s}$.

\myparagraph{Proposal Size and Wire Format}
In the \Goldfish variant of \secref{protocol-goldfish},
for ease of exposition,
proposals include
the proposer's entire \bvtree $\V'$ (\myalgref{protocol}{goldfish-simple-merge1}).
This raises concerns about the resulting communication load.
Proposal messages
would grow over time with the number of blocks,
and could be
inflated by \emph{equivocation spamming}
(\ie, adversary uses one winning lottery ticket
to create many equivocating blocks or votes,
\cf \cite{bandwidth-constrained-consensus,lcfinitebw}).
The following implementation details resolve these concerns.
It suffices for a proposal to only include votes from the latest \phaseVote phase, as
older votes
are already expired anyway.
Another tweak is \emph{equivocation discounting}, \ie, not counting votes during fork-choice from validators who have sent votes for two or more different blocks during the latest \phaseVote phase.
We discuss equivocation discounting in detail, and show it to not compromise security, in \secref{appendix-equivocation-discounting}.
As \emph{any two} equivocating votes suffice as evidence for an honest validator to discount all votes of an equivocating adversary,
the above two measures mean that every proposal needs to include at most two votes per validator eligible to vote in the previous slot.
Notice also that
it suffices
for proposals to include \emph{references} (hashes) to blocks and votes.
In fact, an honest proposer's role in message buffering is only to point validators to messages (\emph{which they already have in their buffer} because at least the honest proposer would have relayed them) that
they should merge into their \bvtree.
Finally,
only blocks with nonzero fork-choice weight need to be referenced, because
blocks with zero weight cannot possibly alter fork-choice regarding the proposer's block.
Nonzero weight blocks are either referenced by votes,
or by a nonzero weight child block.
\emph{Thus, it suffices for proposals to only reference at most two votes per validator eligible to vote in the previous slot.}

Concretely, if \Goldfish is used among $n=1{,}024$ validators without voter subsampling, so $\thresholdVote=1$, with $32\,\mathrm{Byte}$ hashes, then even worst-case a proposal is only of $64\,\mathrm{kByte}$ plus one block. This is representative for a deployment in Ethereum, where votes get aggregated by $1{,}024$ aggregators per slot, \Goldfish's fork-choice would operate on aggregates, and at most two aggregates per aggregator need to be referenced in a proposal.
Comparing $64\,\mathrm{kByte}$
to the
current
block size of
$80\,\mathrm{kByte}$,
message buffering seems feasible
in terms of network load.
\myparagraphlight{Garbage Collection}
Proposals
are discarded
after their slot's \phaseVote phase.
Vote expiry allows to discard votes within two slots.
Blocks
(including `in limbo')
are discarded
once inconsistent
with confirmed blocks (\ie, after at most $\kappa$ slots).

\myparagraph{Block Production Lottery Threshold}
\begin{figure}[tb]
    \centering%
    \begin{tikzpicture}%
        \footnotesize
        \begin{axis}[
            mysimpleplot,
            name=plot1,
            axis y line*=right,
            axis x line=none,
            ylabel={Communication\\{}[kByte/s]},
            xmin=0, xmax=0.0100,
            ymin=0,
            height=0.3\linewidth,
            width=0.85\linewidth,
            grid=major,
            y label style={font=\scriptsize,align=center,myParula05Green},
            y tick label style={font=\scriptsize,myParula05Green},
            ]

            \addplot [myParula05Green,thick,densely dotted,mark=+,mark options={solid}] table [x=pblock,y expr=\thisrow{comms_total_size}/75/4/4/1000]
                {figures/experiment-pblock/n1000_fullparticipation_varyingThrB_repetitions1.txt};
            \label{leg:experiment-thrb-comms-measurement}
            \addplot [myParula05Green,thin,mark=none] table [x=pblock,y expr=5722.911878374327*\thisrow{pblock}+4.472299386947608]
                {figures/experiment-pblock/n1000_fullparticipation_varyingThrB_repetitions1.txt};
            \label{leg:experiment-thrb-comms-theory}

        \end{axis}
        \begin{axis}[
            mysimpleplot,
            name=plot1,
            axis y line*=left,
            xlabel={Block production lottery threshold $\thresholdBlock$},
            ylabel={Ledger growth\\{}rate [blocks/slot]},
            xmin=0, xmax=0.0100,
            ymin=0,
            height=0.3\linewidth,
            width=0.85\linewidth,
            xticklabel style={
                    /pgf/number format/fixed,
                    /pgf/number format/fixed zerofill,
                    /pgf/number format/precision=3
                },
            scaled x ticks=false,
            grid=major,
            y label style={font=\scriptsize,align=center,myParula07Red},
            y tick label style={font=\scriptsize,myParula07Red},
            xtick={0.000,0.002,...,0.010},
            ]

            \addplot [myParula07Red,thick,densely dotted,mark=+,mark options={solid}] table [x=pblock,y expr=\thisrow{ledger_best_len}/75]
                {figures/experiment-pblock/n1000_fullparticipation_varyingThrB_repetitions1.txt};
            \label{leg:experiment-thrb-len-measurement}

            \addplot [myParula07Red,thin,mark=none] table [x=pblock,y expr=1-exp(-\thisrow{pblock}*1000)]
                {figures/experiment-pblock/n1000_fullparticipation_varyingThrB_repetitions1.txt};
            \label{leg:experiment-thrb-len-theory}

        \end{axis}
    \end{tikzpicture}%
    \vspace{-0.75em}%
    \caption[]{Ledger growth rate and average broadcast load of \Goldfish as a function of block production lottery threshold, for experiments (\ref{leg:experiment-thrb-len-measurement}, \ref{leg:experiment-thrb-comms-measurement}) with $n=1000$, $\thresholdVote = 0.1$, $\Delta=4\,\mathrm{s}$, under full honest participation.
    For the block production lottery,
    we expect the number of proposals per slot
    to be binomially distributed
    with mean $n\,\thresholdBlock$.
    The measurements fit the predictions
    for the probability of zero proposals in a given slot
    ($1-e^{-n\,\thresholdBlock}$,
    \ref{leg:experiment-thrb-len-theory}),
    and that the communication load is affine
    ($5723\cdot\thresholdBlock+4.472$ with coefficients to four digit accuracy,
    \ref{leg:experiment-thrb-comms-theory})
    with the constant term accounting for votes.%
    }
    \label{fig:experiment-thrb}
\end{figure}
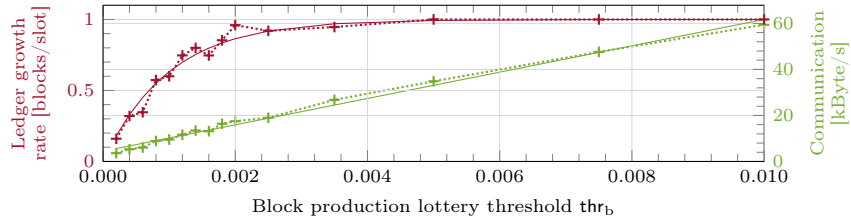
\newcommand{\PlotExperimentTrace}[5]{
    \begin{figure}[tb]
        \centering%
        \begin{tikzpicture}%
            \scriptsize
            \begin{axis}[
                mysimpleplot,
                name=plot1,
                ylabel={Ledger length\\{}[blocks]},
                xmin=16, xmax=#3,
                ymin=0,
                xmajorticks=false,
                legend columns=5,
                height=0.275\linewidth,
                width=0.95\linewidth,
                legend style={
                        at={(0.5,1)},
                        xshift=-2em,
                        anchor=south,
                        draw=none,
                        /tikz/every even column/.append style={
                                column sep=0.3em
                            },
                        cells={
                                align=left
                            }
                    },
                ]

                \addplot [myparula11,mark=none] table [x expr=\thisrow{r}*4,y=partyALWAYSAWAKE_ledger_fast_length]
                    {#1};
                \label{leg:experiment-trace-#2-fast-len}
                \addlegendentry{Fast conf.};

                \addplot [myparula21,mark=none] table [x expr=\thisrow{r}*4,y=partyALWAYSAWAKE_ledger_slow_length]
                    {#1};
                \label{leg:experiment-trace-#2-slow-len}
                \addlegendentry{Slow conf.};

                \addlegendimage{myparula41,mark=none}
                \addlegendentry{Awake};

                \addlegendimage{myparula31,mark=none}
                \addlegendentry{Dreamy};

                \addlegendimage{myParula05Green,no markers,dashed}
                \addlegendentry{Fast live};

                \legend{};

            \end{axis}
            \begin{axis}[
                    mysimpleplot,
                    name=plot2,
                    at=(plot1.south), anchor=north, yshift=-0.5em,
                    ylabel={Age of tip [s]},
                    xmin=16, xmax=#3,
                    ymin=0,
                    xmajorticks=false,
                    height=0.275\linewidth,
                    width=0.95\linewidth,
                ]

                \addplot [myparula11,mark=none] table [x expr=\thisrow{r}*4,y expr=4*(\thisrow{r} - \thisrow{partyALWAYSAWAKE_ledger_fast_age}*4)]
                    {#1};
                \label{leg:experiment-trace-#2-fast-age}

                \addplot [myparula21,mark=none] table [x expr=\thisrow{r}*4,y expr=4*(\thisrow{r} - \thisrow{partyALWAYSAWAKE_ledger_slow_age}*4)]
                    {#1};
                \label{leg:experiment-trace-#2-slow-age}

            \end{axis}
            \begin{axis}[
                    mysimpleplot,
                    name=plot3,
                    at=(plot2.south), anchor=north, yshift=-0.5em,
                    ylabel={Participation},
                    xmin=16, xmax=#3,
                    ymin=0,
                    xmajorticks=false,
                    height=0.275\linewidth,
                    width=0.95\linewidth,
                    extra y ticks={750},
                ]

                \addplot [name path=dacurve,myparula41,mark=none,no marks] table [x expr=\thisrow{r}*4,y=n_honest_awake]
                    {#1};
                \label{leg:experiment-trace-#2-awake}

                \addplot [name path=dacurve,myparula31,mark=none,no marks] table [x expr=\thisrow{r}*4,y=n_honest_dreamy]
                    {#1};
                \label{leg:experiment-trace-#2-dreamy}

                \addplot[name path=threshold,myparula51,mark=none,no marks,densely dotted,domain=-1000:10000] {750};
                \label{leg:experiment-trace-#2-fastquorum}

            \end{axis}
            \begin{axis}[
                mysimpleplot,
                name=plot4,
                at=(plot3.south), anchor=north, yshift=-0.5em,
                xlabel={Time [s]},
                ylabel={Communication\\{}[kByte/s]},
                xmin=16, xmax=#3,
                ymin=0,
                height=0.275\linewidth,
                width=0.95\linewidth,
                ]

                \addplot [name path=dacurve,black,mark=none,thin] table [x expr=\thisrow{r}*4,y expr=\thisrow{partyALWAYSAWAKE_comms_all_size}/4/1000]
                    {#1};
                \label{leg:experiment-trace-#2-comms}

            \end{axis}%
            #5
        \end{tikzpicture}%
        \vspace{-0.75em}%
        \caption[]{#4}%
        \label{fig:experiment-trace-#2}
    \end{figure}
}
\PlotExperimentTrace{figures/experiment-traces/timeline-experiment4-D-0.0030.txt}{exp4-D-0030}{1931}{%
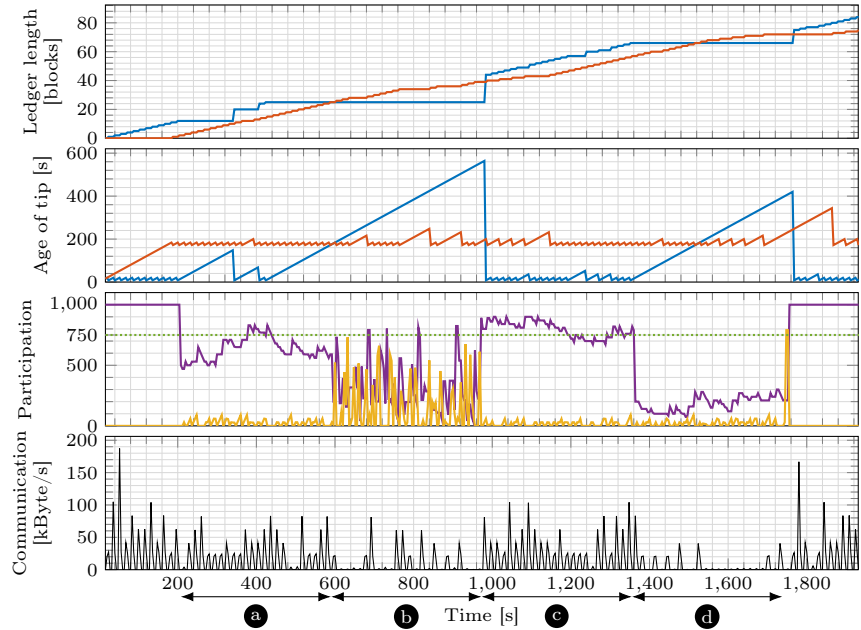
    Ledger length / age of most recently confirmed block
    under fast (\ref{leg:experiment-trace-exp4-D-0030-fast-len})
    / slow (\ref{leg:experiment-trace-exp4-D-0030-slow-len})
    confirmation,
    and resulting broadcast load
    (\ref{leg:experiment-trace-exp4-D-0030-comms}),
    for \Goldfish
    with $n=1000$,
    $\thresholdVote = 0.1$,
    $\Delta=4\,\mathrm{s}$,
    $\thresholdBlock = 3/n$,
    $\kappa = 10$,
    block size $80\,\mathrm{kByte}$,
    different environments of dynamic participation (\ref{leg:experiment-trace-exp4-D-0030-awake}).
    When participation is above the fast-confirmation threshold of approx.\ $3n/4$ (\ref{leg:experiment-trace-exp4-D-0030-fastquorum}),
    transactions are confirmed swiftly ($4\Delta$,
    \ref{leg:experiment-trace-exp4-D-0030-fast-age},
    \circledScriptBlackSlim{a}, \circledScriptBlackSlim{c}),
    otherwise fast confirmation stalls.
    When participation is volatile
    (\cf $600\,\mathrm{s}$ to $950\,\mathrm{s}$,
    \circledScriptBlackSlim{b}),
    many honest validators are \emph{dreamy} (\ref{leg:experiment-trace-exp4-D-0030-dreamy}, \cf \secref{protocol-goldfish-joining-procedure}).
    Then, or when participation is steady
    but at a low level
    (\cf $1400\,\mathrm{s}$ to $1700\,\mathrm{s}$,
    \circledScriptBlackSlim{d}),
    effective participation
    (by honest validators who are neither asleep nor dreamy)
    is low.
    Slow confirmation ($4\Delta\kappa$ base latency,
    \ref{leg:experiment-trace-exp4-D-0030-slow-age})
    takes place throughout,
    but since $\thresholdBlock = 3/n$,
    slow confirmation degrades (\cf \figref{experiment-thrb})
    when effective participation is low
    (\cf slots with no proposal around $650\,\mathrm{s}$ or $1600\,\mathrm{s}$, \ref{leg:experiment-trace-exp4-D-0030-comms}, lead to latency spikes, \ref{leg:experiment-trace-exp4-D-0030-slow-age}).
    Communication load is modest (\ref{leg:experiment-trace-exp4-D-0030-comms}).
}{

    \coordinate (topRight) at (10,-6.1);
    \coordinate (topLeft) at (0,-6.1);
    \draw [Latex-Latex] ($(topLeft)!0.1!(topRight)$) -- ($(topLeft)!0.3!(topRight)$) node [midway,below] {\circledScriptBlackSlim{a}};
    \draw [Latex-Latex] ($(topLeft)!0.3!(topRight)$) -- ($(topLeft)!0.5!(topRight)$) node [midway,below] {\circledScriptBlackSlim{b}};
    \draw [Latex-Latex] ($(topLeft)!0.5!(topRight)$) -- ($(topLeft)!0.7!(topRight)$) node [midway,below] {\circledScriptBlackSlim{c}};
    \draw [Latex-Latex] ($(topLeft)!0.7!(topRight)$) -- ($(topLeft)!0.9!(topRight)$) node [midway,below] {\circledScriptBlackSlim{d}};
}
\secref{analysis} shows
how to tune the vote lottery threshold $\thresholdVote$ so that, w.o.p., all voter committees over 
the
execution horizon have an honest majority.
Given a number $n$ of validators 
and a threshold $\thresholdVote$, the size of a proposal and the communication load resulting from votes are close to constant.
The block production lottery threshold $\thresholdBlock$ is the remaining parameter affecting the overall broadcast load through the expected number of proposals per slot $n\,\thresholdBlock$ (\figref{experiment-thrb}).
For low $\thresholdBlock < 1/n$, communication load is low but ledger growth is impaired because many slots have no proposal.
For high $\thresholdBlock > 1/n$, most slots have more than one proposal, leading to communication overhead but also close-to-optimal ledger growth.
For a reasonable tradeoff
in the non-degraded common case of near-full participation, we tune $\thresholdBlock = 3/n$.

\myparagraph{Behavior under Dynamic Participation}
Based on \figref{experiment-thrb}, we expect a confirmation
performance degradation
under low participation if $\thresholdBlock = 3/n$.
(If good performance is to be ensured
even under very low participation $n_0 \ll n$,
tune $\thresholdBlock$ to $n_0$ rather than to $n$.)
To study the impact of dynamic participation on \Goldfish
with $\thresholdBlock = 3/n$,
we run it (\figref{experiment-trace-exp4-D-0030}) in four different dynamic
participation environments inspired by \cite{lingDA}:
\circledScriptBlackSlim{a}
Stable participation: Starting from $50\%$,
randomly increase
or decrease participation by $3\%$ per
$\Delta$ (unless this would exceed $[10\%,90\%]$).
\circledScriptBlackSlim{b}
Unstable participation: Select a participation level uniformly at random in $[10\%,90\%]$ per $\Delta$.
\circledScriptBlackSlim{c}
High participation: Reset participation to $80\%$, randomly increase
or decrease
by $3\%$ per $\Delta$ (staying in $[70\%,90\%]$).
\circledScriptBlackSlim{d}
Low participation: Reset participation to $20\%$, randomly increase
or decrease
by $3\%$ per $\Delta$ (staying in $[10\%,30\%]$).
Once the participation level was drawn according to this schedule,
from instant to instant the environment selects a random set of asleep (awake)
validators to wake up (put to sleep), respectively,
to meet the participation levels.
A performance-based comparison
of LMD GHOST and \Goldfish 
is apples-to-oranges,
as
LMD GHOST is not secure under dynamic participation,
while \Goldfish is.
That said, 
\Goldfish has a slightly lower
block production rate 
due to extra phases (\cf \figref{lmd-ghost,protocol-horizon,fast-confirmation-horizon}),
at otherwise comparable
confirmation latency and communication load.
The choice of leader election mechanism,
\ie, VRF-based
vs.\ a randomness beacon like Ethereum's RANDAO
(\cf \secref{from-lmdghost-to-goldfish}),
also affects performance of both protocols
equally.

\section*{Acknowledgment}
We thank
Aditya Asgaonkar,
Carl Beekhuizen,
Vitalik Buterin,
Justin Drake,
Dankrad Feist,
Sreeram Kannan,
Georgios Konstantopoulos,
Barnabé Monnot,
Ling Ren,
Dan Robinson,
Danny Ryan,
Caspar Schwarz-Schilling,
Alberto Sonnino,
and
Fan Zhang
for fruitful discussions.
The work of JN
was conducted in part
while at Paradigm.
JN, ENT and DT are supported by a gift from
the
Ethereum Foundation.
JN is supported by
the
Protocol Labs PhD Fellowship and
the
Reed-Hodgson Stanford Graduate Fellowship.
ENT is supported by
the
Stanford Center for Blockchain Research.

\bibliographystyle{splncs04}
\bibliography{references}

\appendix

\section{Protocol Slot Structures}
\label{sec:protocol-slotstructures}

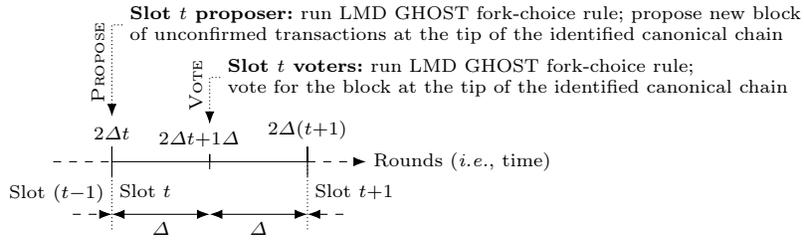
\begin{figure}[tb]
    \centering
    \begin{tikzpicture}[
        x=1.3cm,
        phaselabel/.style = {
                align=left,
                anchor=south west,
                xshift=0.5em,
            },
        phasearrow/.style = {
                -Latex,
                densely dotted,
            },
    ]
    \scriptsize\scriptspacing

    \draw [dashed] (-0.6,0) -- (0,0);
    \draw (0,0) -- (2,0);
    \draw [-Latex,dashed] (2,0) -- (2.6,0) node [right] {Rounds (\ie, time)};

    \draw [] (0,0) -- ++(0,0.2) -- ++(0,-0.4) node [pos=0,above] {$2\Delta t$} node [below right] {Slot $t$} node [below left] {Slot $(t-1)$};
    \draw [] (1,0) -- ++(0,0.1) -- ++(0,-0.2) node [pos=0,above,xshift=-0.5em] {$2\Delta t + 1\Delta$};
    \draw [] (2,0) -- ++(0,0.2) -- ++(0,-0.4) node [pos=0,above] {$2\Delta (t+1)$} node [below right] {Slot $t+1$};

    \draw [densely dotted] (0,0) -- ++(0,-0.9);
    \draw [densely dotted] (2,0) -- ++(0,-0.9);

    \begin{scope}[yshift=-0.7cm]
        \draw [-Latex,dashed] (-0.4,0) -- (0,0);
        \draw [Latex-Latex] (0,0) -- ++(1,0) node [midway,below] {$\Delta$};
        \draw [Latex-Latex] (1,0) -- ++(1,0) node [midway,below] {$\Delta$};
        \draw [Latex-,dashed] (2,0) -- ++(0.4,0);
    \end{scope}

    \node [phaselabel,anchor=west,yshift=4em] (label4) at (1,0) {\textbf{Slot $t$ voters:} run LMD GHOST fork-choice rule;\\vote for the block at the tip of the identified canonical chain};
    \draw [phasearrow] (label4.west) -| (1,0.5) node [anchor=south west,yshift=0.3em,rotate=90] {\phaseVote};

    \node [phaselabel] (label5) at ($(0,0)!(label4.north west)!(0,1)$) {\textbf{Slot $t$ proposer:} run LMD GHOST fork-choice rule;
        propose new block
        \\
        of unconfirmed transactions
        at the tip of the identified canonical chain};
    \draw [phasearrow] (label5.west) -| (0,0.6) node [anchor=south west,yshift=0.3em,rotate=90] {\phasePropose};

\end{tikzpicture}%
    \vspace{-1em}%
    \caption{%
        LMD GHOST has slots with two phases of $\Delta$ duration each. Each slot has a pseudorandomly elected \emph{proposer} and a committee of \emph{voters}.
        \phasePropose: At the start of a slot, the
        proposer runs
        fork-choice
        and proposes a block extending the tip of the identified \emph{canonical chain}.
        \phaseVote: Midway into a slot,
        voters run
        fork-choice
        and vote for the block at the tip of the identified chain.
        *
        For \emph{greedy heaviest observed sub-tree} fork-choice \cite[Alg.~3.1]{gasper} (\cf \algref{ghost}),
        conceptually,
        a validator walks the block tree in its view, starting at the genesis block, and at each block $\varBlock$, the validator proceeds to the child of $\varBlock$ whose subtree is \emph{heaviest}, \ie, received the largest number of votes.
    }
    \label{fig:lmd-ghost}
\end{figure}

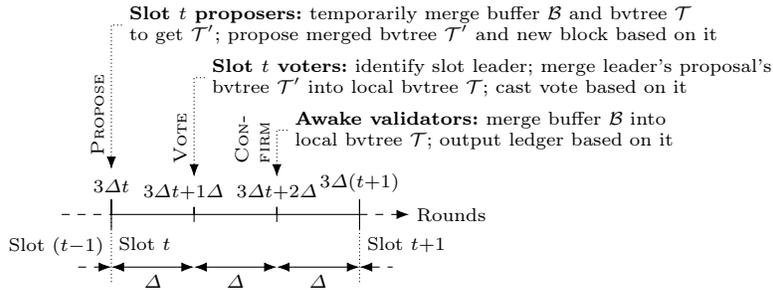
\begin{figure}[tb]
    \centering%
    \begin{tikzpicture}[
        x=1.1cm,
        phaselabel/.style = {
                align=left,
                anchor=south west,
                xshift=0.5em,
            },
        phasearrow/.style = {
                -Latex,
                densely dotted,
            },
    ]
    \scriptsize\scriptspacing

    \draw [dashed] (-0.6,0) -- (0,0);
    \draw (0,0) -- (3,0);
    \draw [dashed,-Latex] (3,0) -- (3.6,0) node [right] {Rounds};

    \draw [] (0,0) -- ++(0,0.2) -- ++(0,-0.4) node [pos=0,above] {$3\Delta t$} node [below right] {Slot $t$} node [below left] {Slot $(t-1)$};
    \draw [] (1,0) -- ++(0,0.1) -- ++(0,-0.2) node [pos=0,above,xshift=-0.5em] {$3\Delta t + 1\Delta$};
    \draw [] (2,0) -- ++(0,0.1) -- ++(0,-0.2) node [pos=0,above] {$3\Delta t + 2\Delta$};
    \draw [] (3,0) -- ++(0,0.2) -- ++(0,-0.4) node [pos=0,above] {$3\Delta (t+1)$} node [below right] {Slot $t+1$};

    \draw [densely dotted] (0,0) -- ++(0,-0.9);
    \draw [densely dotted] (3,0) -- ++(0,-0.9);

    \begin{scope}[yshift=-0.7cm]
        \draw [-Latex,dashed] (-0.4,0) -- (0,0);
        \draw [Latex-Latex] (0,0) -- ++(1,0) node [midway,below] {$\Delta$};
        \draw [Latex-Latex] (1,0) -- ++(1,0) node [midway,below] {$\Delta$};
        \draw [Latex-Latex] (2,0) -- ++(1,0) node [midway,below] {$\Delta$};
        \draw [Latex-,dashed] (3,0) -- (3.4,0);
    \end{scope}

    \node [phaselabel,anchor=west,yshift=4em] (label3) at (2,0) {\textbf{Awake validators:} merge buffer $\B$ into\\local \bvtree $\V$; output ledger based on it};
    \draw [phasearrow] (label3.west) -| (2,0.5) node [anchor=south west,yshift=0.3em,rotate=90,align=left] {\textsc{Con-}\\\textsc{firm}}; %

    \node [phaselabel] (label4) at ($(1,0)!(label3.north west)!(1,1)$) {\textbf{Slot $t$ voters:} identify slot leader; merge leader's proposal's\\\bvtree $\V'$ into local \bvtree $\V$; cast vote based on it};
    \draw [phasearrow] (label4.west) -| (1,0.5) node [anchor=south west,yshift=0.3em,rotate=90] {\phaseVote};

    \node [phaselabel] (label5) at ($(0,0)!(label4.north west)!(0,1)$) {\textbf{Slot $t$ proposers:} temporarily merge buffer $\B$ and \bvtree $\V$\\to get $\V'$; propose merged \bvtree $\V'$ and new block based on it};
    \draw [phasearrow] (label5.west) -| (0,0.6) node [anchor=south west,yshift=0.3em,rotate=90] {\phasePropose};

\end{tikzpicture}%
    \vspace{-1em}%
    \caption{%
        Throughout the execution, validators \emph{buffer} received \proposals and \pieces, and \emph{merge} the blocks and votes contained
        therein into their \bvtrees only as \emph{explicitly} instructed.
        \Goldfish has slots of three phases
        of $\Delta$ rounds each.
        Each slot has proposers (one of which
        will later be recognized as the slot's leader)
        and a committee of voters.
        \phasePropose: At the start of a slot,
        proposers
        temporarily merge their buffers into their local \bvtrees,
        and propose their temporary \bvtrees and a new block based on it.
        \phaseVote: One-thirds into a slot,
        voters identify the slot's leader's proposal,
        merge the proposed \bvtree into their local \bvtrees,
        and cast a vote based on their local \bvtrees.
        \phaseConfirm: Two-thirds into a slot,
        \emph{all awake validators}
        merge their buffers into their local \bvtrees,
        and confirm a ledger based on their local \bvtrees.%
    }%
    \label{fig:protocol-horizon}%
\end{figure}

\begin{figure}[tb]
    \centering%
    \begin{tikzpicture}[
        x=1.2cm,
        phaselabel/.style = {
                align=left,
                anchor=south west,
                xshift=0.5em,
            },
        phasearrow/.style = {
                -Latex,
                densely dotted,
            },
    ]
    \scriptsize\scriptspacing

    \draw [dashed] (-0.5,0) -- (-0,0);
    \draw (0,0) -- (4,0);
    \draw [-Latex,dashed] (4,0) -- (4.5,0) node [right] {Rounds};

    \draw [] (0,0) -- ++(0,0.2) -- ++(0,-0.4) node [pos=0,above] {$4\Delta t$} node [below right] {Slot $t$}; %
    \draw [] (1,0) -- ++(0,0.1) -- ++(0,-0.2) node [pos=0,above] {$4\Delta t + 1\Delta$};
    \draw [] (2,0) -- ++(0,0.1) -- ++(0,-0.2) node [pos=0,above] {$4\Delta t + 2\Delta$};
    \draw [] (3,0) -- ++(0,0.1) -- ++(0,-0.2) node [pos=0,above] {$4\Delta t + 3\Delta$};
    \draw [] (4,0) -- ++(0,0.2) -- ++(0,-0.4) node [pos=0,above] {$4\Delta (t+1)$}; %

    \draw [densely dotted] (0,0) -- ++(0,-0.9);
    \draw [densely dotted] (4,0) -- ++(0,-0.9);

    \begin{scope}[yshift=-0.7cm]
        \draw [Latex-Latex] (0,0) -- ++(1,0) node [midway,below] {$\Delta$};
        \draw [Latex-Latex] (1,0) -- ++(1,0) node [midway,below] {$\Delta$};
        \draw [Latex-Latex] (2,0) -- ++(1,0) node [midway,below] {$\Delta$};
        \draw [Latex-Latex] (3,0) -- ++(1,0) node [midway,below] {$\Delta$};
    \end{scope}

    \node [phaselabel,anchor=west,yshift=4em] (label2) at (3,0) {\textbf{Awake validators:} merge buffer $\B$ into\\local \bvtree $\V$; output ledger based on it};
    \draw [phasearrow] (label2.west) -| (3,0.5);%

    \node [phaselabel] (label3) at ($(2,0)!(label2.north west)!(2,1)$) {\textcolor{NiceBlueColor}{\textbf{Awake validators:} merge buffer $\B$ into local}\\\textcolor{NiceBlueColor}{\bvtree $\V$; run optimistic fast confirmation rule}};
    \draw [phasearrow] (label3.west) -| (2,0.5) node [align=left,anchor=south west,yshift=0.3em,rotate=90] {\textcolor{NiceBlueColor}{\textsc{Fast-}}\\\textcolor{NiceBlueColor}{\textsc{Conf.}}};%

    \node [phaselabel] (label4) at ($(1,0)!(label3.north west)!(1,1)$) {\textbf{Slot $t$ voters:} identify slot leader; merge leader's proposal's\\\bvtree $\V'$ into local \bvtree $\V$; cast vote based on it};
    \draw [phasearrow] (label4.west) -| (1,0.5);%

    \node [phaselabel] (label5) at ($(0,0)!(label4.north west)!(0,1)$) {\textbf{Slot $t$ proposers:} temporarily merge buffer $\B$ and \bvtree $\V$\\to get $\V'$; propose merged \bvtree $\V'$ and new block based on it};
    \draw [phasearrow] (label5.west) -| (0,0.6);%

\end{tikzpicture}%
    \vspace{-1em}%
    \caption{%
        To enable optimistic fast confirmations,
        a \textcolor{NiceBlueColor}{\phaseFastConfirm phase (blue)}
        of $\Delta$ rounds
        is inserted between \phaseVote
        and \phaseConfirm phase
        (\cf \figref{protocol-horizon}).
        \textcolor{NiceBlueColor}{\phaseFastConfirm: Two-fourth into a slot,
            \emph{all awake validators}
            merge their buffers into their local \bvtrees,
            and run the optimistic fast confirmation rule
            based on their local \bvtrees.}%
    }%
    \label{fig:fast-confirmation-horizon}%
\end{figure}
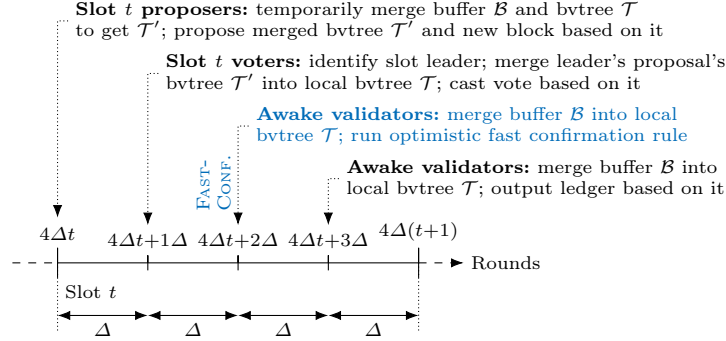

The slot structure of LMD GHOST,
\Goldfish,
and \Goldfish with optimistic fast confirmations
are depicted in
\figref{lmd-ghost},
\figref{protocol-horizon},
and
\figref{fast-confirmation-horizon},
respectively.

\section{Security Proof for \Goldfish}
\label{sec:appendix-analysis-of-goldfish}

In this section, we provide a formal, complete security proof of \Goldfish under a synchronous network in the sleepy model.
For this purpose, we restate and expand on the definitions and theorem statements first presented in \secref{analysis}.

\subsection{Definitions}
In the subsequent analysis,
a valid \proposal $\varProposal$
(\cf~\secref{protocol})
is \emph{for slot $t$} iff $t = \varProposal.\varBlock.t$,
and it \emph{has precedence $p$} iff $p = \lotteryPriority(\varProposal.\varBlock.\rho)$.
A validator $\id$ is
\emph{eligible to propose at slot $t$}
if its ticket $(\id, t)$ is winning
for the lottery $(\tagLotteryBlock,\thresholdBlock)$.
Similarly, a validator $\id$ is
\emph{eligible to vote at slot $t$}
if its ticket $(\id, t)$ is winning
for the lottery
$(\tagLotteryVote,\thresholdVote)$.
Recall that
awake honest validators
consider the \proposal with lowest precendence
received by $3\Delta t + \Delta$ from the leader
of slot $t$ (\myalgref{protocol}{vote-slot-leader}).
We hereafter use blocks
and the sequences of blocks
they induce via the parent-block chain relation
interchangeably.
A block $\varBlock_1$ is a \emph{descendant} (resp., \emph{ancestor}) of block $\varBlock_2$ iff the underlying chains satisfy $\varBlock_2 \preceq \varBlock_1$ (resp., $\varBlock_1 \preceq \varBlock_2$).
Two blocks $\varBlock_1, \varBlock_2$ are \emph{conflicting} if $B_1$ is neither an ancestor nor a descendant of $B_2$.

Let $A_r$ and $H_r$ denote the number of adversary and honest validators awake at round $r$, respectively.
Our security theorems hold
for \emph{compliant executions} that satisfy
the following
relations on $A_r$ and $H_r$:
\begin{definition}%
    \label{def:compliant-execution}
    In the absence of key-evolving cryptographic primitives (signatures and VRFs), an execution is \emph{$(\gamma, \tau)$-compliant} iff:
    \begin{itemize}
        \item $\forall r\colon   \frac{A_r}{A_r+H_{r-\tau}} \leq \beta < \gamma-\epsilon$.
        \item The corruption is mildly adaptive: If the adversary decides to corrupt an honest validator at round $r$, then the validator becomes adversary no earlier than at round $r+\tau$.
    \end{itemize}
    With key-evolving primitives, an execution is \emph{compliant} iff:
    \begin{itemize}
        \item $\forall r\colon   \frac{A_r}{A_r+H_r} \leq \beta < \gamma-\epsilon$.
    \end{itemize}
    Moreover, in both cases, $H_r > \gamma n_0 = \Theta(\kappa)$ for all rounds $r$, and the time horizon $\Thorizon$ of the protocol execution satisfies $\Thorizon = \operatorname{poly}(\kappa)$.
\end{definition}

\subsection{\lemref{honest-leader-almost-everywhere}}
\lemref{honest-leader-almost-everywhere} shows that
in compliant executions,
honest voters outnumber adversary voters
(as long as votes have not yet expired);
and every long interval of slots
contains at least one slot
in which all honest validators recognize the same honest validator
as the slot leader.
\restateHonestLeaderAlmostEverywhere*
\lemref{honest-leader-almost-everywhere}'s proof uses
correctness, uniqueness and pseudorandomness
of VRF-based lotteries along with Chernoff bounds.

\begin{proof}[Proof of \lemref{honest-leader-almost-everywhere}]
    By the \emph{pseudorandomness} property of the VRF-based lottery (\secref{crypto-details}), for any given slot $t$ and validators $\id_1$ and $\id_2$, $\id_1 \neq \id_2$,
    \begin{IEEEeqnarray}{rCl}
        \Prob{
        \lotteryIsWinningTicket^{l_v}((\id, t), \lotteryOpen_{\id_1}^{l_v}(t))
        } &=& \thresholdVote
        \IEEEeqnarraynumspace
        \\
        \Prob{
        \lotteryIsWinningTicket^{l_b}((\id, t), \lotteryOpen_{\id_1}^{l_b}(t))
        } &=& \thresholdBlock
        \IEEEeqnarraynumspace
        \\
        \Prob{
            \lotteryPriority(\lotteryOpen_{\id_1}^{l_b}(t)) < \lotteryPriority(\lotteryOpen_{\id_2}^{l_b}(t))
        } &=& \frac{1}{2},
    \end{IEEEeqnarray}
    where $l_v = (\tagLotteryVote, \thresholdVote)$ and $l_b = (\tagLotteryBlock, \thresholdBlock)$ are the lotteries,
    and
    $\lotteryOpen_{\id_1}^{l_v}(t)$,
    $\lotteryOpen_{\id_1}^{l_b}(t)$,
    $\lotteryOpen_{\id_2}^{l_b}(t)$,
    and
    $\lotteryOpen_{\id_2}^{l_v)}(t)$
    are independent random variables.

    We first consider the protocol without key-evolving primitives.
    By the \emph{uniqueness} property of the lottery (\secref{crypto-details}), w.o.p., for all validators $\id$ and slots $t$, the ticket $(\id,t)$ can be opened at most one unique opening (\myalgref{protocol}{voter-obtain-vrf}).
    Let $\Tilde{H}_t$ denote the number of honest validators awake at round $3\Delta t + \Delta$ and eligible to vote at slot $t$.
    Let $\Tilde{A}_t$ denote the number of adversary validators at round $3\Delta (t+1) + \Delta$ that are eligible to vote at slot $t$.
    Recall that $A_r$ and $H_r$ denote the number of adversary and honest validators awake at round $r$ respectively (note that the honest validators have been awake since the closest round $3\Delta t + 2\Delta$ same as or preceding $r$).
    Let $n_t = H_{3\Delta t + \Delta} + A_{3\Delta (t+1) + \Delta} \geq n_0 = \Theta(\kappa)$.

    By the pseudorandomness property, the adversary cannot predict in advance which honest validators will become eligible to vote or propose at a given slot.
    Moreover, if the adversary decides to corrupt the honest validators eligible to vote at a slot $t$ after learning their identities at round $3\Delta t + \Delta$, it takes over $3\Delta$ rounds for the corruption to take effect, implying that these validators cannot be counted as part of $\Tilde{A}_t$.
    Hence, as $\frac{A_r}{A_r+H_{r-3\Delta}} \leq \beta < \frac{1}{2}-\epsilon$ for all rounds $r$, w.o.p.,
    \begin{IEEEeqnarray*}{rCl}
        \mathbb{E}[\Tilde{H}_t] &=& H_{3\Delta t + \Delta} \thresholdVote \geq (\frac{1}{2}+\epsilon) n_t \thresholdVote  \\
        \mathbb{E}[\Tilde{A}_t] &=& A_{3\Delta (t+1) + \Delta} \thresholdVote \leq (\frac{1}{2}-\epsilon) n_t \thresholdVote
    \end{IEEEeqnarray*}
    By a Chernoff bound,
    \begin{IEEEeqnarray*}{rCl}
        \Prob{ \Tilde{H}_t < \frac{1}{2} n_t \thresholdVote } &\leq& e^{-\frac{\epsilon^2}{1+2\epsilon} n_t \thresholdVote}  \\
        \Prob{ \Tilde{A}_t > \frac{1}{2} n_t \thresholdVote } &\leq& e^{-\frac{\epsilon^2}{1+3\epsilon} n_t \thresholdVote}.
    \end{IEEEeqnarray*}
    Thus, at any given slot $t$, $\Tilde{H}_t > \Tilde{A}_t$, except with probability
    \begin{IEEEeqnarray*}{C}
        2\exp{(-\frac{\epsilon^2}{1+3\epsilon} n_0 \thresholdVote)}.
    \end{IEEEeqnarray*}
    By a union bound, every slot $t$ has more honest validators awake at round $3\Delta t + \Delta$ and eligible to vote at slot $t$ than adversary validators at round $3\Delta (t+1) + \Delta$, eligible to vote at slot $t$ (and more than $\frac{1}{2} n_0 \thresholdVote$ such honest validators), except with probability
    \begin{IEEEeqnarray*}{C}
        2\Thorizon \exp{\left(-\frac{\epsilon^2}{1+3\epsilon} n_0 \thresholdVote\right)} + \negl(\lambda)= \negl(\kappa)+\negl(\lambda),
    \end{IEEEeqnarray*}
    since $n_0 = \Theta(\kappa)$ and $\Thorizon = \Theta(\kappa)$.
    By the same reasoning, w.o.p., every slot $t$ has more honest validators awake and eligible to propose for slot $t$ at round $3\Delta t$ than adversary validators at round $3\Delta t + \Delta$, eligible to propose for slot $t$.

    Finally, for any given slot $t$, each valid slot $t$ \proposal broadcast within rounds $[3\Delta t, 3\Delta t + \Delta]$ has the same probability of achieving the minimum precedence up to terms negligible in $\lambda$.\footnote{We assume that $\poly(\kappa)\negl(\lambda) = \negl(\lambda)$.}
    Now, at a slot $t$, if an honest validator's \proposal achieves the minimum precedence among the valid slot $t$ \proposals broadcast by $\Delta$ rounds into the slot, then that validator is identified as the slot leader by all honest validators awake at round $3\Delta t + \Delta$.
    Taking a fixed $t\geq\kappa$, the probability that no awake honest validator's \proposal has the minimum precedence among the valid slot $s$ \proposals broadcast by $\Delta$ rounds into the slot, during the slots $s \in [t-\kappa,t]$, is upper bounded by $2^{-\kappa} + \negl(\kappa) + \negl(\lambda)$.
    Union bounding over all $\Thorizon$ many such intervals, we find that
    w.o.p.,
    all slot intervals of length $\kappa$
    have at least one slot $t$, where an honest validator is identified as the slot leader by all awake honest validators at round $3\Delta t + \Delta$.

    Now with key-evolving primitives, we define $\Tilde{H}_t = H_{3\Delta t+\Delta}$ and $\Tilde{A}_t = A_{3\Delta t+\Delta}$.
    Similarly, we define $n_t = H_{3\Delta t+\Delta} + A_{3\Delta t+\Delta} \geq n_0 = \Theta(\kappa)$.
    In this case, $\frac{A_r}{A_r+H_{r}} \leq \beta < \frac{1}{2}-\epsilon$ for all rounds $r$.
    Note that the adversary cannot predict in advance which honest validators will become eligible to vote or propose at a given slot due to the pseudorandomness property of the lottery.
    Moreover, if the adversary corrupts the honest validators eligible to vote at a slot $t$ after learning their identities at round $3\Delta t + \Delta$, it cannot make these validators broadcast new valid votes for slot $t$ since the keys for slot $t$ would have been evolved prior to adversary corrruption (\ie, these corrupted validators cannot be counted as part of $\Tilde{A}_t$).
    Hence, the number of valid slot $t$ votes adversary validators can broadcast by round $3\Delta (t+1) + \Delta$ is upper bounded by the number of adversary validators at round $3\Delta t + \Delta$ that are eligible to vote at slot $t$.
    Finally, by the same calculations as above, every slot $t$ has more honest validators eligible to vote and awake at round $3\Delta t + \Delta$ than the adversary validators at round $3\Delta (t+1) + \Delta$ eligible to vote at slot $t$ (and more than $\frac{1}{2} n_0 \thresholdVote$ such honest validators), except with probability
    \begin{IEEEeqnarray*}{C}
        2\Thorizon \exp{\left(-\frac{\epsilon^2}{1+3\epsilon} n_0 \thresholdVote\right)} + \negl(\lambda) = \negl(\kappa)+\negl(\lambda).
    \end{IEEEeqnarray*}
    Similarly, w.o.p., every slot $t$ has more honest validators awake and eligible to propose for slot $t$ at round $3\Delta t$ than adversary validators at round $3\Delta t + \Delta$ eligible to propose for slot $t$.
    Thus, via the same argument, w.o.p., all slot intervals of length $\kappa$
    have at least one slot $t$, where an honest validator is identified as the slot leader by all awake honest validators at round $3\Delta t + \Delta$.
\end{proof}

\subsection{Main Security Results}

The main security results are as follows:
\restateStrongPersistenceOne*
\restateSecurity*
\restateReorgResilience*
We first prove \thmref{security,reorg-resilience}
from \thmref{strong-persistence-1}
and \lemref{honest-leader-almost-everywhere}.
Then, we prove
\thmref{strong-persistence-1}
from the subsequent \lemref{honest-leader-almost-everywhere,view-merge-property,all-honest-voting-together}.

\begin{proof}[Proof of \thmref{security}]
    By \lemref{honest-leader-almost-everywhere}, w.o.p., all slot intervals of length $\kappa$ have at least one slot $t$, where an honest validator with \proposal $\varProposal^*$ is recognized as the slot leader by all awake honest validators at round $3\Delta t + \Delta$,
    and,
    by \thmref{strong-persistence-1},
    $\varProposal^*.\varBlock \preceq B$
    for any $\varBlock$ identified
    in \myalgref{protocol}{fcr-1,fcr-2,fcr-3}
    by any awake honest validator
    in any
    $r \geq 3\Delta t + 2\Delta$.

    \myparagraphlight{Liveness}
    A transaction $\tx$ is input to an honest validator
    at some round $r$.
    At most $6\Delta$ rounds (\ie, $2$ slots)
    later
    the transaction is propagated to all honest validators
    and we have reached the beginning of a slot $t_0$.
    For the next $\kappa$ slots all honest proposers
    will include $\tx$ if they extend a tip whose chain
    does not include $\tx$ yet.
    By the earlier argument, one of these proposals will
    be an ancestor of any
    $\varBlock$ identified
    in \myalgref{protocol}{fcr-1,fcr-2,fcr-3}
    by any awake honest validator
    in any
    $r' \geq 3\Delta (t_0+\kappa) + 2\Delta$.
    From $\kappa$ slots later onwards,
    all awake honest validators include the transaction
    in their ledger
    (\myalgref{protocol}{confirmation}).
    Thus, \Goldfish is live with $\Tconf = 2\kappa + 2$ slots.

    \myparagraphlight{Safety}
    Pick any two honest validators $\id_1$ and $\id_2$,
    and two slots $t_1$ and $t_2 \geq t_1$.
    By the earlier argument,
    there exists a block $\varBlock'$
    proposed (by an honest validator)
    at some slot $t' \in [t_1-\kappa,t_1]$
    such that $\varBlock' \preceq \varBlock$
    for any
    $\varBlock$ identified
    in \myalgref{protocol}{fcr-1,fcr-2,fcr-3}
    by any awake honest validator
    in any
    $r' \geq 3\Delta t' + 2\Delta$.
    As $t' \geq t_1-\kappa$ but by \Goldfish's confirmation rule
    blocks in
    $\chain^{\id_1}_{t_1}$
    are from no later than $t_1-\kappa$,
    $\chain^{\id_1}_{t_1} \preceq \varBlock$.
    Similarly, if $t' \geq t_2-\kappa$, then $\chain^{\id_2}_{t_2} \preceq \varBlock$; otherwise, $\varBlock \preceq \chain^{\id_2}_{t_2}$.
    In both cases, either $\chain^{\id_1}_{t_1} \preceq \chain^{\id_2}_{t_2}$ or $\chain^{\id_2}_{t_2} \preceq \chain^{\id_1}_{t_1}$.
\end{proof}

\begin{proof}[Proof of \thmref{reorg-resilience}]
    By \thmref{strong-persistence-1},
    $\varProposal^*.\varBlock \preceq B$
    for any $\varBlock$ identified
    in \myalgref{protocol}{fcr-1,fcr-2,fcr-3}
    by any awake honest validator
    in any
    $r \geq 3\Delta t + 2\Delta$.
    From $\kappa$ slots later onwards,
    all awake honest validators include the transaction
    in their ledger
    (\myalgref{protocol}{confirmation}).
\end{proof}

Proof of \thmref{strong-persistence-1} follows from
\lemref{honest-leader-almost-everywhere,view-merge-property,all-honest-voting-together},
and is provided at the end of this section.
The structure of the argument is inductive:
\lemref{view-merge-property} shows that in a slot $t$
with honest leader, all honest voters vote for the leader's proposal.
\lemref{all-honest-voting-together} shows that if in slot $t$
all honest voters have voted for a descendant of a certain block,
then in slot $t+1$ all honest voters will vote for a descendant of
that block.

\restateViewMergeProperty*
\begin{proof}
    Let $\V'=\varProposal^*.\V$, and $\B^*$ and $\V^*$ denote the buffer and \bvtree of $\id^*$ at round $3\Delta t$.
    Since $\id^*$ is honest, it must have broadcast $\varProposal^*$ at round $3\Delta t$ with \bvtree $\V' = \Merge(\V^*,\B^*)$ and a new block $\varProposal^*.\varBlock$ with parent $\funOurGhost(\V',t-1)$
    (\myalgref{protocol}{goldfish-simple-merge1,fcr-1,broadcast-prop}).

    By synchrony, any message that a non-asleep honest validator $\id$
    could have added to its \bvtree $\V_\id$
    by $3\Delta (t-1)+2\Delta$,
    is received by $\id^*$ by $3\Delta t$,
    and thus in $\V'$.
    As awake honest validators do not update their \bvtrees
    and no honest validators turn awake
    in the interval $(3\Delta (t-1)+2\Delta,3\Delta t+\Delta)$,
    for any
    honest validator $\id$ awake at round $3 \Delta t + \Delta$,
    $\V_\id \subseteq \V'$
    prior to \myalgref{protocol}{goldfish-simple-merge2}.
    Since $\id^*$ is recognized as the leader of slot $t$ by all awake honest validators at round $3\Delta t + \Delta$, at that round, each awake honest validator $\id$ merges its \bvtree with $\V' \cup \{\varProposal^*.\varBlock\}$
    (\myalgref{protocol}{goldfish-simple-merge2})
    and reaches $\V_\id = \V' \cup \{\varProposal^*.\varBlock\}$.
    Consequently,
    each honest validator $\id$
    awake at round $3\Delta t + \Delta$ and
    eligible to vote at slot $t$
    votes for
    $\varProposal^*.\varBlock$
    due to the recursive structure of the \textOurGhost rule (\algref{ghost}).
\end{proof}

\restateAllHonestVotingTogether*
\begin{proof}
    By \lemref{honest-leader-almost-everywhere}, w.o.p., for every slot $t$,
    the number of adversary validators
    at round $3\Delta (t+1)+\Delta$ and
    eligible to vote at slot $t$
    is less than
    the number of honest validators
    awake at round $3\Delta t + \Delta$ and
    eligible to vote at slot $t$.

    Let $t$ be a slot such that
    all honest validators awake at round $3\Delta t + \Delta$
    and eligible to vote at $t$
    voted for a descendant of $\varBlock$.
    Pick any
    honest validator $\id$
    awake at round $3\Delta (t+1) + \Delta$ and
    eligible to vote at slot $t+1$.
    Since $\id$ must have been awake at least since round $3\Delta t + 2\Delta$, its \bvtree at round $3\Delta t + 2\Delta$ contains all
    votes broadcast by honest validators
    awake at round $3\Delta t + \Delta$
    and eligible to vote at slot $t$ (\myalgref{protocol}{goldfish-simple-merge2}).
    The same is true for its \bvtree at round $3\Delta (t+1) + \Delta$,
    even after $\id$ merges its \bvtree
    with that of any \proposal (\myalgref{protocol}{goldfish-simple-merge1}).
    Moreover, the number of honest validators
    awake at round $3\Delta t + \Delta$
    and eligible to vote at slot $t$
    is greater than
    the number of adversary validators
    at round $3\Delta (t+1)+\Delta$ that are eligible to vote at slot $t$.

    Consequently, upon invoking the \textOurGhost fork-choice rule at round $3\Delta (t+1) + \Delta$ (\myalgref{protocol}{fcr-2}), $\id$ observes
    that at every iteration of the fork choice (\myalgref{ghost}{vote-count}),
    blocks consistent with $\varBlock$ have more votes than blocks
    conflicting with $\varBlock$.
    Thus, at round $3\Delta (t+1) + \Delta$,
    fork choice
    returns a descendant of $\varBlock$, and $\id$ votes for it.
\end{proof}

\begin{proof}[Proof of \thmref{strong-persistence-1}]
    From \lemref{honest-leader-almost-everywhere,view-merge-property,all-honest-voting-together},
    it follows by induction that w.o.p.,
    for all $t' \geq t$,
    all honest validators
    awake at round $3\Delta t' + \Delta$
    and eligible to vote at slot $t'$,
    vote for a descendant of $\varProposal^*.\varBlock$.

    By synchrony, the honest votes of slot $t'$
    reach all honest validators awake at $3\Delta t' + 2 \Delta$ by then,
    when they also merge the votes into their \bvtrees.
    The number of honest validators
    awake at round $3\Delta t' + \Delta$
    and eligible to vote at slot $t'$
    is greater than
    the number of adversary validators
    by round $3\Delta (t'+1)+\Delta$ that are eligible to vote at slot $t'$
    (by \lemref{honest-leader-almost-everywhere}).
    Upon invoking the \textOurGhost rule of
    \myalgref{protocol}{fcr-1,fcr-2,fcr-3}
    at $3\Delta t'+2\Delta$,
    $3\Delta (t'+1)$
    and $3\Delta (t'+1)+\Delta$,
    respectively,
    an awake honest validator $\id$ (who must have been awake since at least
    $3\Delta t'+2\Delta$, due to the joining procedure)
    observes
    that at every iteration of the fork choice (\myalgref{ghost}{vote-count}),
    blocks consistent with $\varProposal^*.\varBlock$ have more votes than blocks
    conflicting with $\varProposal^*.\varBlock$.
    Thus, $\id$'s fork choice
    reaches a descendant of $\varProposal^*.\varBlock$.
\end{proof}

\section{Security Proof of \Goldfish with Fast Confirmation}
\label{sec:appendix-fast-confirmation-analysis}

In the following analysis, we consider a synchronous network in the sleepy model as described in \secref{boilerplate}.
Recall that the total number of validators is $n$ (\cf \secref{boilerplate}).
Since \Goldfish slots consist of $4\Delta$ rounds in the case of fast confirmation, we hereafter assume that the \Goldfish execution is $(\frac{1}{2},4\Delta)$-compliant.
We show that \thmref{security} holds for \Goldfish with fast confirmations (w.o.p.) in compliant executions. To do so, we first prove \thmref{fast-strong-persistence}, an analogue of \thmref{strong-persistence-1} for fast confirmations, showing that fast confirmed blocks are always in the canonical chain of awake validators at later rounds.

Since \Goldfish slots consist of $4\Delta$ rounds in the case of fast confirmation, we state an analogue of \lemref{honest-leader-almost-everywhere} to match the new slot structure:
\begin{lemma}
    \label{lem:fast-honest-leader-almost-everywhere}
    Suppose the \Goldfish execution is $(\frac{1}{2},4\Delta)$-compliant.
    Then, w.o.p., for every slot $t$, the number of adversary validators at round $4\Delta (t+1)+\Delta$, eligible to vote at slot $t$, is less than the number of honest validators, awake at round $4\Delta t + \Delta$ and eligible to vote at slot $t$.
    Also w.o.p., all slot intervals of length $\kappa$
    have at least one slot $t$, where an honest validator is identified as the slot leader by all awake honest validators at round $4\Delta t + \Delta$.
\end{lemma}

Proof of \lemref{fast-honest-leader-almost-everywhere} is analogous to the proof of \lemref{honest-leader-almost-everywhere}, and follows from the same arguments using $(\frac{1}{2},4\Delta)$-compliant executions.

\begin{proposition}
    \label{prop:committee-bound-whp}
    Suppose $\Thorizon = \operatorname{poly}(\kappa)$.
    Then, w.o.p., there can be at most $(1+\epsilon)n\,\thresholdVote$ validators that are eligible to vote at any given slot. If the \Goldfish execution is $(\frac{1}{2}, 4\Delta)$-compliant, then, w.o.p., for all slots $t$, the number of adversary validators at round $4\Delta(t+1)+\Delta$, eligible to vote at slot $t$, is less than $\frac{1}{2}n\,\thresholdVote$.
\end{proposition}
Proof follows from a Chernoff bound.

\begin{lemma}
    \label{lem:fast-base-case}
    Suppose the \Goldfish execution is $(\frac{1}{2}, 4\Delta)$-compliant
    in the synchronous sleepy network model,
    and an honest validator $\id^*$ fast confirms
    a block $\varBlock$ at slot $t$.
    Then, w.o.p, all honest validators
    awake at round $4\Delta (t+1) + \Delta$
    and eligible to vote at slot $t+1$,
    vote for a descendant of $\varBlock$ at slot $t+1$.
\end{lemma}
Proof is stated below and follows from \propref{committee-bound-whp} and a quorum intersection argument.
\begin{proof}[Proof of \lemref{fast-base-case}]
    By \propref{committee-bound-whp}, w.o.p., the number of adversary validators at round $4\Delta(t+1)+\Delta$, eligible to vote at slot $t$, is less than $\frac{1}{2}n\,\thresholdVote$.
    An eligible awake honest validator sends a single slot $t$ vote at round $4\Delta t+\Delta$, implying that over $(\frac{3}{4}+\frac{\epsilon}{2}) n\,\thresholdVote - \frac{1}{2} n\,\thresholdVote = (\frac{1}{4} + \frac{\epsilon}{2}) n\,\thresholdVote$ validators broadcast a single slot $t$ vote by round $4\Delta (t+1)+\Delta$, and that is for a descendant of $\varBlock$.
    By \propref{committee-bound-whp}, w.o.p., for all slots $t$, there can be at most $(1+\epsilon)n\,\thresholdVote$ validators that are eligible to vote at $t$.
    Hence, the number of valid slot $t$ votes for the descendants of any block $\varBlock'$ conflicting with $\varBlock$ must be less than $(1+\epsilon)n\,\thresholdVote -  (\frac{1}{4} + \frac{\epsilon}{2}) n\,\thresholdVote = (\frac{3}{4} + \frac{\epsilon}{2})n\,\thresholdVote$ at any given round.
    The validator $\id^*$ broadcasts $\varBlock$ and over $(\frac{3}{4}+\frac{\epsilon}{2}) n\,\thresholdVote$ valid votes for it (in \pieces) at round $4\Delta t+2\Delta$.
    Each honest validator, awake at round $4\Delta (t+1) + \Delta$ and eligible to vote at slot $t+1$, observes these votes in its \bvtree at the round of voting (\myalgref{fast-protocol}{fast-view-merge2}).
    Upon invoking the \textOurGhost fork-choice rule at any of the rounds $4\Delta t + 3\Delta$, $4\Delta (t+1)$ or $4\Delta (t+1) + \Delta$ (\myalgref{protocol}{fcr-1,fcr-2,fcr-3}), for any awake honest validator $\id$ with \bvtree $\V'$, $\operatorname{\textsc{Votes}}(\V', \varBlock, t) > \operatorname{\textsc{Votes}}(\V', \varBlock', t)$ for any block $\varBlock'$ conflicting with $\varBlock$.
    This implies that all honest validators, awake at round $4\Delta(t+1)+\Delta$ and eligible to vote at slot $t+1$ all vote for $\varBlock$ or one of its descendants at slot $t+1$.
\end{proof}
\begin{theorem}%
    \label{thm:fast-strong-persistence}
    Suppose the \Goldfish execution is $(\frac{1}{2}, 4\Delta)$-compliant in the synchronous sleepy network model, and an honest validator $\id^*$ fast confirms a block $\varBlock$ at slot $t$.
    Then, w.o.p., $\varBlock \preceq B$
    for any $\varBlock$ identified
    in \myalgref{protocol}{fcr-1,fcr-2,fcr-3}
    by any awake honest validator
    in any round
    $r \geq 4\Delta (t+1) + \Delta$.
\end{theorem}
Proof is stated below and follows from \lemref{fast-honest-leader-almost-everywhere, fast-base-case, all-honest-voting-together} and the inductive argument used in the proof of \thmref{strong-persistence-1}.
\begin{proof}[Proof of \thmref{fast-strong-persistence}]
    Follows by \lemref{fast-honest-leader-almost-everywhere, fast-base-case, all-honest-voting-together}, by the same inductive argument used in the proof of \thmref{strong-persistence-1}, in that case following from \lemref{honest-leader-almost-everywhere, view-merge-property, all-honest-voting-together}. Here, \fullVersionRef{\lemref{fast-honest-leader-almost-everywhere}} is the analogue of \lemref{honest-leader-almost-everywhere} with the new slot structure, and \lemref{fast-base-case} provides the base case, substituting \lemref{view-merge-property}.
\end{proof}
\begin{theorem}
    \label{thm:fast-confirmation-safety}
    Suppose the \Goldfish execution is  $(\frac{1}{2},4\Delta)$-compliant.
    Then, \Goldfish with fast confirmations satisfies safety (w.o.p.).
\end{theorem}
Proof is stated below and follows from \thmref{security}.
\begin{proof}[Proof of \thmref{fast-confirmation-safety}]
    If an honest validator fast confirms a block $\varBlock$ at slot $t$, then $\varBlock$ is in the canonical \textOurGhost chain of every awake honest validator at all slots larger than $t$ by \thmref{fast-strong-persistence}.
    Therefore, $\varBlock$ is in the $\kappa$-slots-deep prefix of the canonical \textOurGhost chains of all awake honest validators at slot $t+\kappa$, and thus confirmed by them with the standard confirmation rule. Therefore, \thmref{security} implies the safety of the protocol.
\end{proof}

In $(\frac{1}{2},4\Delta)$-compliant executions, we automatically get liveness of \Goldfish with fast confirmations from the liveness of the standard confirmation rule, since fast confirmation is not needed for a block to be confirmed.
Under optimistic conditions, liveness of fast confirmations holds as well.
We prove that a block within an honest, valid \proposal is immediately fast confirmed within the same slot by the awake honest validators, if there are over $(\frac{3}{4} + \frac{3}{2}\epsilon)n$ awake, honest validators at the voting time of the given slot, implying the liveness of fast confirmations under optimistic conditions.
\begin{theorem}
    \label{thm:fast-confirmations-when-enough-honest}
    Suppose the \Goldfish execution is $(\frac{1}{2},4\Delta)$-compliant.
    Then, \Goldfish with fast confirmations satisfies liveness with $\Tconf = \Theta(\kappa)$ (w.o.p.).

    Consider a slot $t$, such that there are $(\frac{3}{4}+ \frac{3}{2}\epsilon)n\,\thresholdVote$ honest validators eligible to vote at slot $t$ and awake at round $4\Delta t + \Delta$.
    Suppose an honest validator $\id$ with \proposal $\varProposal^*$ is recognized as the leader of a slot $t$ by all awake honest validators at round $4\Delta t + \Delta$ (\myalgref{protocol}{vote-slot-leader}).
    Then all honest validators awake at round $4\Delta t + 2\Delta$ fast confirm $\varProposal^*.\varBlock$ in \myalgref{fast-protocol}{fast-confirmation}.
\end{theorem}
Liveness is stated below and follows from \thmref{security} and fast confirmation from \lemref{view-merge-property}.
\begin{proof}[Proof of \thmref{fast-confirmations-when-enough-honest}]
    Proof of liveness follows from \thmref{security}.

    For the second part of the proof, by \lemref{view-merge-property}, all of eligible and awake honest validators vote for $\varProposal^*.\varBlock$ at slot $t$.
    Then, the buffer of any honest validator awake at round $4\Delta t + 2\Delta$ contains at least $(\frac{3}{4} + \frac{\epsilon}{2})n\,\thresholdVote$ votes (by Chernoff bound) for the block $\varProposal^*.\varBlock$, implying that all honest validators awake at rounds $4\Delta t + 2\Delta$ fast confirm $\varProposal^*.\varBlock$ at the respective slots.
\end{proof}

\section{\Goldfish with Overlay Gadgets}
\label{sec:aa-analysis}

Two properties desired for Ethereum consensus as a whole, besides security under dynamic participation, fast confirmation, and reorg resilience, are \emph{security under partial synchrony} and \emph{accountable safety}.
However, it is impossible for all these properties to be satisfied by a single ledger~\cite{cap,lewispye2020resource,ebbandflow,aadilemma,sankagiri_clc}.
For this reason,
Ethereum's consensus protocol
(Gasper~\cite{gasper}, \figref{gasper-diagram})
consists of an \emph{overlay} finality/accountability gadget (Casper FFG~\cite{casper})
which provides accountable safety under asynchrony, on top of an \emph{underlay}
chain 
which should be secure under synchrony and dynamic participation,
and provide fast confirmations and reorg resilience.
The desiderata for
Gasper
were formalized by
\emph{ebb-and-flow}~\cite{ebbandflow,aadilemma,sankagiri_clc}.
The objective is, slightly more abstractly,
to design a flexible two-ledger consensus protocol,
which supports
a full dynamically available ledger in conjunction with a finalized and accountable prefix ledger. 
The finalized ledger falls behind the full ledger when the network partitions or participation is low,
but catches up when the network heals.
Clients adopt the ledger that provides the property
(accountable safety under network partition,
or liveness under dynamic participation)
which they value more.

In this section,
we show that 
\Goldfish can indeed be used as an underlay
chain in conjunction with an overlay
finality/accountability gadget,
and that the so composed protocol satisfies
the design goal for Gasper, 
the ebb-and-flow formulation~\cite{ebbandflow,aadilemma}.
Specifically, for ease of exposition,
since the focus of this paper is on designing
a new underlay, \Goldfish,
rather than designing the 
gadget/overlay and the checkpointing interaction between underlay and overlay,
and since we are not aware of 
any formal work showing how to apply
Casper to an underlay chain to obtain a secure
ebb-and-flow protocol,
we instead reuse finality/accountability gadgets
from the literature~\cite{ebbandflow,sankagiri_clc,aadilemma}.
Given their earlier analyses, the primary job left to do
for us as designers of the \Goldfish underlay,
is to show that \Goldfish `heals' (to be made precise below)
after network partition and in conjunction with the gadget
(\ie, despite the gadget's influence over the underlay).

\secref{appendix-model-ebb-and-flow} presents the formal model and problem formulation.
\secref{protocol-gadgets} describes the composition of \Goldfish with accountability gadgets~\cite{aadilemma}.
A security proof for the composition is presented in \secref{overview-of-analysis-ebb-and-flow,appendix-chainava-security,appendix-chainacc-liveness,appendix-ebb-and-flow-security}.

\subsection{Model}
\label{sec:appendix-model-ebb-and-flow}

\subsubsection{Partial Synchrony}

Security under a partially synchronous network captures the resilience of the consensus protocol against network partitions.
A \emph{partially synchronous network in the sleepy model} \cite{ebbandflow} is characterized by a global stabilization time (\GST), a global awake time (\GAT), and a delay upper-bound $\Delta$.
$\GST$ and $\GAT$ are constants
unknown to the honest validators
chosen \emph{adaptively} by the adversary,
\ie, as causal functions
of the execution,
whereas $\Delta$ is a constant known to the validators.
Before $\GST$, message delays are arbitrarily adversary (\emph{asynchronous}).
After $\GST$, message delays are subject to the delay upper bound $\Delta$ (\emph{synchronous}).
Similarly, before $\GAT$,
the adversary can set the sleep schedule for honest validators.
After $\GAT$, all honest validators are awake.

\subsubsection{Security}
We next formalize the notion of security \emph{after a certain time}, generalizing \defref{security}.
Security is parameterized by $\kappa$, which, for longest-chain protocols and \Goldfish, determines the confirmation delay for transactions (\ie, these protocols come with a security--latency trade-off).
We consider a finite time horizon $\Thorizon$ that is polynomial in $\kappa$.
We denote a consensus protocol's output
ledger, \eg, the \Goldfish ledger, in the view of a validator $i$ at round $r$ by $\chain^{i}_{r}$.
We write $\chain_1 \preceq \chain_2$ to express that the ledger $\chain_1$ is
a prefix of (or the same as) ledger $\chain_2$.

\begin{definition}[Security]
  \label{def:security-after-new}
  Let $\Tconf$ be a polynomial function of the security parameter $\kappa$.
  A state machine replication protocol
  that outputs a ledger $\chain{}{}$ is \emph{secure after time $\Tafter$}, and has transaction confirmation time $\Tconf$, iff:

  \noindent
  \textbf{Safety:} For any two rounds $r, r' \geq \Tafter$, and any two honest validators $i, j$
  awake at rounds $r$ and $r'$, respectively, either $\chain^{i}_{r} \preceq \chain^{j}_{r'}$ or $\chain^{j}_{r'} \preceq \chain^{i}_{r}$.

  \noindent
  \textbf{Liveness:} If a transaction has been received by some awake honest validator by some round $r \geq \Tafter$, then for any round $r' \geq r+\Tconf$ and any honest validator $i$ awake at round $r'$, the transaction will be included in $\chain^{i}_{r'}$.
\end{definition}
The protocol satisfies \emph{$\bar{f}$-safety} (\emph{$\bar{f}$-liveness}) if it satisfies safety (liveness) as long as the number of adversary validators $f$ stays below $\bar{f}$ for all rounds.
It satisfies \emph{$1/2$-safety} (\emph{$1/2$-liveness}) if it satisfies safety (liveness) if the fraction of adversary validators $\beta$
is bounded above away from $1/2$
for all rounds.

\subsubsection{Accountable Safety}
Accountable safety provides a \emph{trust-mini\-mizing} strengthening of safety, with the aim to hold validators accountable for their actions.
In a protocol with accountable safety resilience $\bar{f}>0$, after a safety violation, one can, upon collecting evidence from sufficiently many honest validators, generate a cryptographic proof that identifies $\bar{f}$ adversary validators as protocol violators~\cite{forensics,aadilemma}.
By definition, the proof does not falsely accuse any honest validator, except with negligible probability.

\subsubsection{The Ebb-and-Flow Formulation}

As \Goldfish outputs a \emph{dynamically available} ledger (\ie, live under dynamic participation), by the availability-accountability dilemma \cite{aadilemma}, its output ledger cannot satisfy accountable safety.
Similarly, it cannot satisfy safety under a partially synchronous network (\ie, \emph{finality}), by an analogue of the CAP theorem~\cite{cap,lewispye2020resource}.
However, \Goldfish can be used as an underlay
composed with an accountability gadget as overlay
(\cf \figref{gasper-diagram,block-diagram})
in order to obtain a separate prefix ledger that attains accountable safety under partial synchrony while staying consistent with the output of \Goldfish~\cite{aadilemma}.
Denoting the output of \Goldfish as the available ledger $\chainava$ and that of the accountability gadget as the accountable final prefix ledger $\chainacc$,
the desiderata are captured in the \emph{ebb-and-flow formulation}~\cite{ebbandflow}:
\begin{definition}[Ebb-and-flow formulation~\cite{ebbandflow,aadilemma}]
  \label{def:ebbandflow}
  \noindent
  \begin{enumerate}
    \item (\textbf{P1: Accountability and finality})
          Under a partially synchronous network in the sleepy model, the accountable final prefix ledger $\chainacc$ has accountable safety resilience $n/3$ at all times,
          (except w.p.\ $\negl(\lambda)$),
          and there exists a constant $\mathbf{C}$ such that $\chainacc$ provides $n/3$-liveness with confirmation time $\Tconf$ after round $\max(\GST,\GAT)+\mathbf{C} \cdot \kappa$ (w.o.p.).

    \item (\textbf{P2: Dynamic availability})
          Under a synchronous network in the sleepy model (\ie, for $\GST=0$), the available ledger $\chainava$ provides $1/2$-safety and $1/2$-liveness at all times (w.o.p.).

    \item (\textbf{Prefix})
          For each honest $\id$ and round $r$,
          $\chainaccPARAM{\id}{r} \preceq \chainavaPARAM{\id}{r}$.
  \end{enumerate}
\end{definition}
The accountable final prefix ledger $\chainacc$ can experience liveness violations before $\GST$ or $\GAT$, due to lack of timely communication among sufficiently many honest validators, but $\chainacc$ remains accountably safe throughout.
The available ledger $\chainava$ can experience safety violations before $\GST$, but remains live throughout.
When conditions improve,
$\chainacc$ catches up with
$\chainava$.
This ebb-and-flow behavior lends the formulation its name.
Providing the irreconcilable properties
in two separate but \emph{consistent} ledgers
provides a user-dependent resolution to the CAP theorem~\cite{cap,lewispye2020resource}.

\subsection{\Goldfish with Accountability Gadgets}
\label{sec:protocol-gadgets}

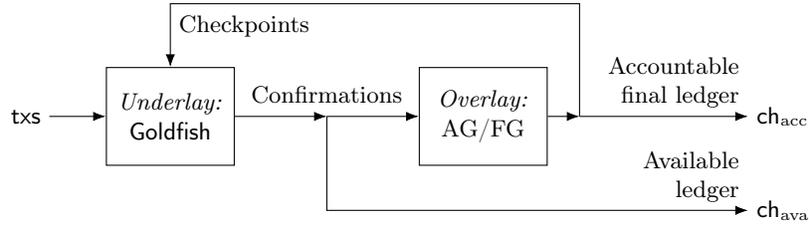
\begin{figure}[tb]
  \centering%
  \begin{tikzpicture}[x=1.6cm,y=1cm]
    \footnotesize

    \node (txs) at (-1.5,0) {$\txs$};

    \node [draw,align=center,minimum width=1.7cm,minimum height=1.3cm] (lmdghost) at (-0.3,0) {\emph{Underlay:}\\\Goldfish};

    \coordinate (logavailable) at (1,0) {};

    \node [draw,align=center,minimum width=1.7cm,minimum height=1.3cm] (casperffg) at (2.3,0) {\emph{Overlay:}\\AG/FG};

    \coordinate (logaccountable) at (3.1,0) {};
    \coordinate (logaccountableOUT) at (4.5,0) {};

    \draw [-Latex] (txs) -- (lmdghost);
    \draw [-Latex] (lmdghost) -- (logavailable) node [above,pos=1,yshift=2pt] {Confirmations};
    \draw [-Latex] (logavailable) -- (casperffg);
    \draw [-Latex] (casperffg) -- (logaccountable);
    \draw [-Latex] (logaccountable) -- (logaccountableOUT) node [pos=1,above,align=right,anchor=south east] {Accountable\\final ledger} node [pos=1,right] {$\chainacc$};
    \draw [-Latex] (logaccountable) -- ++(0,1.5) -- ++(-3.4,0) -- (lmdghost) node [pos=0.35,right] {Checkpoints};
    \draw [-Latex] (logavailable) -- ++(0,-1.25) -- ++(2,0) -- ++(1.5,0) node [pos=1,above,align=right,anchor=south east] {Available\\ledger} node [pos=1,right] {$\chainava$};
\end{tikzpicture}%
  \vspace{-1em}%
  \caption{%
    An accountability/finality gadget (AG/FG, a.k.a.\ \emph{overlay}, like Casper)
    checkpoints decisions of the dynamically available protocol \Goldfish (a.k.a.\ \emph{underlay}) (\cf \figref{gasper-diagram}).
    A feedback loop ensures that
    \Goldfish respects earlier checkpoints.
    This construction
    satisfies
    the \emph{ebb-and-flow} design objective
    of Ethereum,
    to
    produce an available full ledger that is secure
    under dynamic participation of validators,
    and a prefix ledger that is accountably secure under network partition~\cite{ebbandflow,aadilemma}.%
  }%
  \label{fig:block-diagram}%
\end{figure}

\begin{algorithm}[tb]
    \caption{\textOurGhost
        (\cf~\algref{ghost})
        \textcolor{NiceGreenColor}{modified (green) to respect the latest checkpoint $\varBlock$}.
        See \algref{ghost} for
        $\operatorname{\textsc{Children}}$
        and
        $\operatorname{\textsc{Votes}}$.%
    }
    \label{alg:modified-ghost}
    \begin{algorithmic}[1]
        \scriptsize
        \Function{\sc GHOST-Eph}{$\V, t, \textcolor{NiceGreenColor}{\varBlock}$}
            \CommentLine{\textcolor{NiceGreenColor}{Start fork-choice from latest checkpoint $\varBlock$}}
            \label{line:cgenesis}
            \Forever%
                \CommentLine{Choose the \emph{heaviest} subtree rooted (breaking ties deterministically) at one of the children blocks $\varBlock'$ of $\varBlock$, by number of validators that have cast a vote in slot $t$ into the subtree rooted at $B'$; $\varBlock' = \bot$ if $\operatorname{\textsc{Children}}(\V,\varBlock) = \emptyset$}
                \Let{\varBlock'}{\argmax_{\varBlock'\in\operatorname{\textsc{Children}}(\V,\varBlock)} \textsc{Votes}(\V, \varBlock',t)}
                \If{$\varBlock' = \bot$}
                    \Return $\varBlock$
                \EndIf
                \Let{\varBlock}{\varBlock'}
            \EndForever
        \EndFunction
    \end{algorithmic}
\end{algorithm}

\begin{algorithm}[tb]
    \caption{%
    Composition of \Goldfish and accountability gadget
    (\cf~\figref{block-diagram}, \cite[Alg.~1]{aadilemma}),
    executed by validator $\id$.
    Here, \Goldfish (\cf~\algref{protocol})
    uses a modified \textOurGhost 
    rule (\algref{modified-ghost}),
    starting the recursion
    from the latest checkpoint, \ie, the last block of $\chainacc^\id$.
    Throughout,
    \Goldfish maintains the available chain $\chainava^\id$.
    \textsc{RunAccountabilityGadget} attempts the next iteration
    of the gadget,
    where valid checkpoint candidates
    are determined using
    $\chainava^\id$.
    Iterations may fail ($\bot$), \eg, if the gadget invokes a malicious leader.
    }
    \label{alg:full-protocol}
    \begin{algorithmic}[1]
        \scriptsize
        \Let{\chainacc^\id}{\varBlock_0}
        \Comment{`Zero-th' checkpoint: \Goldfish's genesis block}
        \For{$c=1,2,\ldots$}
            \Comment{Checkpoint iterations}
                \Let{\mathsf{checkpoint}}{\textsc{RunAccountabilityGadget}(\chainava^{\id})}   \label{line:accgadget}
                \If{$\mathsf{checkpoint} \neq \bot$}
                    \Let{\chainacc^\id}{\mathsf{checkpoint}}  \label{line:modified}
                    \Comment{Update latest checkpoint}
                    \Sleep{\Tcheckpoint}
                \EndIf
        \EndFor
    \end{algorithmic}
\end{algorithm}

For the composition of \Goldfish with accountability gadgets, we follow the construction of~\cite{aadilemma,sankagiri_clc}
(\figref{block-diagram}, \algref{full-protocol}).
In this construction,
a partially synchronous accountably-safe consensus protocol such as Streamlet, Tendermint, or HotStuff~\cite{streamlet,tendermint,yin2018hotstuff,snapandchat},
with accountable safety resilience of $n/3$ out of $n$ validators,
is used to determine checkpoints of \Goldfish's output ledger.
To ensure that \Goldfish respects earlier checkpoints,
its fork-choice rule is modified to respect
earlier checkpoint decisions
(\cf~\algref{modified-ghost}).
The most recent checkpoint forms
the accountably-safe finalized prefix ledger $\chainacc$,
while \Goldfish's output forms the dynamically available full ledger $\chainava$
(\cf~ebb-and-flow, \defref{ebbandflow}).
As \Goldfish now respects checkpoints, $\chainacc \preceq \chainava$ holds.

The full protocol proceeds in checkpointing iterations
(\cf~\algref{full-protocol}).
Iterations may fail, \eg, when the consensus protocol of the gadget
invokes a malicious leader, or during asynchrony before $\GST$,
or while many validators are asleep before $\GAT$.
Successful checkpoint iterations are separated
by at least
$\Tcheckpoint$ rounds of inactivity of the gadget.
In the following sections,
we apply the techniques of earlier analyses~\cite{aadilemma,sankagiri_clc}
to the combination of \Goldfish and the accountability gadget,
to
show how to tune
$\Tcheckpoint$ as a function of the network delay $\Delta$
and the confirmation parameter $\kappa$,
and to formally prove that the combination satisfies the ebb-and-flow desiderata:

\begin{theorem}[Ebb-and-flow property]
  \label{thm:ebb-and-flow-formalized}
  \Goldfish combined with accountability gadgets (\cf~\secref{protocol-gadgets}) satisfies the ebb-and-flow property of \defref{ebbandflow}.
\end{theorem}

Proof of \thmref{ebb-and-flow-formalized} is provided in \secref{overview-of-analysis-ebb-and-flow,appendix-chainava-security,appendix-chainacc-liveness,appendix-ebb-and-flow-security}.
It follows the same blueprint as the original construction of accountability gadgets in \cite[Appendices B, C]{aadilemma}.

\myparagraph{Fast Confirmation Rule and Accountability Gadgets}
When composing accountability gadgets and \Goldfish with the fast confirmation rule, we stipulate that the validators input to the gadget only those blocks confirmed via the standard confirmation rule ($\funOurGhost(\V,t)^{\lceil \kappa}$) in their view.
This is necessary to ensure that all honest validators promptly agree on the confirmation status of the blocks input to the gadget for checkpointing, which in turn is a prerequisite for the liveness of the accountable final prefix ledger $\chainacc$.
Otherwise, it is possible that a block fast confirmed by one honest validator might not become confirmed in the view of another honest validator until after $\kappa$ slots, stalling the checkpointing process of the accountability gadget for that block.
Thus, the fast confirmation rule is primarily for reducing the latency of the available ledger $\chainava$, and does not affect the time for a block to enter the accountable final prefix ledger $\chainacc$.

\subsection{Overview of the Analysis}
\label{sec:overview-of-analysis-ebb-and-flow}

Recall that in \Goldfish with accountability gadgets, the fork-choice rule of \Goldfish is modified to respect earlier checkpoint decisions (\algref{modified-ghost}).
This modification requires adjustments of the analysis of \Goldfish,
because
it opens up the possibility that for a \proposal $\varProposal^*$ by an honest leader, $\varProposal^*.\varBlock \preceq \varBlock$ no longer holds for all blocks~$\varBlock$ identified in \myalgref{protocol}{fcr-1,fcr-2,fcr-3} by awake honest validators at future rounds, due to a new checkpoint conflicting with $\varProposal^*.\varBlock$.

In the synchronous sleepy network model, \secref{analysis} implies that $\chainava$ remains secure until the first checkpoint is determined.
Therefore, checkpoints cannot undermine its security since only confirmed blocks in $\chainava$ are approved for checkpointing by honest validators (\secref{appendix-chainava-security} for formal analysis).
However, when $\GST>0$, one cannot directly rely on the analysis of Goldfish under synchrony.
In this case, to prevent checkpoints from undermining the security,
and rigorously argue security for the combination despite
the modified fork-choice rule,
the framework of accountability gadgets~\cite{aadilemma,sankagiri_clc} relies on two principles:
\begin{itemize}
  \item \textbf{Gap property:} After a (successful) checkpointing iteration with a new checkpoint, honest validators wait for $\Tcheckpoint = \Theta(\kappa)$ rounds before participating in the next iteration.
  \item \textbf{Recency property:} For checkpointing, honest validators suggest and approve only the blocks that were \emph{recently} confirmed as part of $\chainava$.
\end{itemize}
We prove that once the network heals and honest validators become awake at round $\max(\GST,\GAT)$, $\chainava$ regains its security with the help of these properties, a feature called the \emph{healing property}.
The healing property, together with the liveness of the accountability gadget's consensus protocol imply the liveness of $\chainacc$ in the partially synchronous sleepy network model (\secref{appendix-chainacc-liveness} for formal analysis).
Finally, accountable safety of $\chainacc$ follows from the accountable safety of the gadget.
Security of $\chainava$ under the synchronous sleepy network model, accountable safety of $\chainacc$ and its liveness after $\max(\GST,\GAT)$ together imply the ebb-and-flow property, \ie Theorem~\thmref{ebb-and-flow-formalized} (\secref{appendix-ebb-and-flow-security}).

We now formally prove the ebb-and-flow property for \Goldfish combined with accountability gadgets (\figref{block-diagram}).
The following analysis extensively refers to the details of the accountability gadgets described in \cite[Section 4]{aadilemma}.
To distinguish the votes cast by validators as part of the accountability gadget iterations from those broadcast within \Goldfish, we will refer to the former as \emph{gadget votes}.
Similarly, to distinguish the leaders of accountability gadget iterations from the leaders of \Goldfish slots, we will refer to the former as the \emph{iteration leaders}.
We refer the reader to \cite{aadilemma} for the accountability gadget specific definitions of the timeout parameter $\Ttimeout$ and the confirmation delay $\Tbft$ of the BFT protocol.
We highlight that honest iteration leaders propose only the blocks $\varBlock^*$ that are \emph{confirmed} in their view of $\chainava$, \ie, $\varBlock^* \preceq \varBlock^{\lceil \kappa}$ for $\varBlock$ identified in \myalgref{protocol}{fcr-1,fcr-2,fcr-3} run using $\chainava$.
Similarly, honest validators send accepting gadget votes only for the checkpointing proposals that are \emph{confirmed} in their view of $\chainava$.
We set $\Tcheckpoint$, the time gap between the accountability gadget iterations, to be at least $6\Delta(\kappa+1)+\Ttimeout+\Tbft$ (this is necessary for proving the ebb-and-flow property as will be evident in the following proofs).
This makes the upper bound $\Tupper$ on the total duration of an iteration $\Tcheckpoint+\Ttimeout+\Tbft = 6\Delta(\kappa+1) + 2(\Ttimeout + \Tbft) = \Theta(\kappa)$.

\subsection{Security of $\chainava$ under the Synchronous Sleepy Network Model}
\label{sec:appendix-chainava-security}

We first show that $\chainava$ remains secure under synchrony in the sleepy network model, despite the added gadget.

\begin{proposition}
  \label{prop:do-not-alter}
  Suppose a $(\frac{1}{2},3\Delta)$-compliant execution of \Goldfish in the synchronous sleepy network model of \secref{model}.
  If a block $\varBlock$ is observed to be checkpointed by an honest validator for the first time at some round $r$, then $\varBlock$ is in the common prefix of the chains identified in \myalgref{protocol}{fcr-1,fcr-2,fcr-3} right before round $r$ by all awake honest validators.
\end{proposition}

\begin{proof}
  Since the execution is $(\frac{1}{2},3\Delta)$-compliant, for a block to become checkpointed, at least one honest validator must have sent an accepting gadget vote for that block.
  Let $\varBlock_i$ denote the sequence of checkpointed blocks listed in the order of the rounds $r_i$ at which, an awake honest validator observed $\varBlock_i$ to be checkpointed for the first time.
  Proof is by induction on these blocks' indices.

  \myparagraphlight{Induction Hypothesis} $\varBlock_i$ is in the common prefix of the chains identified in \myalgref{protocol}{fcr-1,fcr-2,fcr-3} right before round $r_i$ by all awake honest validators, and stays so until at least round $r_{i+1}$.

  \myparagraphlight{Base Case} Since an honest validator sends an accepting gadget vote only for a confirmed block (\ie, $\kappa$ slots deep), $\varBlock_1$ must have been confirmed by an honest validator at some slot $t_1$ before round $r_1$.
  As all honest validators start the fork-choice at the genesis block prior to $r_1$ and $\varBlock_1$ is confirmed in an honest view, it is in the prefix of a block proposed by an honest leader by \lemref{honest-leader-almost-everywhere} and \thmref{strong-persistence-1}.
  Hence, $\varBlock_1$ is in the common prefix of the chains identified in \myalgref{protocol}{fcr-1,fcr-2,fcr-3} right before round $r_1$ by all awake honest validators.
  It also stays in the common prefix until at least round $r_2$.

  \myparagraphlight{Inductive Step} By the induction hypothesis, checkpointing of the blocks $\varBlock_1, \ldots, \varBlock_{i-1}$ does not alter the fork-choice rule at \myalgref{modified-ghost}{cgenesis} for any awake honest validator.
  Hence, by the same reasoning above, $\varBlock_i$ is in the common prefix of the chains identified in \myalgref{protocol}{fcr-1,fcr-2,fcr-3} right before round $r_i$ by all awake honest validators, and stays so until at least round $r_{i+2}$.
\end{proof}

\begin{lemma}[Safety and liveness of $\chainava$ under synchrony]
  \label{lem:p2}
  Suppose a $(\frac{1}{2},3\Delta)$-compliant execution of \Goldfish in the synchronous sleepy network model of \secref{model}.
  Then, w.o.p., the available ledger $\chainava$ satisfies $1/2$-safety and $1/2$-liveness (at all times).
\end{lemma}

\begin{proof}
  By \propref{do-not-alter}, checkpointing of blocks does not alter the fork-choice rule at \myalgref{modified-ghost}{cgenesis} for any awake honest validator.
  Concretely, if the honest validators started the fork-choice rule from the genesis block at all rounds instead of the latest checkpoint in view, then they would end up with the same execution.
  Thus, the security of $\chainava$ follows from \thmref{security}.
\end{proof}

\subsection{Liveness of $\chainacc$ after $\max(\GST,\GAT)$}
\label{sec:appendix-chainacc-liveness}

We next demonstrate the liveness of $\chainacc$ after $\max(\GST,\GAT)$.
In the subsequent analysis, the total number of validators is denoted by $n$ (\cf \secref{boilerplate}).
The accountability gadget is instantiated with a BFT protocol that has an accountable safety resilience of $n/3$.

\begin{proposition}[Prop.~2 of \cite{aadilemma}]
  \label{prop:bft-live}
  The BFT protocol satisfies $n/3$-liveness after $\max(\GST,\GAT)$ with transaction confirmation time $\Tbft<\infty$.
\end{proposition}

\begin{proposition}[Prop.~3 of \cite{aadilemma}]
  \label{prop:entering-iter-after-gst}
  Consider a $(\frac{1}{3},3\Delta)$-compliant execution of \Goldfish in the partially synchronous sleepy network model of \secref{model}.
  Suppose a block from iteration $c$ was checkpointed in the view of an honest validator at round $r$.
  Then, every honest validator enters iteration $c+1$ by round $\max(\GST,\GAT,r)+\Delta$.

  Let $c'$ be the largest iteration such that a block $\varBlock$ was checkpointed in the view of some honest validator before $\max(\GAT,\GST)$. (Let $c'=0$ and $\varBlock$ be the genesis block if there does not exist such an iteration.)
  If an honest validator enters an iteration $c'' > c'$ at some round $r \geq \max(\GAT,\GST) + \Delta + \Tcheckpoint$, every honest validator enters iteration $c''$ by round $r + \Delta$.
\end{proposition}

Proof of \propref{entering-iter-after-gst} follows from the proof of \cite[Prop.~3]{aadilemma}.
\begin{proof}
  Suppose a block $\varBlock$ from iteration $c$ was checkpointed in the view of an honest validator $\id$ at round $r$.
  Then, there are over $2n/3$ accepting gadget votes for $\varBlock$ from iteration $c$ on $\LOGbft{\id}{r}$, the output ledger of the BFT protocol in $\id$'s view at round $r$.
  All gadget votes and BFT protocol messages observed by $\id$ by round $r$ are delivered to all other honest validators by round $\max(\GST,\GAT,r)+\Delta$.
  Hence, by the safety of the BFT protocol when $f<n/3$, for any honest validator $\id'$, the ledger $\LOGbft{\id}{r}$ is the same as or a prefix of the ledger observed by $\id'$ at round $\max(\GST,\GAT,r)+\Delta$.
  Thus, for any honest validator $\id'$, there are over $2n/3$ accepting gadget votes for $\varBlock$ from iteration $c$ on $\LOGbft{}{}$ at round $\max(\GST,\GAT,r)+\Delta$.
  This implies every honest validator enters iteration $c+1$ by round $\max(\GST,\GAT,r)+\Delta$.

  Finally, by the reasoning above, all honest validators enter iteration $c'+1$ by round $\max(\GAT,\GST)+\Delta$.
  Thus, entrance time of the honest validators to subsequent iterations have become synchronized by round $\max(\GAT,\GST) + \Delta + \Tcheckpoint$:
  If an honest validator enters an iteration $c'' > c'$ at some round $r \geq \max(\GAT,\GST) + \Delta + \Tcheckpoint$, every honest validator enters iteration $c''$ by round $r + \Delta$.
  Similarly, if a block from iteration $c''$ is first checkpointed in the view of an honest validator at some round after $\max(\GAT,\GST) + \Delta + \Tcheckpoint$, then it is checkpointed in the view of all honest validators within $\Delta$ rounds.
\end{proof}

\begin{lemma}[Liveness of $\chainacc$, analogue of Thm.~4 of \cite{aadilemma}]
  \label{lem:liveness-lemma}
  Consider a $(\frac{1}{3},3\Delta)$-compliant execution of \Goldfish in the partially synchronous sleepy network model of \secref{model}.
  Suppose $\chainava$ is secure (safe and live) after some round $\Theal \geq \max(\GST,\GAT)+\Delta+\Tcheckpoint$.
  Then, w.o.p., $\chainacc$ satisfies $n/3$-liveness after round $\Theal$ with transaction confirmation time $\Tconf = \Theta(\kappa^2)$.
\end{lemma}

Proof of \lemref{liveness-lemma} follows from the proof of \cite[Thm.~4]{aadilemma}.
\begin{proof}
  By \propref{bft-live}, $\LOGbft{}{}$ is live with transaction confirmation time $\Tbft$ after $\max(\GST,\GAT)$, a fact we will use subsequently.

  Let $c'$ be the largest iteration such that a block $\varBlock$ was checkpointed in the view of some honest validator before $\max(\GAT,\GST)$ (Let $c'=0$ and $\varBlock$ be the genesis block if there does not exist such an iteration).
  Then, by \propref{entering-iter-after-gst}, entrance times of the honest validators to subsequent iterations become synchronized by round $\max(\GAT,\GST)+\Delta+\Tcheckpoint$: If an honest validator enters an iteration $c > c'$ at some round $r \geq \max(\GAT,\GST) + \Delta + \Tcheckpoint$, every honest validator enters iteration $c$ by round $r + \Delta$.

  Suppose an iteration $c>c'$ has an honest iteration leader $\ld{c}$, which sends a checkpoint proposal, denoted by $\bprop{c}$, at some round $r > \Theal + \Tcheckpoint$.
  The proposal $\bprop{c}$ is received by every honest validator by round $r+\Delta$.
  Since the entrance times of the validators are synchronized by $\Theal \geq \max(\GST,\GAT)+\Delta+\Tcheckpoint$, every honest validator sends a gadget vote by round $r+\Delta$.
  By \lemref{healing}, $\bprop{c} \preceq \varBlock^{\lceil \kappa}$ for any $\varBlock$ identified in \myalgref{protocol}{fcr-1,fcr-2,fcr-3} by any awake honest validator after $r$.
  Moreover, $\bprop{c}$ is a descendant all of the checkpoints seen by the honest validators until then.
  Consequently, at iteration $c$, every honest validator sends a gadget vote accepting $\bprop{c}$ by round $r+\Delta$, all of which appear within $\LOGbft{}{}$ in the view of every honest validator by round $r+\Delta+\Tbft$.
  Thus, $\bprop{c}$ becomes checkpointed in the view of every honest validator by round $r+\Delta+\Tbft$.
  (Here, we assume that $\Ttimeout$ was chosen large enough for $\Ttimeout>\Delta+\Tbft$ to hold.)

  Since $r>\Theal+\Tcheckpoint$, by \lemref{healing}, $\bprop{c}$ contains at least one honest block since an earlier checkpointed block in its prefix from before iteration $c$.
  This implies that the prefix of $\bprop{c}$ contains at least one fresh honest block that enters $\chainacc$ by round $r+\Delta+\Tbft$.

  Next, we show that an adversary leader cannot make an iteration last longer than $\Delta+\Ttimeout+\Tbft$ for any honest validator after the initial $\Tcheckpoint$ period elapsed.
  Indeed, if an honest validator $\id$ enters an iteration $c$ at round $r-\Tcheckpoint$, by round $r+\Ttimeout$, either it sees a block (potentially $\bot$) become checkpointed for iteration $c$, or it sends a reject vote for iteration $c$.
  In the first case, every honest validator sees a block checkpointed for iteration $c$ by round at most $r+\Ttimeout+\Delta$.
  In the second case, rejecting gadget votes from over $2n/3 > n/3$ validators appear in $\LOGbft{}{}$ in the view of every honest validator by round at most $r+\Ttimeout+\Delta+\Tbft$.
  Hence, a new checkpoint, potentially $\bot$, is output in the view of every honest validator by round $r+\Ttimeout+\Delta+\Tbft$.

  Finally, we observe that except with probability $(1/3)^\kappa$, there exists a checkpoint iteration with an honest leader within $\kappa$ consecutive iterations.
  Since an iteration lasts at most $\max(\Delta+\Ttimeout+\Tbft,\Delta+\Tcheckpoint+\Tbft) \leq \Delta+\Tcheckpoint+\Ttimeout+\Tbft = \Theta(\kappa)$ rounds, and a new checkpoint containing a fresh honest block in its prefix appears when an iteration has an honest leader (\lemref{healing}), w.o.p., any transaction received by an honest validator at round $t$ appears within $\chainacc$ in the view of every honest validator by round at most $t+\kappa(\Delta+\Ttimeout+\Tbft+\Tcheckpoint)$.
  Hence, via a union bound over the total number of iterations (which is a polynomial in $\kappa$), we observe that if $\chainava$ satisfies security after some round $\Theal$, then w.o.p., $\chainacc$ satisfies liveness after $\Theal$ with a transaction confirmation time $\Tconf = \Theta(\kappa^2)$.
\end{proof}

The latency expression $\Tconf=\Theta(\kappa^2)$ stated in \lemref{liveness-lemma} is a \emph{worst-case} latency to guarantee that an honest block enters the accountable, final prefix ledger $\chainacc$ with overwhelming probability.
In the expression, the first $\kappa$ term comes from the requirement to have $\Tcheckpoint = \Theta(\kappa)$ slots in between the accountability gadget iterations, and the second $\kappa$ term comes from the fact that it takes $\Theta(\kappa)$ iterations for the accountability gadget to have an honest iteration leader except with probability $\negl(\kappa)$.
The accountability gadget protocol asks honest validators to wait for $\Tcheckpoint = \Theta(\kappa)$ slots in between iterations to ensure the security of the protocol, reasons for which will be evident in the proof of \lemref{healing}.

Unlike the worst-case latency, the expected latency for an honest block to enter $\chainacc$ after $\chainava$ regains its security would be $\Theta(\kappa)$ as each checkpointing iteration has an honest leader with probability at least $2/3$.
In this context, the latency of $\Theta(\kappa)$ is purely due to the requirement to have $\Tcheckpoint = \Theta(\kappa)$ slots in between the accountability gadget iterations.
Here, waiting for $\Tcheckpoint$ slots in between iterations guarantees the inclusion of a new honest block in $\chainava$, which in turn appears in the prefix of the next checkpoint, implying a liveness event whenever there is an honest iteration leader.

\lemref{liveness-lemma} requires the available ledger $\chainava$ to eventually regain security under partial synchrony when there are less than $n/3$ adversary validators.
Towards this goal, we first analyze the gap and recency properties, the core properties that must be satisfied by the accountability gadget for recovery of security of $\chainava$.
The gap property states that the blocks are checkpointed sufficiently apart in time, controlled by the parameter $\Tcheckpoint$:
\begin{proposition}[Gap property, analogue of Prop.~4 of \cite{aadilemma}]
  \label{prop:checkpoint-gap}
  Consider a $(\frac{1}{3},3\Delta)$-compliant execution of \Goldfish in the partially synchronous sleepy network model of \secref{model}.
  Given any round interval of size $\Tcheckpoint$, no more than a single block can be checkpointed in the interval in the view of any honest validator.
\end{proposition}
Proof of \propref{checkpoint-gap} follows from the fact that upon observing a new checkpoint that is not $\bot$ for an iteration, honest validators wait for $\Tcheckpoint$ rounds before sending gadget votes for the checkpoint proposal of the next iteration, and there cannot be two conflicting checkpoints for the same iteration in the view of any honest validator.

As in \cite{aadilemma} and \cite{sankagiri_clc}, we state that a block $\varBlock^*$ checkpointed at iteration $c$ and round $r>\max(\GST,\GAT)$ in the view of an honest validator $\id$ is $\Trecent$-recent if $\varBlock^* \preceq \varBlock^{\lceil \kappa}$ for $\varBlock$ identified in \myalgref{protocol}{fcr-3} by $\id'$ at some round within $[r-\Trecent,r]$.
Then, we can express the recency property as follows:
\begin{lemma}[Recency property, analogue of Lem.~1 of \cite{aadilemma}]
  \label{lem:checkpoint-recency}
  Consider a $(\frac{1}{3},3\Delta)$-compliant execution of \Goldfish in the partially synchronous sleepy network model of \secref{model}.
  Every checkpointed block proposed after $\max(\GST,\GAT)$ is $\Trecent$-recent for $\Trecent = \Delta+\Ttimeout+\Tbft$.
\end{lemma}
\begin{proof}
  By the proof of \lemref{liveness-lemma}, if a block $\varBlock$ proposed after $\max(\GST,\GAT)$ is checkpointed in the view of an honest validator at some round $r$, it should have been proposed after round $r-(\Delta+\Ttimeout+\Tbft)$.
  Moreover, over $2n/3$ validators must have sent accepting gadget votes for $\varBlock$ by round $r$.
  Let $\id$ denote such an honest validator.
  It would vote for $\varBlock$ only after it sees the checkpoint proposal for iteration $c$, \ie, after round $r-\Trecent = r-(\Delta+\Ttimeout+\Tbft)$, and only if the proposal is confirmed in its view.
  Hence, $\varBlock$ must be $\kappa$ slots deep in the chain returned at \myalgref{protocol}{fcr-3} by validator $\id$ at some round within $[r-\Trecent,r]$.
  This concludes the proof that every checkpointed block proposed after $\max(\GST,\GAT)$ is $\Trecent$-recent.
\end{proof}

\begin{lemma}[Healing property, analogue of Thm.~5 of \cite{aadilemma}]
  \label{lem:healing}
  Consider a $(\frac{1}{3},3\Delta)$-compliant execution of \Goldfish in the partially synchronous sleepy network model of \secref{model}.
  Then, $\chainava$ is secure with transaction confirmation time $\Tcheckpoint + \Ttimeout + \Tbft = \Theta(\kappa)$ after round $\max(\GAT,\GST) + \Delta + 2\Tcheckpoint$.

  Moreover, for the iteration proposal $\bprop{c}$ of an honest iteration leader broadcast at round $r$, it holds that $\bprop{c} \preceq \varBlock^{\lceil \kappa}$ for any $\varBlock$ identified in \myalgref{protocol}{fcr-1,fcr-2,fcr-3} by any awake honest validator after $r$, and $\bprop{c}$ contains a fresh honest block that is not in the prefix of any checkpoint from before iteration $c$.
\end{lemma}

Proof of \lemref{healing} is different from the proof of \cite[Thm.~5]{aadilemma} since the accountability gadget is applied to a longest chain protocol in \cite{aadilemma}, whereas it is applied to \Goldfish in our case.
Therefore, the full proof is presented below.
\begin{proof}
  By \cite[Thm.~3]{aadilemma}, $\chainacc$ provides accountable safety with resilience $n/3$ except with probability $\negl(\lambda)$ in the partially synchronous sleepy network model.
  As the execution is $(\frac{1}{3},3\Delta)$-compliant, w.o.p., no two checkpoints observed by awake honest validators conflict.

  Let $c$ be the largest iteration such that a block $\varBlock$ was checkpointed in the view of some honest validator before $\max(\GAT,\GST)$.
  (Let $c=0$ and $\varBlock$ be the genesis block if there does not exist such an iteration.)
  Then, by \propref{entering-iter-after-gst}, if an honest validator enters an iteration $c' > c$ at some round $r \geq \max(\GAT,\GST) + \Delta + \Tcheckpoint$, every honest validator enters iteration $c$ by round $r + \Delta$.
  Let $c'$ be the first iteration such that the first honest validator to enter $c'$ enters it after round $\max(\GAT,\GST) + \Delta + \Tcheckpoint$ (\eg, at some round $r$ such that $\max(\GAT,\GST) + \Delta + \Tcheckpoint < r < \max(\GAT,\GST) + \Delta + 2\Tcheckpoint$).
  Then, all honest validators enter iteration $c'$ and agree on the last checkpointed block within $\Delta$ rounds.
  Subsequently, the honest validators wait for $\Tcheckpoint$ rounds before casting any gadget vote for a checkpoint proposal of iteration $c'$, during which no block can be checkpointed (\propref{checkpoint-gap}, gap property).

  By \lemref{honest-leader-almost-everywhere}, w.o.p., the slot interval of length $\kappa$ starting after round $r+\Delta$ contains a slot $t$ with an honest leader and \proposal $\varProposal^*$.
  After round $r \geq \GST$, all messages broadcast by honest validators are received by all honest validators within $\Delta$ rounds.
  As honest validators agree on the last checkpointed block during the interval $[r+\Delta,r+\Tcheckpoint]$, by the absence of new checkpoints, the \textOurGhost fork-choice rule starts at the same last checkpointed block for all honest validators during the interval (\myalgref{ghost}{cgenesis}).
  Then, by \lemref{honest-leader-almost-everywhere}, w.o.p., $\varProposal^*.\varBlock \preceq \varBlock$ for any $\varBlock$ identified in \myalgref{protocol}{fcr-1,fcr-2,fcr-3} by any awake honest validator in any round after $3\Delta t + 2\Delta$, until at least a new block is checkpointed in the view of an honest validator.

  By \lemref{checkpoint-recency} (recency property), the next block checkpointed in the view of an honest validator (which happens earliest at some iteration $c'' \geq c'$ and round $r' \geq r+\Tcheckpoint$ by \propref{checkpoint-gap}, the gap property) must have been confirmed by some honest validator $\id$ at some round within $[r'-\Trecent,r']$, where $r'-\Trecent \geq r + 6\Delta\kappa + 4\Delta$.
  Hence, the new checkpointed block is $\kappa$ slots deep in the chains identified in \myalgref{protocol}{fcr-1,fcr-2,fcr-3} by $\id$, and is a descendant of $\varProposal^*.\varBlock$.
  This implies $\varProposal^*.\varBlock \preceq \varBlock$ for any $\varBlock$ identified in \myalgref{protocol}{fcr-1,fcr-2,fcr-3} by any awake honest validator in any round after $3\Delta t + 2\Delta$ \emph{indefinitely}.

  Note that if the iteration leader was honest, for its proposal $\bprop{c}$ broadcast at some round $r''$, it holds that $\bprop{c} \preceq \varBlock^{\lceil \kappa}$ for any $\varBlock$ identified in \myalgref{protocol}{fcr-1,fcr-2,fcr-3} by any awake honest validator after round $r$.
  Moreover, $\varProposal^*.\varBlock \preceq \bprop{c}$, implying that honest checkpoint proposals contain fresh honest blocks in their prefixes.

  Finally, we extend the above argument to future checkpoints by induction.
  Let $\varBlock_n$ denote the sequence of checkpointed blocks, ordered by their iteration numbers $c_n \geq c'$, $c_1 = c''$.
  The rounds $r_n$, at which the blocks $\varBlock_n$ are first checkpointed in the view of an honest validator satisfy the relation $r_{n+1} \geq r_n + \Tcheckpoint$ and $r_1 = r''$.
  Via the inductive assumption and the reasoning above, w.o.p., in each interval $[r_n+\Delta,r_{n+1}-\Trecent]$, there exists a slot $t_n$ with an honest leader and \proposal $\varProposal_n$ such that $\varProposal_n.\varBlock \preceq \varBlock$ for any $\varBlock$ identified in \myalgref{protocol}{fcr-1,fcr-2,fcr-3} by any awake honest validator in any round after $3\Delta t_n + 2\Delta$ \emph{indefinitely}.
  Hence, for a sufficiently large confirmation time exceeding the maximum possible iteration length (\ie, $\Tconf \geq \Tcheckpoint + \Ttimeout + \Tbft$), these honest blocks imply the security of the \Goldfish protocol after round $\max(\GAT,\GST) + \Delta + 2\Tcheckpoint$.
\end{proof}
\thmref{strong-persistence-1} holds for the honest blocks proposed in intervals $[r_n+\Delta,r_{n+1}-\Trecent]$ as all honest validators agree on the latest checkpoint during these intervals.

\subsection{Proof of \thmref{ebb-and-flow-formalized}}
\label{sec:appendix-ebb-and-flow-security}
\begin{proof}[Proof of \thmref{ebb-and-flow-formalized}]
  We first show the property \textbf{P1}, namely, the accountable safety and liveness of the accountable, final prefix ledger $\chainacc$ under partial synchrony in the sleepy model.
  By \cite[Thm.~3]{aadilemma}, $\chainacc$ provides accountable safety with resilience $n/3$ except with probability $\negl(\lambda)$ under partial synchrony in the sleepy model.
  By \lemref{healing}, under the same model, the available ledger $\chainava$ is secure after round $\max(\GAT,\GST) + \Delta + 2\Tcheckpoint$.
  Using this fact and \lemref{liveness-lemma}, we can state that, w.o.p., $\chainacc$ satisfies liveness after round $\max(\GAT,\GST) + \Delta + 2\Tcheckpoint$ with transaction confirmation time $\Tconf = \Theta(\kappa^2)$.

  Finally, the property \textbf{P2} follows from \lemref{p2}, and \textbf{Prefix} follows by construction of the ledgers $\chainacc$ and $\chainava$.
  This concludes the proof of the ebb-and-flow property.
\end{proof}

\section{Equivocation Discounting to Mitigate Spamming}
\label{sec:appendix-equivocation-discounting}

For ease of exposition, we have presented a version of \Goldfish which deals with equivocating votes simply by accepting all of them, but counting at most one per subtree (\myalgref{ghost}{vote-count}).
This approach is vulnerable to spamming attacks, because it requires validators to accept all the votes they receive.
Even a single adversarially controlled validator can be used to create an arbitrarily large number of equivocating votes at a slot, with the goal of creating network congestion and making it impossible for honest validators to download all of the other votes in time, which can result in a loss of safety.

Equivocations are attributable faults, punishable by slashing \emph{a posteriori}, but this does not prevent the attack vector \emph{a priori}, given that only one validator is required for it. %
To mitigate it, we introduce equivocation discounting.
This general technique is already present in the current implementation of Ethereum, but the ephemerality of votes in \Goldfish allows for a simpler rule, with clear bounds on the number of messages required for honest views to converge.
This is particularly important in order to have guarantees about the functioning of the vote buffering technique, and in turn about the security of the whole protocol, which relies on reorg resilience.
We formalize the simple equivocation discounting rule here, as a combination of a modification to the \textOurGhost fork-choice, a download rule, and a validity condition for proposals.

\myparagraph{Equivocation Discounting}
\begin{enumerate}[(a)]
    \item \textbf{Fork-choice discounting:}\ \ When running the \textOurGhost fork-choice rule at slot $t$,, only count the valid slot $t-1$ votes from those validators for which your \bvtree contains a single valid slot $t-1$ vote, \ie, those which are not viewed to have equivocated at slot $t-1$.

    \item \textbf{Download rule:}\ \ Only download (or forward as part of the peer-to-peer gossip layer) votes from the current and prior slots, and at most two votes per eligible validator (\ie, the opened ticket $(\id, t)$ for the validator $\id$ is winning for the tag $(\tagLotteryVote,\thresholdVote)$, \cf \secref{analysis}).

    \item \textbf{Validity condition for proposals:}\ \ A \proposal whose \bvtree contains more than two valid votes for the same slot from some validator is invalid, and so is one which contains any invalid vote.
\end{enumerate}
The download rule and validity condition ensure that a validator only ever needs to download at most two votes per subsampled validator of the current and previous slot.
Setting the subsampling parameters so that this is manageable, we can ensure that equivocations cannot succeed at creating network congestion sufficient to prevent the functioning of vote buffering.
Previously, this meant guaranteeing that an honest proposer's \bvtree be a superset of honest validators' \bvtrees.
Instead, the success of vote buffering now only requires that a leader's view of votes from voters which have not equivocated in the last slot is a superset of the validators' views of such votes, and so is its view of the list of equivocators from the previous slot.
Agreement on these two is sufficient for agreement on the fork-choice output, \ie, \lemref{view-merge-property} still holds.
Note that the leader still only needs to include its \bvtree in the proposal message, because following the download rule guarantees that it will contain exactly all valid votes from validators which have not equivocated in the previous slot, together with a pair of votes, \ie, equivocation evidence, for validators which have.

The security analysis for \Goldfish with equivocation discounting is then the same as that for vanilla \Goldfish.
Vote buffering implies that all honest validators vote together when the \proposal with the minimum precedence is honest, as in \lemref{view-merge-property}, and all honest validators voting together implies that the proposal is never reorged, as in \lemref{all-honest-voting-together}.
The latter is not affected by equivocation discounting, because it relies on the valid votes of honest validators, which do not equivocate.
From these two properties, we obtain reorg resilience as in \thmref{reorg-resilience}, and from reorg resilience, we eventually obtain safety and liveness.

Optimistic fast confirmations are also compatible with equivocation discounting, without any loss of resilience.
Liveness and fast confirmation of honest proposals follow from \thmref{fast-confirmations-when-enough-honest}, since equivocation discounting plays no role in it.
For safety, the key ingredient is \lemref{fast-base-case}, from which \thmref{fast-confirmation-safety} follows unchanged.
We thus prove \lemref{fast-base-case} here for \Goldfish with equivocation discounting, by making a very small modification to the argument:

\begin{proof}[Proof of \lemref{fast-base-case} with equivocation discounting]
    By \propref{committee-bound-whp}, w.o.p., the number of adversary validators at round $4\Delta(t+1)+\Delta$, eligible to vote at slot $t$, is less than $\frac{1}{2}n\,\thresholdVote$.
    An eligible awake honest validator sends a single slot $t$ vote at round $4\Delta t+\Delta$, implying that over $(\frac{3}{4}+\frac{\epsilon}{2}) n\,\thresholdVote - \frac{1}{2} n\,\thresholdVote = (\frac{1}{4} + \frac{\epsilon}{2}) n\,\thresholdVote$ validators broadcast a single slot $t$ vote by round $4\Delta (t+1)+\Delta$, and that is for a descendant of $\varBlock$.
    By \propref{committee-bound-whp}, w.o.p., for all slots $t$, there can be at most $(1+\epsilon)n\,\thresholdVote$ validators that are eligible to vote at $t$.
    Hence, the number of valid slot $t$ votes for the descendants of any block $\varBlock'$ conflicting with $\varBlock$, and which are from validators which have not also cast one of the $(\frac{3}{4} + \frac{\epsilon}{2})n\,\thresholdVote$ votes for $B$, must be less than $(1+\epsilon)n\,\thresholdVote - (\frac{3}{4} + \frac{\epsilon}{2})n\,\thresholdVote = (\frac{1}{4} + \frac{\epsilon}{2}) n\,\thresholdVote $ at any given round.
    The validator $\id^*$ broadcasts $\varBlock$ and over $(\frac{3}{4}+\frac{\epsilon}{2}) n\,\thresholdVote$ valid votes for it (in \pieces) at round $4\Delta t+2\Delta$.
    Each honest validator, awake at round $4\Delta (t+1) + \Delta$ and eligible to vote at slot $t+1$, observes these votes in its \bvtree at the round of voting (\myalgref{fast-protocol}{fast-view-merge2}).
    Upon invoking the \textOurGhost fork-choice rule at any of the rounds $4\Delta t + 3\Delta$, $4\Delta (t+1)$ or $4\Delta (t+1) + \Delta$, using only the votes from validators which are not seen to be equivocating at slot $t-1$, the votes for the descendants of any block $\varBlock'$ conflicting with $\varBlock$ are then less than $(\frac{1}{4} + \frac{\epsilon}{2}) n\,\thresholdVote$, and the votes for descendants of $\varBlock$ are over $(\frac{1}{4} + \frac{\epsilon}{2}) n\,\thresholdVote$.
    This implies that all honest validators, awake at round $4\Delta(t+1)+\Delta$ and eligible to vote at slot $t+1$, all vote for $\varBlock$ or one of its descendants at slot $t+1$.
\end{proof}

\section{From LMD GHOST to \Goldfish}
\label{sec:comparison}

In this section, we outline the shortcomings of LMD GHOST in comparison to \Goldfish, then discuss how \Goldfish could replace it in the Ethereum protocol.

\subsection{Limitations of Gasper}
\label{sec:unavailability-interaction}

In the first iteration of Gasper's LMD GHOST, ex-ante reorgs and balancing attacks~\cite{ethresearch-balancing-attack,ethresearch-balancing-attack2,3attacks} prevent security even in the full participation setting and without subsampling. The proposer boost technique~\cite{mitigationlmdghostbalancingattacks} mitigates these issues, but is itself not compatible with dynamic participation, and it entails a lower adversary tolerance ($\frac{1}{4}$) than what is obtained with message buffering ($\frac{1}{2}$). Moreover, ex-ante reorgs~\cite{3attacks} are still possible with subsampling, compromising reorg resilience, and the latest message rule (LMD) itself is not compatible with dynamic participation. Both of these issues are due to considering votes from older slots, and \Goldfish solves them through vote expiry. In the following, we give a more detailed account of all of these limitations.

\myparagraph{Interaction of LMD GHOST and Casper~FFG}
The combination of \Goldfish with the accountability gadget in \secref{protocol} follows the generic construction of \cite{aadilemma}, which is proven to be secure for any appropriately secure dynamically available protocol and accountable BFT protocol.
On the other hand, the combination of LMD GHOST and Casper~FFG in HLMD GHOST, the hybrid fork-choice rule of \cite{gasper}, is ad-hoc and complicated to reason about.
Firstly, it is known to be susceptible to a \emph{bouncing attack}~\cite{ethresearch-bouncing-attack-analysis}.
Instead of LMD GHOST starting its fork-choice iteration from the last block \emph{finalized} by Casper~FFG, it starts from the last \emph{justified} block, in the terminology of Casper~FFG, \ie, the last block that has been the target of FFG votes by a supermajority of all $n$ validators.
This is sufficient to ensure accountable safety of the finalized checkpoints; however, it hinders safety of the available ledger $\chainava$ (after $\max(\GST,\GAT)$) under partial synchrony in the sleepy model, in particular negating the healing property (\fullVersionRef{\lemref{healing}}) of $\chainava$, preventing us from proving the ebb-and-flow property.
The current mitigation for the bouncing attack causes other problems such as the splitting attack~\cite{ethresearch-bouncing-attack-prevention}, akin to the balancing attacks~\cite{ebbandflow}.
Another problematic interaction stems from the fact that the FFG votes at any Ethereum epoch point at the epoch boundary block of that epoch, regardless of its confirmation status by the underlying LMD GHOST rule.
(In fact, there is no confirmation rule specified for LMD GHOST.) The accountability gadget can then in principle interfere with the available chain,  jeopardizing its standalone security properties.
Finally, the FFG voting schedule is staggered throughout an epoch, as FFG votes are cast together with LMD GHOST votes, so it is not clear how to ensure that the views of honest validators when casting FFG votes are consistent, which would at least ensure liveness of the accountable chain.

\myparagraph{Stale Votes in LMD GHOST}
Without vote expiry, the votes of honest asleep validators can be weaponized by an adversary controlling a small fraction of the validator set to execute an arbitrarily long reorg. This implies that the protocol is not dynamically available with \emph{any} confirmation rule with finite confirmation time $\Tconf$.
Consider for example a validator set of size $n = 2m+1$, and a partition of the validator set into three sets, $V_1$, $V_2$, $V_3$, with $|V_1| = |V_2| = m$ and $|V_3| = 1$.
The validators in $V_1$, $V_2$ are all honest, while the one in $V_3$ is adversary. Suppose that the adversary validator in $V_3$ is the leader of slots $t$, and that it broadcasts two proposals, with conflicting blocks $\varBlock_1$ and $\varBlock_2$. It does so in such a way that validators in $V_1$ see only $\varBlock_1$ before voting, and validators in $V_2$ only $\varBlock_2$. Validators in $V_1$ then vote for $\varBlock_1$, and so does the adversary validator, while validators in $V_2$ vote for $\varBlock_2$. $\varBlock_1$ becomes canonical, since it has received $m+1$ votes. The adversary then puts all validators in $V_2$ to sleep, and they do not become awake for the remainder of the protocol. The adversary validator does not cast any more votes for a while. Meanwhile, validators in $V_1$, keep voting for descendants of $\varBlock_1$.
After waiting for $>\Tconf$ slots, the adversary validator votes for $\varBlock_2$. Since the $m$ latest votes of the validators in $V_2$ are still for $\varBlock_2$, it now has $m+1$ votes and becomes canonical, resulting in all awake honest validators experiencing a reorg of all blocks confirmed after slot $t$. If there are no such blocks, liveness is violated, and otherwise safety is violated.

\myparagraph{Proposer Boost}
Proposer boost is not compatible with dynamic
participation, because the artificial fork-choice weight it temporarily provides to proposals is independent of participation: the lower the participation, the more powerful the boost is relative to the weight of real attestations from awake validators, and thus the more it can be exploited by the adversary.
When the weight of awake honest validators is less than the boost, the adversary has complete control of the fork-choice during the slots in which it is elected as the leader.

\myparagraph{Reorg Resilience}
Even in the setting of full participation, where the adversary cannot take advantage of votes of asleep validators, LMD GHOST lacks reorg resilience.
This is firstly due to subsampling without vote expiry, because it allows the adversary to accumulate fork-choice weight by withholding blocks and attestations, \ie, to execute ex ante reorgs~\cite{3attacks}.
Without subsampling, LMD GHOST is indeed reorg resilient in the full participation setting, \emph{if proposer boost is replaced by vote buffering}.
In fact, \thmref{reorg-resilience} obtains reorg resilience as a consequence of two properties, \lemref{view-merge-property,all-honest-voting-together}, respectively the properties that all honest awake validators vote for an honest \proposal, and all honest validators voting together guarantee the inclusion of honest blocks in the canonical \textOurGhost chain, both of which also hold for LMD GHOST with vote buffering.

With proposer boost, LMD GHOST is not reorg resilient for $\beta \geq \frac{1}{4}$, even in the full participation setting and without subsampling, because those two properties are in conflict for such $\beta$, for any boost value $\Wp$.
The first property only holds if $\Wp > 2\beta$, because the adversary can otherwise still conclude an ex ante reorg by revealing later votes, which move all adversary weight $\beta$ from the proposer's branch to a conflicting one, and outweigh the proposer boost $\Wp$.
On the other hand, the second property only holds if $\Wp + \beta < 1-\beta$, because otherwise an adversary proposer can make use of boost to conclude an ex post reorg.
Therefore, we can only have reorg resilience when $3\beta <  \Wp + \beta < 1 - \beta$, \ie, for $\beta < \frac{1}{4}$, by setting $\Wp = \frac{1}{2}$.

\subsection{Replacing LMD GHOST with \Goldfish in Gasper}
\label{sec:from-lmdghost-to-goldfish}

For \Goldfish to be used as a drop-in replacement for LMD GHOST in Ethereum, only a few adjustments are required. Most importantly, vote expiry and message buffering would have to be introduced, with the latter replacing proposer boost.
In principle, the proposer selection mechanism does not need to be overhauled, as \Goldfish can operate with RANDAO, the proposer selection mechanism of LMD GHOST. RANDAO always selects a unique proposer, which reduces the communication load, when compared to a VRF lottery. On the other hand, it is not compatible with adaptive security, because the selected proposer is publicly known in advance, and moreover the selection is biasable. A VRF lottery also enables
the confirmation time to be independent of participation.

Finally, in order to benefit from the security guarantees of \Goldfish in its combination with an accountability gadget, the interaction with Casper~FFG would have to be modified to fit the construction from~\cite{aadilemma}, which we have also employed in this work.

\section{Cryptographic Preliminaries}
\label{sec:crypto-details}

\subsection{Digital Signatures}
\label{sec:crypto-details-boilerplate-sigs}
\begin{definition}[Informal, \cf~\cite{DBLP:books/crc/KatzLindell2014,boneh-shoup-book}]%
      A \emph{signature} scheme $\sigSCHEME = (\sigGenBLANK, \sigSignBLANK, \sigVerifyBLANK)$ consists of probabilistic poly-time (PPT) algorithms so that:
      \begin{itemize}
            \item $(\sigSk, \sigPk) \gets \sigGen(1^\lambda)$
                  creates a secret/public key pair.
            \item $\sigma \gets \sigSign(\sigSk, m)$
                  creates a signature on a message.
            \item $\{\FALSE, \TRUE\} \gets \sigVerify(\sigPk, m, \sigma)$
                  verifies a signature.
            \item \textbf{Correctness:}
                  With overwhelming probability, for all messages,\\
                  $\sigVerify(\sigPk, m, \sigSign(\sigSk, m)) = \TRUE$.
            \item \textbf{Security (existential unforgeability):}
                  An adversary with access to $\sigPk$ and
                  to a signing oracle $\sigSign(\sigSk, .)$
                  cannot produce a valid $(m, \sigma)$
                  other than via the oracle.
      \end{itemize}
\end{definition}

\subsection{Verifiable Random Functions}
\label{sec:crypto-details-boilerplate-vrfs}
A verifiable random function (VRF)~\cite{vrf} is used for leader election and subsampling of the validators within the \Goldfish protocol.
\begin{definition}[Informal, \cf~{\cite[Sec.~3.2, Fig.~2]{david2018ouroboros}, \cite{DBLP:conf/pkc/DodisY05,algorand}}]%
      A \emph{verifiable random function} (VRF) scheme $\vrfSCHEME = (\vrfGenBLANK, \vrfProveBLANK, \vrfVerifyBLANK)$ consists of PPT algorithms so that:
      \begin{itemize}
            \item $(\vrfSk, \vrfPk) \gets \vrfGen(1^\lambda)$
                  samples a VRF with associated secret/public key pair for evaluation/verification.
            \item $(y, \pi) \gets \vrfProve(\vrfSk, x)$
                  obtains the output $y$ of the VRF at input $x$,
                  and the evaluation proof $\pi$.
            \item $\{\FALSE, \TRUE\} \gets \vrfVerify(\vrfPk, x, (y, \pi))$
                  verifies an evaluation.
            \item \textbf{Correctness:}
                  With overwhelming probability, for all inputs,
                  \\
                  $\vrfVerify(\vrfPk, x, \vrfProve(\vrfSk, x)) = \TRUE$.
            \item \textbf{Uniqueness:}
                  Per input $x$, there is only one output $y$:
                  if
                  \\
                  $\vrfVerify(\vrfPk, x, (y, \pi)) = \TRUE$
                  for $(y, \pi) = (y_1, \pi_1)$
                  and $(y, \pi) = (y_2, \pi_2)$,
                  then $y_1 = y_2$.
            \item \textbf{`Pseudorandomness':}
                  Conceptually,
                  the VRF behaves like a random oracle
                  that is \emph{unpredictable} (\ie,
                  without knowledge of $\vrfSk$,
                  the VRF output cannot be distinguished
                  from a random string)
                  and \emph{verifiable} (\ie, given $\vrfPk$,
                  an alleged output of the VRF can be verified).
                  For a formal definition, see
                  \cite[Sec.~3.2, Fig.~2]{david2018ouroboros}.

      \end{itemize}
\end{definition}

\end{document}